\documentclass[aps,prx,reprint,groupedaddress,superscriptaddress,floatfix]{revtex4-2}

\usepackage[T1]{fontenc}
\usepackage{newtxtext}
\usepackage{newtxmath}

\usepackage{amsmath}
\usepackage{mathtools}
\usepackage{amsfonts}
\usepackage{bbold}

\usepackage{color}
\usepackage{xcolor}
\usepackage[colorlinks,citecolor=blue,linkcolor=blue,urlcolor=blue]{hyperref}

\usepackage{tabularx}
\usepackage{makecell}

\usepackage{graphicx}
\usepackage[labelfont=bf,justification=Justified]{caption}
\usepackage[all]{hypcap}

\usepackage{enumitem}

\makeatletter
\def\@ssect@ltx#1#2#3#4#5#6[#7]#8{%
  \def\H@svsec{\phantomsection}%
  \@tempskipa #5\relax
  \@ifdim{\@tempskipa>\z@}{%
    \begingroup
      \interlinepenalty \@M
      #6{%
       \@ifundefined{@hangfroms@#1}{\@hang@froms}{\csname @hangfroms@#1\endcsname}%
       {\hskip#3\relax\H@svsec}{#8}%
      }%
      \@@par
    \endgroup
    \@ifundefined{#1smark}{\@gobble}{\csname #1smark\endcsname}{#7}%
  }{%
    \def\@svsechd{%
      #6{%
       \@ifundefined{@runin@tos@#1}{\@runin@tos}{\csname @runin@tos@#1\endcsname}%
       {\hskip#3\relax\H@svsec}{#8}%
      }%
      \@ifundefined{#1smark}{\@gobble}{\csname #1smark\endcsname}{#7}%
      \addcontentsline{toc}{#1}{\protect\numberline{}#8}%
    }%
  }%
  \@xsect{#5}%
}%
\makeatother

\begin{document}

\title{Quantum-Enhanced Greedy Combinatorial Optimization Solver}

\author{Maxime Dupont}
\email[Corresponding author:~]{mdupont@rigetti.com}
\affiliation{Rigetti Computing, Berkeley, California 94710, USA}

\author{Bram Evert}
\affiliation{Rigetti Computing, Berkeley, California 94710, USA}

\author{Mark J. Hodson}
\affiliation{Rigetti Computing, Berkeley, California 94710, USA}

\author{Bhuvanesh Sundar}
\affiliation{Rigetti Computing, Berkeley, California 94710, USA}

\author{Stephen Jeffrey}
\affiliation{Rigetti Computing, Berkeley, California 94710, USA}

\author{Yuki Yamaguchi}
\affiliation{Rigetti Computing, Berkeley, California 94710, USA}

\author{Dennis Feng}
\affiliation{Rigetti Computing, Berkeley, California 94710, USA}

\author{Filip B. Maciejewski}
\affiliation{QuAIL, NASA Ames Research Center, Moffett Field, California 94035, USA}
\affiliation{USRA Research Institute for Advanced Computer Science, Mountain View, California 94035, USA}

\author{Stuart Hadfield}
\affiliation{QuAIL, NASA Ames Research Center, Moffett Field, California 94035, USA}
\affiliation{USRA Research Institute for Advanced Computer Science, Mountain View, California 94035, USA}

\author{M. Sohaib Alam}
\affiliation{QuAIL, NASA Ames Research Center, Moffett Field, California 94035, USA}
\affiliation{USRA Research Institute for Advanced Computer Science, Mountain View, California 94035, USA}

\author{Zhihui Wang}
\affiliation{QuAIL, NASA Ames Research Center, Moffett Field, California 94035, USA}
\affiliation{USRA Research Institute for Advanced Computer Science, Mountain View, California 94035, USA}

\author{Shon Grabbe}
\affiliation{QuAIL, NASA Ames Research Center, Moffett Field, California 94035, USA}

\author{P. Aaron Lott}
\affiliation{QuAIL, NASA Ames Research Center, Moffett Field, California 94035, USA}
\affiliation{USRA Research Institute for Advanced Computer Science, Mountain View, California 94035, USA}

\author{Eleanor G. Rieffel}
\affiliation{QuAIL, NASA Ames Research Center, Moffett Field, California 94035, USA}

\author{Davide Venturelli}
\affiliation{QuAIL, NASA Ames Research Center, Moffett Field, California 94035, USA}
\affiliation{USRA Research Institute for Advanced Computer Science, Mountain View, California 94035, USA}

\author{Matthew J. Reagor}
\email[Corresponding author:~]{matt@rigetti.com}
\affiliation{Rigetti Computing, Berkeley, California 94710, USA}

\begin{abstract}
    Combinatorial optimization is a broadly attractive area for potential quantum advantage, but no quantum algorithm has yet made the leap. Noise in quantum hardware remains a challenge, and more sophisticated quantum-classical algorithms are required to bolster their performance. Here, we introduce an iterative quantum heuristic optimization algorithm to solve combinatorial optimization problems. The quantum algorithm reduces to a classical greedy algorithm in the presence of strong noise. We implement the quantum algorithm on a programmable superconducting quantum system using up to 72 qubits for solving paradigmatic Sherrington-Kirkpatrick Ising spin glass problems. We find the quantum algorithm systematically outperforms its classical greedy counterpart, signaling a quantum enhancement. Moreover, we observe an absolute performance comparable with a state-of-the-art semidefinite programming method. Classical simulations of the algorithm illustrate that a key challenge to reaching quantum advantage remains improving the quantum device characteristics.
\end{abstract}

\maketitle

A handful of promising classes of quantum algorithms have been advanced for combinatorial optimization problems, such as quantum adiabatic evolution algorithms~\cite{Farhi2001}, variational quantum algorithms~\cite{Farhi2014,Farhi2014b,Cerezo2021}, as well as others~\cite{PhysRevResearch.2.013056,Alexandru_2020}. In all cases, the problem takes the form of an objective function to extremize, which can be interpreted as an Ising-type Hamiltonian whose ground state is the global extremum of the problem. Solving a generic Ising model is NP-hard~\cite{Barahona1982}, and it remains open whether quantum computers can indeed provide a practical advantage over classical methods. A tremendous amount of work has been dedicated to quantum annealers, which leverage adiabatic evolution~\cite{Boixo2014,Ronnow2014,PhysRevX.5.031040}, while the focus for gate-based quantum computers has been mainly on parameterized quantum circuits like the quantum approximate optimization algorithm (QAOA)~\cite{Farhi2014,Farhi2014b,Farhi2016,Hadfield2019,Otterbach2017,Pagano2020,Harrigan2021}.

The QAOA has been implemented on several experimental platforms to solve a range of combinatorial optimization problems~\cite{Otterbach2017,Pagano2020,Harrigan2021,ebadi2022quantum,graham2021demonstration,Pelofske2023,moses2023race,Shaydulin2023,Sack2023,Maciejewski2023}. However, a major challenge in these demonstrations has been the stringent technical requirement of reducing hardware noise in order to provide good quality solutions that are well-separated from trivial classical approaches such as random sampling. In particular, loading arbitrary graph problems, beyond the native topology of the quantum computer, often demands an additional overhead, increasing noise and lowering the performance of the quantum algorithm. For example, while an early implementation of the QAOA on a superconducting quantum system for solving hardware-native maximum cut problems on $19$ qubits already showed performance better than random sampling~\cite{Otterbach2017}, an implementation for solving high-dimensional graph problems beyond the hardware-native topology on $23$ qubits found results only as good as random guessing on contemporary devices~\cite{Harrigan2021}, with similar trends for Rydberg atoms and trapped ions~\cite{ebadi2022quantum,Pagano2020,graham2021demonstration,zhu2022multi,Shaydulin2023}. Yet, it is unavoidable for quantum computers to tackle more intricate problem instances en-route to solving universal and real-world problems.

Indeed, to date, noise has made even the most straightforward classical optimization approaches better candidates for solving real-world optimization. For instance, classical greedy algorithms, which iteratively build a solution by making the locally optimal choice at each stage, are intuitive, easy to implement, and will most likely outperform a modern noisy quantum computer. This raises the question: Can one design algorithms utilizing current quantum technologies to their advantage with performance guarantees, making them realistically competitive against classical ones for arbitrary problems at scale?

\begin{figure*}[!ht]
    \centering
    \includegraphics[width=1.0\textwidth]{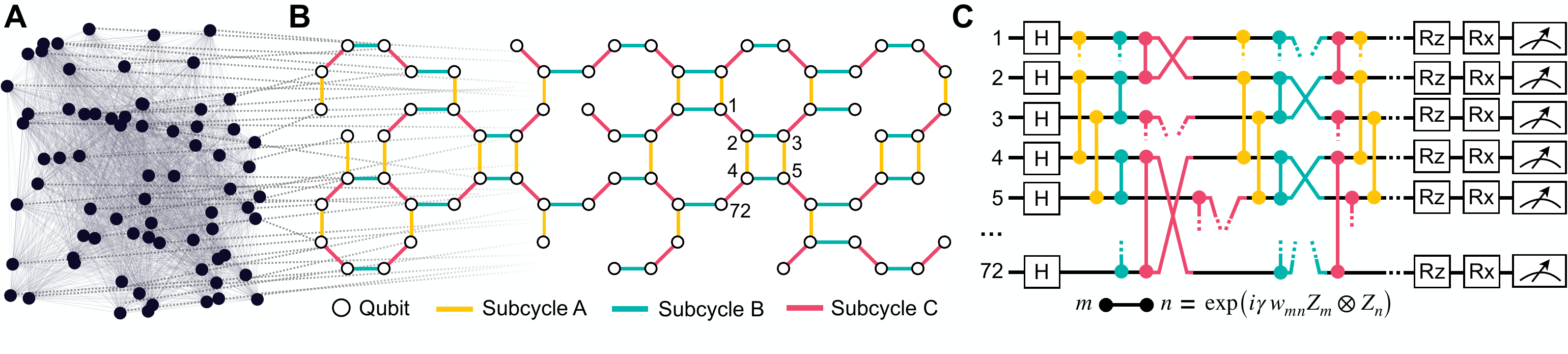}
    \caption{\textbf{Problem mapping and quantum circuit.} \textsf{\textbf{A.}} Binary optimization problem with $N=72$ variables [Eq.~\eqref{eq:objective_f}]. \textsf{\textbf{B.}} Binary variables are randomly mapped to a qubit of the square-octagon topology of the Rigetti Aspen-M-3 quantum processor. The edges are divided into three independent sets $A$, $B$, and $C$ on which two-qubit gates can be executed in parallel. \textsf{\textbf{C.}} QAOA quantum circuit [Eq.~\eqref{eq:circuit_qaoa}]. We show a truncated version run on hardware: We apply $R_{zz}$ gates on the yellow edges in subcycle A, then $R_{zz}$ gates on the teal edges in subcycle B, then $R_{zz}$ and SWAP gates on the magenta edges in subcycle C, and so on as shown above.}
    \label{fig:introduction}
\end{figure*}

Overall, our goal is to minimize the objective function,
\begin{equation}
    C=u+\sum\nolimits_{i=1}^N v_i Z_i+\sum\nolimits_{j<i}^N w_{ij}Z_i Z_j,
    \label{eq:objective_f}
\end{equation}
where $u$, $v_i$, and $w_{ij}$ are problem-specific scalar parameters, and $Z_i\in\{-1,+1\}$ are Ising spin variables with corresponding bit values $B_i=1/2-Z_i/2\in\{0,1\}$. The goal is to find a bit string $\mathbf{B}=(B_1,B_2,\ldots,B_N)$ minimizing Eq.~\eqref{eq:objective_f}. Substituting the Ising variables for Pauli $Z$ operators (we use the same symbol for both when clear from context) in Eq.~\eqref{eq:objective_f}, the minimization can be achieved on a quantum computer using quantum adiabatic evolution or variational algorithms. The latter, such as the QAOA~\cite{Farhi2014,Farhi2014b}, are better suited for near-term digital devices with their theoretical performance at least on par with a discretized adiabatic evolution with a fixed number of layers. A QAOA circuit with $p$ layers reads,
\begin{equation}
    \bigl\vert{\boldsymbol{\gamma},\boldsymbol{\beta}}\bigr\rangle=\left[\prod\nolimits_{d=1}^pU_\textrm{M}(\beta_d)U_\textrm{PS}(\gamma_d)\right]H^{\otimes N}\vert{0}\rangle^{\otimes N},
    \label{eq:circuit_qaoa}
\end{equation}
where $H$ is the one-qubit Hadamard gate. The unitaries $U_\textrm{PS}$ and $U_\textrm{M}$ are called the phase separator and the mixer, respectively, and are parametrized with the real-valued angles $\gamma_d$ and $\beta_d$, respectively. They are defined as,
\begin{equation}
    U_\textrm{PS}(\gamma_d)=e^{i\gamma_d C},~~~\textrm{and}~~~~U_\textrm{M}(\beta_d)=e^{i\beta_d\sum\nolimits_{j=1}^NX_j},
    \label{eq:definition_unitaries}
\end{equation}
where $C$ is the operator corresponding to the objective function of Eq.~\eqref{eq:objective_f} and where $X_j$ is the Pauli operator on qubit $j$. The search for optimal angles $\boldsymbol{\gamma}^*$ and $\boldsymbol{\beta}^*$ is done in a quantum-classical hybrid fashion by minimizing the expectation value $\langle{C}\rangle=\langle{\boldsymbol{\gamma},\boldsymbol{\beta}}\vert{C}\vert{\boldsymbol{\gamma},\boldsymbol{\beta}}\rangle$. Candidate bit string solutions $\{\mathbf{B}\}$ are obtained by sampling the quantum state $\vert{\boldsymbol{\gamma}^*,\boldsymbol{\beta}^*}\rangle$.

The quality of a solution is ranked by its cost value $C^*$ with respect to the maximum (worst) and minimum (optimal) ones, $C_\textrm{max}$ and $C_\textrm{min}$, respectively, through the approximation ratio,
\begin{equation}
    r=\bigl(C_\textrm{max}-C^*\bigr)\Bigr/\bigl(C_\textrm{max}-C_\textrm{min}\bigr),
    \label{eq:approximation_ratio}
\end{equation}
which is equal to $1$ for an optimal solution $C^*=C_\textrm{min}$. On difficult problem instances, it is proven to be NP-hard to achieve an ensemble-average approximation ratio greater than a given value $r^\star$, and intensive ongoing theoretical efforts~\cite{Farhi2014,Farhi2014b,Farhi2016,Farhi2017,PhysRevA.97.022304,Marwaha2021localclassicalmax,PhysRevA.103.042612,basso_et_al,Farhi2022quantumapproximate,marwaha2022bounds} are attempting to establish whether QAOA with $p$ layers can lead to an average approximation ratio larger than the best-known classical methods on some problem classes. Because of noise, when QAOA is run on current generation (noisy) quantum hardware, it leads to approximation ratios much smaller than theoretical bounds and those obtained on small-scale classical emulations~\cite{Otterbach2017,Pagano2020,Harrigan2021}. In fact, in a strong noise regime where the quantum state tends to be described as a maximally mixed state, the expected performance is that of a random bit string sampling. Under the assumption that the spectrum of $C$ [Eq.~\eqref{eq:objective_f}] is symmetric about $u$, this corresponds to $r=1/2$.

In the following, we develop a hybrid classical-quantum algorithm performing as good as a randomized classical greedy algorithm in the presence of strong noise. For instance, the classical greedy approach has a performance $r\simeq 0.848497...$ for Sherrington-Kirkpatrick (SK) Ising spin glass problems (see Methods), much larger than random guessing and a very noisy vanilla QAOA execution. We implement the quantum algorithm on Rigetti Aspen-M-3 programmable superconducting quantum system using up to $72$ qubits and find that it systematically outperforms its classical greedy counterpart, signaling a quantum enhancement.

We introduce an iterative algorithm for solving discrete optimization problems which bears similarities with divide-and-conquer methods~\cite{Ayanzadeh2022}, and more closely with other iterative/recursive techniques referred as RQAOA~\cite{PhysRevLett.125.260505,Wagner2023} or greedy decompositions in quantum annealing~\cite{Ramin2022}---the main difference being the freezing procedure. At each iteration, a set of variables are frozen to their classical values depending on the output returned by a quantum computer---although, as discussed later, the approach works with any sampleable distribution over bit strings, quantum or classical. These variables are removed, and an updated, smaller optimization problem is generated. The procedure is repeated until all variables are frozen or until the remaining problem is small enough for brute force.

A main difference between the quantum-enhanced greedy algorithm that we develop, and prior iterative approaches, is its robustness to noise, which is key when executing quantum algorithms at scale on current quantum hardware. Indeed, in a strong depolarizing noise regime, our quantum algorithm maps to a classical randomized greedy algorithm for which one can analytically estimate its average performance for problems such as Sherrington-Kirkpatrick Ising spin glasses. In the same strong noise regime, as expectation values, such as two-point correlations $\langle Z_i Z_j\rangle$, tend towards zero, other existing iterative algorithms ~\cite{PhysRevLett.125.260505,Ayanzadeh2022,Ramin2022,Wagner2023} would perform as well as a random sampling strategy in the absence of a mapping to a classical greedy baseline.

\begin{figure*}[!ht]
    \centering
    \includegraphics[width=1.0\textwidth]{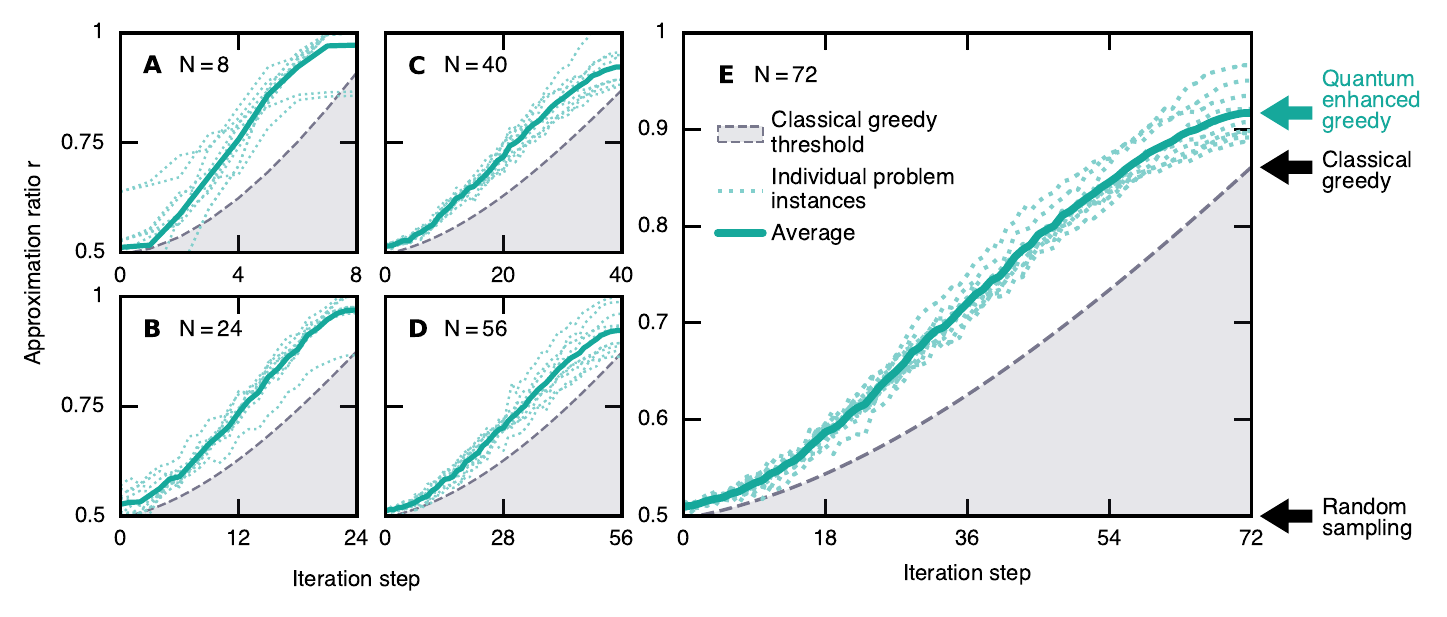} 
    \caption{\textbf{Approximation ratio versus iteration step.} Each panel corresponds to a Sherrington-Kirkpatrick problem instance of different size: \textsf{\textbf{A.}} $N=8$, \textsf{\textbf{B.}} $N=24$, \textsf{\textbf{C.}} $N=40$, \textsf{\textbf{D.}} $N=56$, and \textsf{\textbf{E.}} $N=72$. A random sampling strategy leads to an average approximation ratio $r=0.5$. The performance of the classical greedy baseline is shown by the shaded region with a final average approximation ratio $r\simeq 0.848497...$ for $N\to+\infty$ (see Methods). The quantum-enhanced greedy data show the expectation value of the approximation ratio for $10$ randomly generated problem instances (see Fig.~\ref{fig:introduction}). The performance at iteration step $0$ is that of the truncated one-layer QAOA. The performance at iteration step $N$ is that of the quantum-enhanced algorithm, also reported in Fig.~\ref{fig:performance_boost}. Because we display all individual instances, we omit the error bar for the average case.}
    \label{fig:approximation_ratio_versus_step}
\end{figure*}

Each iteration $\ell$ of our quantum-enhanced greedy algorithm follows the steps,
\begin{enumerate}[label=\textbf{\#\arabic*},leftmargin=*,nosep]
    \item Obtain a list of $M^{(\ell)}$ bit strings $\{\mathbf{B}^{(\ell)}\}$ that encode candidate solutions to the problem. In the quantum version, these bit strings are sampled from the output of the quantum computer, where the quantum circuit optimizes the objective function $C^{(\ell)}$ [Eq.~\eqref{eq:objective_f}]. In the classical randomized greedy version, these bit strings are sampled from a uniform distribution of all bit strings.

    \item Find a set $\{k\}$ of $K$ variables to freeze. Different heuristics can be envisioned~\cite{PhysRevLett.125.260505,Wagner2023}, including a majority vote based on one-body expectation values, i.e., $\max_i\vert\langle Z_i\rangle\vert$, or based on two-body expectation values, as well as many others. In the following, we use a two-body expectation strategy developed in the Methods to select $K=1$ variable.

    \item Find the frozen value of each variable selected in \#2: For each variable in $\{k\}$, loop over all the possible values $\{s\}$ of the variable (i.e., $s=0$ or $1$ for a binary variable), substitute them for all bit string of the list such that $\{\mathbf{B}^{(\ell)}_k\leftarrow s\}$, and compute the expectation value of the cost based on the modified $\{\mathbf{B}^{(\ell)}\}$. The assignment $s$ leading to the best expectation value of the cost is taken as the variable's frozen value (see Methods).

    \item Update the problem $C^{(\ell)}$ by replacing each of the operators in $\{Z_k\}$ by a constant based on the expectation value of $\langle{Z_k}\rangle$. The scalars are absorbed into $u$, $v_i$, and $w_{ij}$ in Eq.~\eqref{eq:objective_f} to create a new problem $C^{(\ell+1)}$ with at least $K$ fewer variables (see Methods).
\end{enumerate}

Other strategies can be implemented in the third step ($\#3$). The essential point is that for random bit strings $\{\mathbf{B}^{(\ell)}\}$, the freezing decisions are locally optimal with respect to the objective function, independently of the selected variables. As such, the average approximation ratio from random bit strings and optimal freezing is $r \simeq 0.848497\cdots$ (see Methods). Also, if $\{\mathbf{B}^{(\ell)}\}$ is replaced by optimal bit strings with respect to the objective function, then the above algorithm will preserve an optimal solution, and yield $r=1$. Therefore, the intuition is that for bit strings which are between random and optimal, there should be a performance boost with respect to the classical greedy baseline. Better-than-random bit strings, on average, should help make better-informed decisions for the selection and thus guide an otherwise randomized greedy process.

The complexity of the above algorithm is $O[(N/K)N_\textrm{edges}]$ with $N_\textrm{edges}$ the number of two-body terms in the graph problem. Taking $K\sim O(1)$ and $N_\textrm{edges}\sim O(N^2)$ for the Sherrington-Kirkpatrick instances considered in the following leads to $O(N^3)$ complexity. We note that the complexity of the classical randomized version of the algorithm can be reduced to $O(N^2)$ when not working with explicit bit strings and considering that all expectation values average to zero.

We implement the quantum-enhanced greedy algorithm on Sherrington-Kirkpatrick problem instances~\cite{PhysRevLett.35.1792} by setting $u=v_i=0$ in Eq.~\eqref{eq:objective_f}, and draw the parameters $w_{ij}$ uniformly from $\{+1,-1\}$. SK models correspond to paradigmatic Ising spin glasses. Although it was recently proven that the ground state energy of SK models can be efficiently approximated with an approximation ratio $(1-\varepsilon)$ by an approximate message passing algorithm~\cite{Montanari2018}, SK models remain a relevant benchmark for combinatorial optimization methods.

We run the algorithm on Rigetti's superconducting quantum processor Aspen-M-3 with a planar square-octagon topology of $79$ qubits (see the Methods for the parameters used in practice and more details). The limited connectivity of the hardware, displayed in Fig.~\ref{fig:introduction}, requires an extensive swap network to cover two-qubit gates between arbitrary qubits~\cite{Weidenfeller2022}. Consequently, implementation of the phase separator unitary of Eq.~\eqref{eq:circuit_qaoa} is not practical (due to noise) for SK problems with large $N$ [Figs.~\ref{fig:introduction}A and~\ref{fig:introduction}B]. Instead, we use a truncated one-layer QAOA ansatz: At each iteration of the algorithm, the problem is randomly mapped to the hardware-native architecture, and only gates involving qubits connected within $2$ swap cycles are considered, with the others dismissed, as exemplified in Fig.~\ref{fig:introduction}C. The circuit [Eq.~\eqref{eq:circuit_qaoa}] is compiled into hardware-native gates, resulting in about $400$ native $\sqrt{i\textrm{SWAP}}$ two-qubit gates for the largest problems considered, as detailed in the Methods. We collect a total of $M^{(\ell)}=256$ bit strings for all steps $\ell$.

We use the quantum-enhanced greedy algorithm to solve a set of $10$ random SK problem instances for sizes $N=8$, $24$, $40$, $56$, and $72$. We freeze one variable ($K=1$) at a time and use two-body expectation values to inform the selection process. We report the estimated expectation value of the approximation ratio $r\equiv\langle{r}\rangle_{\gamma^*,\beta^*}$ by computing the expectation value of the corresponding objective function over all sampled candidate bit string solutions at angles $\gamma^*$ and $\beta^*$. For $N\leq 24$, we use brute force to compute $C_\textrm{max}$ and $C_\textrm{min}$ [Eq.~\eqref{eq:approximation_ratio}] for a given problem instance. For larger $N$, we rely on the fact that the cost of the optimal solution is self-averaging, known exactly for $N\to+\infty$, and that finite-size corrections have also been studied over an ensemble of random instances~\cite{Boettcher2005}. This gives access to a proxy for approximating $r$, assuming one is not interested in the performance for an individual problem but that of an ensemble. Additional detail is given in Methods.

\begin{figure}[!t]
    \centering
    \includegraphics[width=1\columnwidth]{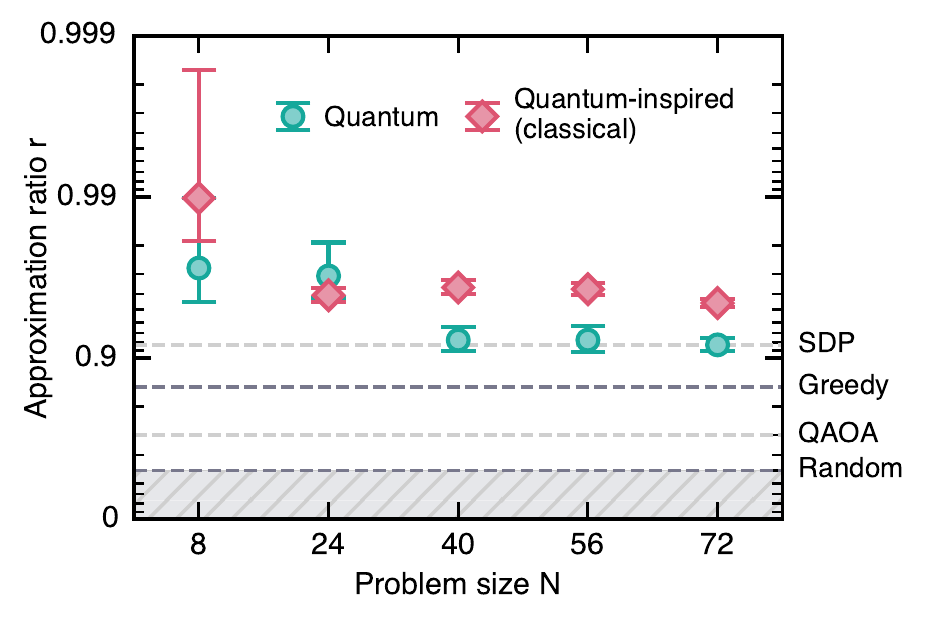} 
    \caption{\textbf{Performance boost.} Approximation ratio based on different bit string generators: The hardware-run QAOA of Fig.~\ref{fig:approximation_ratio_versus_step}, averaged over $10$ problem instances, and quantum-inspired classical tensor network simulations (see Methods), averaged over $100$ random SK problem instances. The different horizontal dashed lines correspond to the average performance of various algorithms in the limit $N\to+\infty$. From bottom to top: Random sampling $r=1/2$, a perfectly executed single-layer, nontruncated, QAOA circuit $r=1/2+1/4\mathsf{P}\sqrt{e}\simeq 0.698688...$~\cite{Farhi2022quantumapproximate}, the classical greedy baseline $r=1/2+\sqrt{2/\pi}/3\mathsf{P}\simeq 0.848497...$ (see Methods), a classical semidefinite programming (SDP) approach $r=1/2+1/\pi\mathsf{P}\simeq 0.917090...$~\cite{Aizenman1987,Montanari2015,Bandeira2019} where $\mathsf{P}$ is the Parisi constant~\cite{PhysRevLett.43.1754,Schmidt2008}. Error bars indicate one standard deviation. Numerical values are tabulated in the Supplementary Information.}
    \label{fig:performance_boost}
\end{figure}

We show the obtained approximation ratio as a function of the iteration step in Fig.~\ref{fig:approximation_ratio_versus_step}. Iteration step $0$ corresponds to a truncated one-layer QAOA ansatz run on the initial problem of $N$ variables. For all sizes, this is slightly above the $r=1/2$ random sampling bar, emphasizing that the QAOA displays a low average performance on current hardware. The last step corresponds to the final solution of the quantum-enhanced greedy algorithm. Its average performance is systematically above that of the classical greedy baseline. We note that the approximation ratio in both the classical greedy and the quantum-enhanced algorithms have a distribution around the average. This is more evident at small $N$, for example $N=8$ and $N=24$ in Fig.~\ref{fig:approximation_ratio_versus_step}, where the quantum greedy results for some problem instances dip below the average classical greedy results. The estimated approximation ratio at the last step is displayed in Fig.~\ref{fig:performance_boost} as a function of problem size $N$. We observe a decrease in performance with increasing problem size, which has two primary causes. First, the larger the size, the larger the quantum circuit, leading to a higher error rate. Second, the phase separator unitary of the truncated QAOA ansatz only covers an $O(1/N)$ density of edges of the graph problem, which accounts for a vanishing fraction of two-body terms in the SK problem instances (about $3\%$ for $N=72$). For comparison, a noiseless simulation of a nontruncated standard single-layer QAOA circuit leads to an average approximation ratio $r=1/2+1/4\mathsf{P}\sqrt{e}\simeq 0.698688...$ on large SK problem instances~\cite{Farhi2022quantumapproximate}, where $\mathsf{P}$ is the Parisi constant~\cite{PhysRevLett.43.1754,Schmidt2008}.

Embedding the QAOA into the quantum-enhanced algorithm greatly enhances the quality of the end result. From step $0$ to step $N$ of the iterative loop [Fig.~\ref{fig:approximation_ratio_versus_step}], we experimentally obtain a $6$ to $7$ fold increase in the average approximation ratio for $N=72$ where $1-r\simeq 0.49\to 0.08$ [Fig.~\ref{fig:performance_boost}]. A fairer comparison point is the classical greedy baseline which runs the same algorithm with random bit strings as input instead. For the SK problem instances considered here, the expected approximation ratio of this classical heuristic is $r\simeq 0.848497...$ as $N\to+\infty$ (see Methods). It is a high absolute bar to pass, much higher than what we obtain from a noisy and truncated one-layer QAOA run at iteration step $0$. For reference, it requires at least a perfectly executed nontruncated four- to five-layer QAOA circuit to meet this classical performance (see Supplementary Information and Ref.~\cite{Farhi2022quantumapproximate}). Here, the quantum-enhanced greedy algorithm run on noisy quantum hardware improves upon the average approximation ratio of its classical counterpart by about a factor $2$ ($1-r\simeq 0.151503...\to 0.08$) for the largest problem size $N=72$ [Fig.~\ref{fig:performance_boost}], empirically confirming the intuition that better-than-random bit strings should, on average, help make better-informed decisions in the freezing process. An explanation is that while bit strings look close to random as a whole because of noise, they might still locally retain relevant information. Here, we look for the the most correlated variable (see Methods), as this suggests a well-defined value for the corresponding bit, making it a good candidate for freezing. We observe an absolute performance comparable with state-of-the-art semidefinite programming method (SDP), corresponding to a spectral relaxation rounding to $\pm 1$ each entry of the leading eigenvector of the adjacency matrix of the graph problem~\cite{Aizenman1987,Montanari2015,Bandeira2019}.

While our empirical results demonstrate that the quantum-enhanced algorithm can outperform the classical greedy threshold, signaling a quantum boost, proving this rigorously remains as future work. Indeed, all the iterative algorithm needs is a bit string generator at step $\#1$, analogous to classical methods such as genetic algorithms and Monte Carlo methods. We highlight the work ahead for reaching a quantum advantage by using noiseless classical tensor network simulations to generate bit strings from a truncated one-layer QAOA circuit with two swap cycles embedded into a one-dimensional lattice (see Methods). This leads to an average approximation ratio $r\simeq 0.95$ for the largest problems considered [Fig.~\ref{fig:performance_boost}], higher than what we obtained with the quantum runs. We, therefore, expect that the performance of the quantum-enhanced greedy algorithm will continue to improve with advances in hardware fidelity. Interestingly, the algorithm can use expectation-based error mitigation techniques~\cite{Cai2022} for improving the freezing decisions---and therefore, the final solution---, which are otherwise not usable for more traditional quantum optimization methods when one is typically interested in enhancing an individual bit string. Indeed, further classical post-processing methods~\cite{Dupont2023} may be leveraged to develop and explore more sophisticated variable freezing procedures, such as post-selecting only bit strings that do at least as well as random guessing. Ultimately, the backbone remains the quantum device, and the overarching goal should be to improve its characteristics. Improving those will allow running deeper QAOA circuits as well as other state-of-the-art quantum circuits, leading to an overall performance increase through better-informed decisions (see Supplementary Information). This would warrant additional, comprehensive benchmarks against a panoply of state-of-the-art classical algorithms and techniques, such as Ising machines~\cite{Mohseni2022}, to assess on the existence of a practical quantum speedup or advantage.

The freezing decisions have a classical component that can be adapted to deal with some hard constraints; such as post-selecting on valid solutions. Hard constraints are ubiquitous in real-world optimization problems and are notoriously difficult to handle in practice. A possible strategy is to design quantum circuits working within the in-constraint space with dynamics restricted to the subspace of feasible solutions~\cite{Hadfield2019,marsh2019quantum,bartschi2020grover}, but these methods may require greater quantum resources. Another typical approach uses penalty terms that will disfavor the appearance of out-of-constraint bit strings, but implementing them on near-term devices and tuning their strength can be similarly challenging. Here, the idea is to make only freezing decisions which do not violate any constraints. Note that this is only possible for some classes of constraints, as a general satisfiability problem is NP-complete. We investigate a proof of concept of the modified algorithm on a constrained binary portfolio optimization problem in the Supplementary Information, comparing it to the state of the art~\cite{Herman2022}.

In practice, scaling to hundreds of qubits will require freezing variables simultaneously to keep the runtime under control. We suggest the following modification~\cite{PhysRevLett.125.260505}: After selecting $K$ variables in step $\#2$, the decision in step $\#3$ attempts the $2^K$ substitutions in $\{\mathbf{B}^{(\ell)}\}$ and keeps the best one.

Finally, it is interesting to think of iterative hybrid classical-quantum setups with performance guarantees on noisy hardware in the context of quantum problems rather than classical ones. For instance, can one enhance the performance of variational quantum eigensolvers~\cite{Cerezo2021,TILLY20221} for chemistry and other quantum many-body problems by embedding ideas from real-space renormalization group methods, such as linked-cluster expansions or the contractor renormalization group technique~\cite{PhysRevD.54.4131}?

\appendix

\section*{Methods}

\subsection*{Rigetti Aspen-M-3 Superconducting Platform}

Rigetti's Aspen-M-3 is a programmable and universal superconducting quantum computer based on transmon qubits. There are $79$ qubits arranged on a planar square-octagon topology. We make use of one-qubit rotation gates about the $x$ axis $\textrm{Rx}(\phi\in\mathbb{Z})=\exp(-iX\phi\pi/4)$, one-qubit rotation gates about the $z$ axis $\textrm{Rz}(\theta\in\mathbb{R})=\exp(-iZ\theta/2)$, and the two-qubit gate $\sqrt{i\textrm{SWAP}}=\exp[i\pi(XX+YY)/8]$~\cite{Abrams2020}. $X$, $Y$, and $Z$ are Pauli operators. The qubits have an average relaxation time $T_1=25(2)~\mu$s,  an average dephasing time $T_2=28(2)~\mu$s, an average readout fidelity of $94.6(7)\%$, and an average one-qubit $\textrm{Rx}$ fidelity of $99.4(2)\%$ estimated by randomized benchmarking~\cite{PhysRevA.77.012307}.

\begin{figure}[!t]
    \centering
    \includegraphics[width=1\columnwidth]{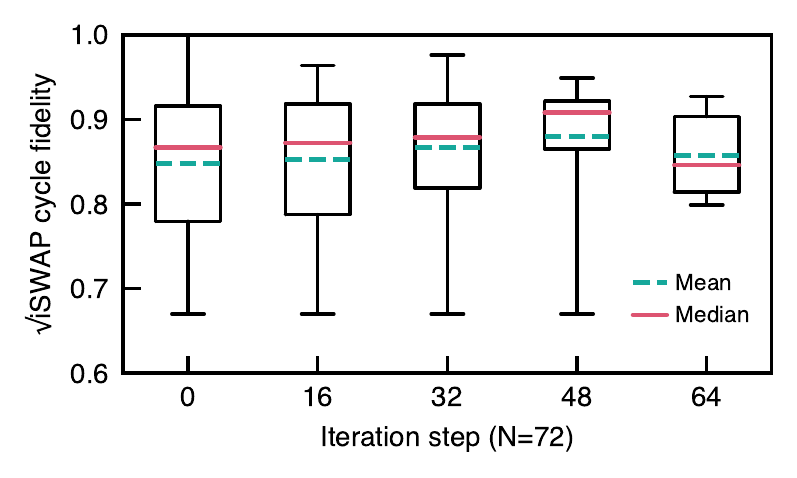} 
    \caption{\textbf{Cycle fidelity for two-qubit gates.} Statistics for the cycle fidelity of $\sqrt{i\textrm{SWAP}}$ at different iteration steps for a problem size $N=72$. The boxes cover the first to the third quartile. The whiskers cover the whole range of the data. With each iteration step of the algorithm, the size of the problem is reduced and we successively target smaller and smaller subsets of qubits of higher and higher overall quality resulting in an increasing mean and median with decreasing spread.}
    \label{fig:sqiswap_fidelity}
\end{figure}

To estimate the fidelity of the operations $\sqrt{i\textrm{SWAP}}$, we use cycle benchmarking~\cite{erhard_characterizing_2019}. In the circuit, cycles were constructed by splitting the three independent edge groups of Fig.~\ref{fig:introduction}C each into two clock cycles (for a total of six), where the $\sqrt{i\textrm{SWAP}}$ gates are separated by at least one idle qubit. This is done to minimize crosstalk between the $\sqrt{i\textrm{SWAP}}$ gates. Cycle benchmarking provides a measurement of fidelity that is less forgiving, but more realistic, than isolated randomized benchmarking. The cycles are benchmarked the same way they are played in the application circuit, accumulating additional error from decoherence while idling and from the single-qubit gates which precede each two-qubit gate in the cycle. We report in Fig.~\ref{fig:sqiswap_fidelity} the marginal process fidelity contributed by each pair of qubits playing an entangling gate in the cycle. In general, the majority of the infidelity is contributed by the entangling operations. We expect further refinements to the cycle calibration process to reduce control errors and yield better cycle performance.

As the quantum-enhanced greedy algorithm iteratively reduces the size of the problem and thus the number of qubits at each iteration step, we successively target smaller and smaller subsets of qubits of higher and higher overall quality. This is visible from the general trends of Fig.~\ref{fig:sqiswap_fidelity} where the mean and median fidelities increase and the spread decreases with each iteration step.

\subsection*{Hardware-Native Quantum Circuits}

The unitaries of the QAOA in Eq.~\eqref{eq:definition_unitaries} are expressed through one- and two-qubit gates exclusively, as pictured in Fig.~\ref{fig:introduction}C. These gates are further decomposed into the hardware native gate set. For instance, the one-qubit gate $\textrm{Rx}(\phi)$ is implemented for arbitrary angles using the standard ``ZXZXZ'' decomposition~\cite{barenco1995elementary}. We use the hardware-native two-qubit $\sqrt{i\textrm{SWAP}}$ gate to express $\textrm{Rzz}(\varphi)=\exp(-i\varphi ZZ /2)$ and $\textrm{SWAP}$. Because we always precede a SWAP gate by $\textrm{Rzz}(\varphi)$, we decompose directly the combination of these two gates. Both $\textrm{Rzz}$ and $\textrm{Rzz}\times\textrm{SWAP}$ can be implemented using hardware-native one-qubit gates and at most $2$ and $3$ two-qubit $\sqrt{i\textrm{SWAP}}$ gates~\cite{huang2021quantum}, respectively, as shown in Fig.~\ref{fig:gate_decomposition}; the precise decomposition is given in the Supplementary Information. While this may not be the optimal synthesis in general, it is optimal in terms of number of pulses for the Rigetti Aspen-M-3 quantum processor with respect to alternatives.

\begin{figure}[!t]
    \centering
    \includegraphics[width=1\columnwidth]{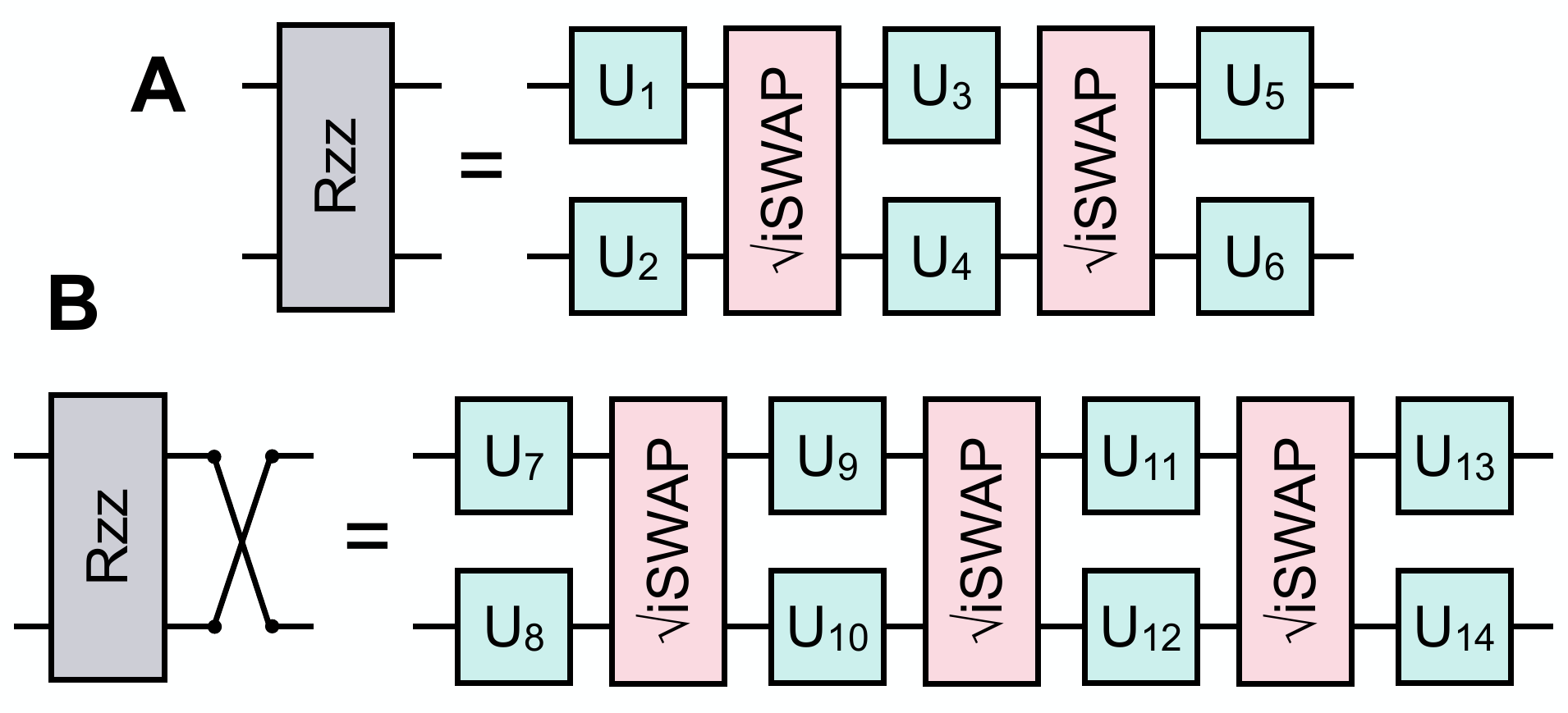} 
    \caption{\textbf{Two-qubit gate decompositions.} \textsf{\textbf{A.}} Decomposition of the parametric two-qubit gate $\textrm{Rzz}(\varphi)=\exp(-i\varphi ZZ /2)$ using $2$ hardware-native $\sqrt{i\textrm{SWAP}}=\exp[i\pi(XX+YY)/8]$ gates. $X$, $Y$, and $Z$ are Pauli operators. \textsf{\textbf{B.}} Decomposition of the parametric two-qubit gate $\textrm{Rzz}(\varphi)$ directly followed by a $\textrm{SWAP}$ using $3$ hardware-native $\sqrt{i\textrm{SWAP}}$ gates. In each case, the one-qubit gates $U_{1...14}$ carry the angle $\varphi$ and are decomposed explicitly as a function of $\textrm{Rx}$ and $\textrm{Rz}$ gates (see Supplementary Information for details).}
    \label{fig:gate_decomposition}
\end{figure}

\subsection*{Optimization of QAOA Circuits and Sampling Bit Strings}

A QAOA circuit with $p$ layers is defined in Eq.~\eqref{eq:circuit_qaoa}. It is parameterized by $2p$ angles, which should be such that the resulting quantum state minimizes the expectation value of the objective function $C$ of Eq.~\eqref{eq:objective_f}. In this work, we consider truncated one-layer QAOA circuits [Fig.~\ref{fig:introduction}B] with only two angles $\gamma$ and $\beta$. As such, we optimize them through a $16\times 16$ grid search in the range $\gamma\in[0,2\pi]$ and $\beta\in[0,\pi]$. For each of the $16\times 16=256$ pairs of angles, we collect $M=256$ bit strings $\{\mathbf{B}\}$ by sampling the quantum state $\vert{\gamma,\beta}\rangle$ to compute the expectation value,
\begin{equation}
    \langle C\rangle_{\gamma,\beta}=\langle{\gamma,\beta}\vert{C}\vert{\gamma,\beta}\rangle=\frac{1}{M}\sum\nolimits_{\{\mathbf{B}\}}\langle\mathbf{B}\vert{C}\vert\mathbf{B}\rangle.
\end{equation}
The angles leading to the minimum expectation value of $\langle C\rangle_{\gamma,\beta}$ are $\gamma^*$ and $\beta^*$. The $M$ bit strings that led to the minimum expectation value $\langle C\rangle_{\gamma^*,\beta^*}$ during the grid search are used in the first step ($\#1$) of the iterative quantum-enhanced greedy algorithm. Examples of $\langle C\rangle_{\gamma,\beta}$ are shown in the Supplementary Information.

\subsection*{Iterative Process}

We provide additional detail on our implementation of the iterative steps $\#2$, $\#3$, and $\#4$ of the algorithm. At this stage, we have already obtained a list of bit strings $\{\mathbf{B}\}$ sampled from the quantum state of the optimized QAOA circuit at angles $\gamma^*$ and $\beta^*$. At step $\#2$, we find $K=1$ variable to freeze based on a two-body expectation strategy. Precisely, for each active node $k$ of the graph, we compute:
\begin{align}
    F_k=\frac{1}{M}\Biggl[\sum\nolimits_{i\in\{\textrm{active}\}\neq k}\,\Bigl\vert w_{ik}\sum\nolimits_{\{\mathbf{B}\}}\langle\mathbf{B}\vert{Z_iZ_k}\vert\mathbf{B}\rangle\Bigr\vert\nonumber\\
    + \Bigl\vert v_{k}\sum\nolimits_{\{\mathbf{B}\}}\langle\mathbf{B}\vert{Z_k}\vert\mathbf{B}\rangle\Bigr\vert\Biggr],
\end{align}
where we find $k$ such that $\max_kF_k$. The function $F_k$ corresponds to the expectation value of the objective function at the given iteration step where all the terms involving the node $k$ have been individually sandwiched by a modulus symbol. We now move to step $\#3$ and generate two modified versions of the bit strings $\{\mathbf{B}\}$ by setting the bit $k$ on all the $M$ bit strings to either $0$ or $1$. We then compute the expectation value of the cost $\langle C_0\rangle=\sum\nolimits_{\{\mathbf{B}_k\leftarrow 0\}}\langle\mathbf{B}\vert{C}\vert\mathbf{B}\rangle/M$ and $\langle C_1\rangle=\sum\nolimits_{\{\mathbf{B}_k\leftarrow 1\}}\langle\mathbf{B}\vert{C}\vert\mathbf{B}\rangle/M$. The value $B_k=0$ or $1$ which provided the smallest of $\langle C_0\rangle$ and $\langle C_1\rangle$ will be used as the variable's frozen value $\sigma_k=(-1)^{B_k}$. Following the freezing decision of the variable $k$, the problem is updated as follows,
\begin{align}
    w_{ik}Z_iZ_k&\longrightarrow v_i'Z_i~\,\forall i~~\textrm{with}~~v'_i=v_i+w_{ik}\sigma_k,\nonumber\\
    v_{k}Z_k&\longrightarrow u'=u+v_k\sigma_k,
\end{align}
where the notation of Eq.~\eqref{eq:objective_f} is used for scalar parameters $u$, $v_i$, and $w_{ij}$.

\subsection*{Classical Simulations of Quantum Circuits}

We supplement the quantum experimental results with classical simulations of the circuits. We employ two methods throughout this work. The first one, used in the Supplementary Information, is based on a state vector approach where the quantum state for $N$ qubits is represented as a complex vector of $2^N$ components. It is an exact method. The second one, used for generating the data of Fig.~\ref{fig:performance_boost}, is a tensor network approach based on matrix product states~\cite{PhysRevLett.93.040502}. The circuits we simulate with matrix product states are a truncated one-layer QAOA circuit with two swap cycles embedded into a one-dimensional lattice with open boundary conditions. The circuits are shallow enough to be executed exactly with a relatively low bond dimension (the bond dimension is a control parameter of a matrix product state simulator), independent of the number of qubits involved, thus enabling exact classical simulations at $N=72$. Precisely, the circuits have a brick wall pattern of two-qubit gates with a total of $4$ layers in addition to one-qubit gates. Additional details on matrix product states are given in the Supplementary Information.

\subsection*{Optimal Cost of Sherrington-Kirkpatrick Instances}

Computing the approximation ratio of Eq.~\eqref{eq:approximation_ratio} requires the extremum (best $C_\textrm{min}$ and worst $C_\textrm{max}$) cost values for the SK problem instance of interest. For problems $N\leq 24$, we compute them through brute-force by enumerating the $2^N$ bit strings. For larger $N$, we rely on the fact that the cost of the optimal solution is self-averaging, known exactly for $N\to+\infty$, and that finite-size corrections have also been studied over an ensemble of random instances~\cite{Boettcher2005}. In particular, one has,
\begin{equation}
    C_\textrm{min}\times N^{-3/2}\simeq-\mathsf{P} + aN^{-\omega},
    \label{eq:sk_value}
\end{equation}
where $\mathsf{P}=0.763166726566547...$ is a universal constant known as the Parisi value~\cite{PhysRevLett.43.1754,Schmidt2008}, $\omega=2/3$ is a universal exponent accounting for finite-$N$ corrections, and $a\simeq 0.70(1)$ is a nonuniversal constant~\cite{Boettcher2005}. Moreover, on average $C_\textrm{max}=-C_\textrm{min}$ since it corresponds to solving an equivalent problem with flipped signs for the parameters $w_{ij}$ [Eq.~\eqref{eq:objective_f}] (see Supplementary Information). Hence, for an ensemble of SK problems leading to an average cost value of $C^*$, we approximate the obtained approximation ratio as,
\begin{equation}
    r\simeq\frac{1}{2}\left(1+\frac{C^*}{C_\textrm{min}}\right),
    \label{eq:approx_approximation_ratio}
\end{equation}
where $C_\textrm{min}$ is evaluated numerically using Eq.~\eqref{eq:sk_value}. For randomly generated bit strings, $C^*$ is symmetrically distributed between $C_\textrm{min}$ and $C_\textrm{max}$, i.e., $C^*=0$ in expectation, and the corresponding average approximation ratio is $r\simeq1/2$.

\subsection*{Classical Greedy Baseline}

The classical greedy algorithm iteratively builds a solution by making the locally optimal choice at each stage. There are as many iteration steps as variables in the problem. We note $Z_i=\pm 1$ the contribution to the cost function of the frozen variable $i$. We consider the SK problem instances of the main text [Eq.~\eqref{eq:objective_f} with $w_{ij}=\pm 1$, $u=v_i=0$] for which the freezing decisions go as follows. Iteration step $1$, freeze $Z_1$ arbitrarily to $\pm 1$. Iteration step $2$, freeze $Z_2$ such that $w_{12}Z_1Z_2$ is minimized, i.e., $\min_{Z_2}[w_{12}Z_1Z_2]$. Iteration step $3$, freeze $Z_3$ such that $\min_{Z_3}[Z_3(w_{13}Z_1+w_{23}Z_2)]$. At iteration step $\ell$, freeze $Z_\ell$ such that $Z_\ell\sum_{i=1}^{\ell-1}w_{i\ell}Z_i$ is minimized. This is repeated until iteration step $N$. The final cost value is,
\begin{equation}
    C_\textrm{greedy}=-\sum\nolimits_{\ell=2}^N\Bigl\vert\sum\nolimits_{i=1}^{\ell-1}w_{i\ell}Z_i\Bigr\vert.
    \label{eq:cost_greedy}
\end{equation}
The absolute value terms in Eq.~\eqref{eq:cost_greedy} can be seen as individual random walks containing $\ell-1$ steps of length $\pm 1$. As such, for $\ell\to+\infty$, their average contribution will be $\simeq\sqrt{2(\ell-1)/\pi}$. This step is analogous to averaging over different freezing orders for the variables. Under the hood, this corresponds to running the algorithm of the main text with a finite number of random bit strings from which statistical fluctuations will lead to different freezing selections in $\#2$ from one run to the next. Hence, one finds that, on average,
\begin{equation}
    C_\textrm{greedy}\simeq-\sqrt{2/\pi}\sum\nolimits_{\ell=1}^{N-1}\sqrt{\ell}.
    \label{eq:cost_avg_greedy}
\end{equation}
As $N\to+\infty$, an asymptotic expansion based on Euler-Maclaurin formula leads to $\sum\nolimits_{\ell=1}^{N-1}\sqrt{\ell}\sim (2/3)N^{3/2}+...$, which makes it possible to evaluate the average approximation ratio of the classical greedy algorithm using Eqs.~\eqref{eq:sk_value} and~\eqref{eq:approx_approximation_ratio} for an infinite-size SK problem instance,
\begin{equation}
    \frac{C_\textrm{greedy}}{C_\textrm{min}}=\frac{2\sqrt{2/\pi}}{3\mathsf{P}}~\Rightarrow~r=\frac{1}{2}+\frac{\sqrt{2/\pi}}{3\mathsf{P}}\simeq 0.848497...
    \label{eq:approx_ratio_greedy}
\end{equation}
The case of finite $N$ is studied numerically in Supplementary Information. By definition, this is also the average approximation ratio of the quantum-enhanced algorithm with random input. Physically, this would mean the quantum computer generates a maximally mixed state $\rho=\mathsf{I}/2^N$, as the result of, e.g., strong depolarizing noise: Here $\rho$ is the density matrix describing $N$ qubits and $\mathsf{I}$ the Identity matrix of dimensions $2^N\times 2^N$.

For comparison, other algorithms yield the following average performance: Random sampling $r=1/2$, a perfectly executed single-layer, nontruncated, QAOA circuit $r=1/2+1/4\mathsf{P}\sqrt{e}\simeq 0.698688...$~\cite{Farhi2022quantumapproximate}, a classical semidefinite programming approach $r=1/2+1/\pi\mathsf{P}\simeq 0.917090...$~\cite{Aizenman1987,Montanari2015,Bandeira2019}.

We extend nonexhaustively the analytical analysis of the classical greedy baseline to other problems in the Supplementary Information.

\section*{Acknowledgments}

This work is supported by the Defense Advanced Research Projects Agency (DARPA) under Agreement No. HR00112090058 and IAA 8839, Annex 114. Authors from USRA also acknowledge support under NASA Academic Mission Services under contract No. NNA16BD14C.

\section*{Author contributions}

M. Dupont conceived the project with support from B. Evert, M. Hodson, M. Reagor, E. Rieffel, and D. Venturelli. M. Dupont, B. Evert, M. Hodson, S. Jeffrey, B. Sundar, and Y. Yamaguchi developed the code, ran the experiments, and collected the data. M. Dupont wrote the manuscript with contributions from B. Evert, S. Hadfield, M. Reagor, B. Sundar, and input from all the co-authors. All co-authors contributed to the discussions leading to the completion of this project. The experiments were performed through Rigetti Computing's Quantum Cloud Services QCS\textsuperscript{TM} using the superconducting quantum processor Rigetti Computing Aspen-M-3 that was developed, fabricated, and operated by Rigetti Computing Inc.

\section*{Competing interests}

M. Dupont, B. Evert, D. Feng, M. Hodson, S. Jeffrey, M. Reagor, and B. Sundar are, have been, or may in the future be participants in incentive stock plans at Rigetti Computing Inc. M. Dupont is inventor on two pending patent applications related to this work (No. 63/381,831 and No. 63/487,898). The other authors declare that they have no competing interests.

\section*{Data availability}

All experimental data are publicly available at \href{https://doi.org/10.5281/zenodo.7709803}{doi.org/10.5281/zenodo.7709803}.

\let\oldaddcontentsline\addcontentsline
\renewcommand{\addcontentsline}[3]{}
\bibliography{references}

\begin{thebibliography}{63}%
\makeatletter
\providecommand \@ifxundefined [1]{%
 \@ifx{#1\undefined}
}%
\providecommand \@ifnum [1]{%
 \ifnum #1\expandafter \@firstoftwo
 \else \expandafter \@secondoftwo
 \fi
}%
\providecommand \@ifx [1]{%
 \ifx #1\expandafter \@firstoftwo
 \else \expandafter \@secondoftwo
 \fi
}%
\providecommand \natexlab [1]{#1}%
\providecommand \enquote  [1]{``#1''}%
\providecommand \bibnamefont  [1]{#1}%
\providecommand \bibfnamefont [1]{#1}%
\providecommand \citenamefont [1]{#1}%
\providecommand \href@noop [0]{\@secondoftwo}%
\providecommand \href [0]{\begingroup \@sanitize@url \@href}%
\providecommand \@href[1]{\@@startlink{#1}\@@href}%
\providecommand \@@href[1]{\endgroup#1\@@endlink}%
\providecommand \@sanitize@url [0]{\catcode `\\12\catcode `\$12\catcode
  `\&12\catcode `\#12\catcode `\^12\catcode `\_12\catcode `\%12\relax}%
\providecommand \@@startlink[1]{}%
\providecommand \@@endlink[0]{}%
\providecommand \url  [0]{\begingroup\@sanitize@url \@url }%
\providecommand \@url [1]{\endgroup\@href {#1}{\urlprefix }}%
\providecommand \urlprefix  [0]{URL }%
\providecommand \Eprint [0]{\href }%
\providecommand \doibase [0]{https://doi.org/}%
\providecommand \selectlanguage [0]{\@gobble}%
\providecommand \bibinfo  [0]{\@secondoftwo}%
\providecommand \bibfield  [0]{\@secondoftwo}%
\providecommand \translation [1]{[#1]}%
\providecommand \BibitemOpen [0]{}%
\providecommand \bibitemStop [0]{}%
\providecommand \bibitemNoStop [0]{.\EOS\space}%
\providecommand \EOS [0]{\spacefactor3000\relax}%
\providecommand \BibitemShut  [1]{\csname bibitem#1\endcsname}%
\let\auto@bib@innerbib\@empty
\bibitem [{\citenamefont {Farhi}\ \emph {et~al.}(2001)\citenamefont {Farhi},
  \citenamefont {Goldstone}, \citenamefont {Gutmann}, \citenamefont {Lapan},
  \citenamefont {Lundgren},\ and\ \citenamefont {Preda}}]{Farhi2001}%
  \BibitemOpen
  \bibfield  {author} {\bibinfo {author} {\bibfnamefont {E.}~\bibnamefont
  {Farhi}}, \bibinfo {author} {\bibfnamefont {J.}~\bibnamefont {Goldstone}},
  \bibinfo {author} {\bibfnamefont {S.}~\bibnamefont {Gutmann}}, \bibinfo
  {author} {\bibfnamefont {J.}~\bibnamefont {Lapan}}, \bibinfo {author}
  {\bibfnamefont {A.}~\bibnamefont {Lundgren}},\ and\ \bibinfo {author}
  {\bibfnamefont {D.}~\bibnamefont {Preda}},\ }\href
  {https://doi.org/10.1126/science.1057726} {\bibfield  {journal} {\bibinfo
  {journal} {Science}\ }\textbf {\bibinfo {volume} {292}},\ \bibinfo {pages}
  {472} (\bibinfo {year} {2001})}\BibitemShut {NoStop}%
\bibitem [{\citenamefont {Farhi}\ \emph
  {et~al.}(2014{\natexlab{a}})\citenamefont {Farhi}, \citenamefont
  {Goldstone},\ and\ \citenamefont {Gutmann}}]{Farhi2014}%
  \BibitemOpen
  \bibfield  {author} {\bibinfo {author} {\bibfnamefont {E.}~\bibnamefont
  {Farhi}}, \bibinfo {author} {\bibfnamefont {J.}~\bibnamefont {Goldstone}},\
  and\ \bibinfo {author} {\bibfnamefont {S.}~\bibnamefont {Gutmann}},\ }\href
  {https://arxiv.org/abs/1411.4028} {\bibfield  {journal} {\bibinfo  {journal}
  {arXiv:1411.4028}\ } (\bibinfo {year} {2014}{\natexlab{a}})}\BibitemShut
  {NoStop}%
\bibitem [{\citenamefont {Farhi}\ \emph
  {et~al.}(2014{\natexlab{b}})\citenamefont {Farhi}, \citenamefont
  {Goldstone},\ and\ \citenamefont {Gutmann}}]{Farhi2014b}%
  \BibitemOpen
  \bibfield  {author} {\bibinfo {author} {\bibfnamefont {E.}~\bibnamefont
  {Farhi}}, \bibinfo {author} {\bibfnamefont {J.}~\bibnamefont {Goldstone}},\
  and\ \bibinfo {author} {\bibfnamefont {S.}~\bibnamefont {Gutmann}},\ }\href
  {https://arxiv.org/abs/1412.6062} {\bibfield  {journal} {\bibinfo  {journal}
  {arXiv:1412.6062}\ } (\bibinfo {year} {2014}{\natexlab{b}})}\BibitemShut
  {NoStop}%
\bibitem [{\citenamefont {Cerezo}\ \emph {et~al.}(2021)\citenamefont {Cerezo},
  \citenamefont {Arrasmith}, \citenamefont {Babbush}, \citenamefont {Benjamin},
  \citenamefont {Endo}, \citenamefont {Fujii}, \citenamefont {McClean},
  \citenamefont {Mitarai}, \citenamefont {Yuan}, \citenamefont {Cincio},\ and\
  \citenamefont {Coles}}]{Cerezo2021}%
  \BibitemOpen
  \bibfield  {author} {\bibinfo {author} {\bibfnamefont {M.}~\bibnamefont
  {Cerezo}}, \bibinfo {author} {\bibfnamefont {A.}~\bibnamefont {Arrasmith}},
  \bibinfo {author} {\bibfnamefont {R.}~\bibnamefont {Babbush}}, \bibinfo
  {author} {\bibfnamefont {S.~C.}\ \bibnamefont {Benjamin}}, \bibinfo {author}
  {\bibfnamefont {S.}~\bibnamefont {Endo}}, \bibinfo {author} {\bibfnamefont
  {K.}~\bibnamefont {Fujii}}, \bibinfo {author} {\bibfnamefont {J.~R.}\
  \bibnamefont {McClean}}, \bibinfo {author} {\bibfnamefont {K.}~\bibnamefont
  {Mitarai}}, \bibinfo {author} {\bibfnamefont {X.}~\bibnamefont {Yuan}},
  \bibinfo {author} {\bibfnamefont {L.}~\bibnamefont {Cincio}},\ and\ \bibinfo
  {author} {\bibfnamefont {P.~J.}\ \bibnamefont {Coles}},\ }\href
  {https://doi.org/10.1038/s42254-021-00348-9} {\bibfield  {journal} {\bibinfo
  {journal} {Nat. Rev. Phys.}\ }\textbf {\bibinfo {volume} {3}},\ \bibinfo
  {pages} {625} (\bibinfo {year} {2021})}\BibitemShut {NoStop}%
\bibitem [{\citenamefont {Montanaro}(2020)}]{PhysRevResearch.2.013056}%
  \BibitemOpen
  \bibfield  {author} {\bibinfo {author} {\bibfnamefont {A.}~\bibnamefont
  {Montanaro}},\ }\href {https://doi.org/10.1103/PhysRevResearch.2.013056}
  {\bibfield  {journal} {\bibinfo  {journal} {Phys. Rev. Res.}\ }\textbf
  {\bibinfo {volume} {2}},\ \bibinfo {pages} {013056} (\bibinfo {year}
  {2020})}\BibitemShut {NoStop}%
\bibitem [{\citenamefont {Alexandru}\ \emph {et~al.}(2020)\citenamefont
  {Alexandru}, \citenamefont {Bridgett-Tomkinson}, \citenamefont {Linden},
  \citenamefont {MacManus}, \citenamefont {Montanaro},\ and\ \citenamefont
  {Morris}}]{Alexandru_2020}%
  \BibitemOpen
  \bibfield  {author} {\bibinfo {author} {\bibfnamefont {C.-M.}\ \bibnamefont
  {Alexandru}}, \bibinfo {author} {\bibfnamefont {E.}~\bibnamefont
  {Bridgett-Tomkinson}}, \bibinfo {author} {\bibfnamefont {N.}~\bibnamefont
  {Linden}}, \bibinfo {author} {\bibfnamefont {J.}~\bibnamefont {MacManus}},
  \bibinfo {author} {\bibfnamefont {A.}~\bibnamefont {Montanaro}},\ and\
  \bibinfo {author} {\bibfnamefont {H.}~\bibnamefont {Morris}},\ }\href
  {https://doi.org/10.1088/2058-9565/abb003} {\bibfield  {journal} {\bibinfo
  {journal} {Quantum Sci. Technol.}\ }\textbf {\bibinfo {volume} {5}},\
  \bibinfo {pages} {045014} (\bibinfo {year} {2020})}\BibitemShut {NoStop}%
\bibitem [{\citenamefont {Barahona}(1982)}]{Barahona1982}%
  \BibitemOpen
  \bibfield  {author} {\bibinfo {author} {\bibfnamefont {F.}~\bibnamefont
  {Barahona}},\ }\href {https://doi.org/10.1088/0305-4470/15/10/028} {\bibfield
   {journal} {\bibinfo  {journal} {J. Phys. A Math. Theor.}\ }\textbf {\bibinfo
  {volume} {15}},\ \bibinfo {pages} {3241} (\bibinfo {year}
  {1982})}\BibitemShut {NoStop}%
\bibitem [{\citenamefont {Boixo}\ \emph {et~al.}(2014)\citenamefont {Boixo},
  \citenamefont {R{\o}nnow}, \citenamefont {Isakov}, \citenamefont {Wang},
  \citenamefont {Wecker}, \citenamefont {Lidar}, \citenamefont {Martinis},\
  and\ \citenamefont {Troyer}}]{Boixo2014}%
  \BibitemOpen
  \bibfield  {author} {\bibinfo {author} {\bibfnamefont {S.}~\bibnamefont
  {Boixo}}, \bibinfo {author} {\bibfnamefont {T.~F.}\ \bibnamefont
  {R{\o}nnow}}, \bibinfo {author} {\bibfnamefont {S.~V.}\ \bibnamefont
  {Isakov}}, \bibinfo {author} {\bibfnamefont {Z.}~\bibnamefont {Wang}},
  \bibinfo {author} {\bibfnamefont {D.}~\bibnamefont {Wecker}}, \bibinfo
  {author} {\bibfnamefont {D.~A.}\ \bibnamefont {Lidar}}, \bibinfo {author}
  {\bibfnamefont {J.~M.}\ \bibnamefont {Martinis}},\ and\ \bibinfo {author}
  {\bibfnamefont {M.}~\bibnamefont {Troyer}},\ }\href
  {https://doi.org/10.1038/nphys2900} {\bibfield  {journal} {\bibinfo
  {journal} {Nat. Phys.}\ }\textbf {\bibinfo {volume} {10}},\ \bibinfo {pages}
  {218} (\bibinfo {year} {2014})}\BibitemShut {NoStop}%
\bibitem [{\citenamefont {R{\o}nnow}\ \emph {et~al.}(2014)\citenamefont
  {R{\o}nnow}, \citenamefont {Wang}, \citenamefont {Job}, \citenamefont
  {Boixo}, \citenamefont {Isakov}, \citenamefont {Wecker}, \citenamefont
  {Martinis}, \citenamefont {Lidar},\ and\ \citenamefont
  {Troyer}}]{Ronnow2014}%
  \BibitemOpen
  \bibfield  {author} {\bibinfo {author} {\bibfnamefont {T.~F.}\ \bibnamefont
  {R{\o}nnow}}, \bibinfo {author} {\bibfnamefont {Z.}~\bibnamefont {Wang}},
  \bibinfo {author} {\bibfnamefont {J.}~\bibnamefont {Job}}, \bibinfo {author}
  {\bibfnamefont {S.}~\bibnamefont {Boixo}}, \bibinfo {author} {\bibfnamefont
  {S.~V.}\ \bibnamefont {Isakov}}, \bibinfo {author} {\bibfnamefont
  {D.}~\bibnamefont {Wecker}}, \bibinfo {author} {\bibfnamefont {J.~M.}\
  \bibnamefont {Martinis}}, \bibinfo {author} {\bibfnamefont {D.~A.}\
  \bibnamefont {Lidar}},\ and\ \bibinfo {author} {\bibfnamefont
  {M.}~\bibnamefont {Troyer}},\ }\href
  {https://doi.org/10.1126/science.1252319} {\bibfield  {journal} {\bibinfo
  {journal} {Science}\ }\textbf {\bibinfo {volume} {345}},\ \bibinfo {pages}
  {420} (\bibinfo {year} {2014})}\BibitemShut {NoStop}%
\bibitem [{\citenamefont {Venturelli}\ \emph {et~al.}(2015)\citenamefont
  {Venturelli}, \citenamefont {Mandr\`a}, \citenamefont {Knysh}, \citenamefont
  {O'Gorman}, \citenamefont {Biswas},\ and\ \citenamefont
  {Smelyanskiy}}]{PhysRevX.5.031040}%
  \BibitemOpen
  \bibfield  {author} {\bibinfo {author} {\bibfnamefont {D.}~\bibnamefont
  {Venturelli}}, \bibinfo {author} {\bibfnamefont {S.}~\bibnamefont
  {Mandr\`a}}, \bibinfo {author} {\bibfnamefont {S.}~\bibnamefont {Knysh}},
  \bibinfo {author} {\bibfnamefont {B.}~\bibnamefont {O'Gorman}}, \bibinfo
  {author} {\bibfnamefont {R.}~\bibnamefont {Biswas}},\ and\ \bibinfo {author}
  {\bibfnamefont {V.}~\bibnamefont {Smelyanskiy}},\ }\href
  {https://doi.org/10.1103/PhysRevX.5.031040} {\bibfield  {journal} {\bibinfo
  {journal} {Phys. Rev. X}\ }\textbf {\bibinfo {volume} {5}},\ \bibinfo {pages}
  {031040} (\bibinfo {year} {2015})}\BibitemShut {NoStop}%
\bibitem [{\citenamefont {Farhi}\ and\ \citenamefont
  {Harrow}(2016)}]{Farhi2016}%
  \BibitemOpen
  \bibfield  {author} {\bibinfo {author} {\bibfnamefont {E.}~\bibnamefont
  {Farhi}}\ and\ \bibinfo {author} {\bibfnamefont {A.~W.}\ \bibnamefont
  {Harrow}},\ }\href {https://arxiv.org/abs/1602.07674} {\bibfield  {journal}
  {\bibinfo  {journal} {arXiv:1602.07674}\ } (\bibinfo {year}
  {2016})}\BibitemShut {NoStop}%
\bibitem [{\citenamefont {Hadfield}\ \emph {et~al.}(2019)\citenamefont
  {Hadfield}, \citenamefont {Wang}, \citenamefont {O'Gorman}, \citenamefont
  {Rieffel}, \citenamefont {Venturelli},\ and\ \citenamefont
  {Biswas}}]{Hadfield2019}%
  \BibitemOpen
  \bibfield  {author} {\bibinfo {author} {\bibfnamefont {S.}~\bibnamefont
  {Hadfield}}, \bibinfo {author} {\bibfnamefont {Z.}~\bibnamefont {Wang}},
  \bibinfo {author} {\bibfnamefont {B.}~\bibnamefont {O'Gorman}}, \bibinfo
  {author} {\bibfnamefont {E.~G.}\ \bibnamefont {Rieffel}}, \bibinfo {author}
  {\bibfnamefont {D.}~\bibnamefont {Venturelli}},\ and\ \bibinfo {author}
  {\bibfnamefont {R.}~\bibnamefont {Biswas}},\ }\bibfield  {journal} {\bibinfo
  {journal} {Algorithms}\ }\textbf {\bibinfo {volume} {12}},\ \href
  {https://doi.org/10.3390/a12020034} {10.3390/a12020034} (\bibinfo {year}
  {2019})\BibitemShut {NoStop}%
\bibitem [{\citenamefont {Otterbach}\ \emph {et~al.}(2017)\citenamefont
  {Otterbach}, \citenamefont {Manenti}, \citenamefont {Alidoust}, \citenamefont
  {Bestwick}, \citenamefont {Block}, \citenamefont {Bloom}, \citenamefont
  {Caldwell}, \citenamefont {Didier}, \citenamefont {Fried}, \citenamefont
  {Hong}, \citenamefont {Karalekas}, \citenamefont {Osborn}, \citenamefont
  {Papageorge}, \citenamefont {Peterson}, \citenamefont {Prawiroatmodjo},
  \citenamefont {Rubin}, \citenamefont {Ryan}, \citenamefont {Scarabelli},
  \citenamefont {Scheer}, \citenamefont {Sete}, \citenamefont {Sivarajah},
  \citenamefont {Smith}, \citenamefont {Staley}, \citenamefont {Tezak},
  \citenamefont {Zeng}, \citenamefont {Hudson}, \citenamefont {Johnson},
  \citenamefont {Reagor}, \citenamefont {da~Silva},\ and\ \citenamefont
  {Rigetti}}]{Otterbach2017}%
  \BibitemOpen
  \bibfield  {author} {\bibinfo {author} {\bibfnamefont {J.~S.}\ \bibnamefont
  {Otterbach}}, \bibinfo {author} {\bibfnamefont {R.}~\bibnamefont {Manenti}},
  \bibinfo {author} {\bibfnamefont {N.}~\bibnamefont {Alidoust}}, \bibinfo
  {author} {\bibfnamefont {A.}~\bibnamefont {Bestwick}}, \bibinfo {author}
  {\bibfnamefont {M.}~\bibnamefont {Block}}, \bibinfo {author} {\bibfnamefont
  {B.}~\bibnamefont {Bloom}}, \bibinfo {author} {\bibfnamefont
  {S.}~\bibnamefont {Caldwell}}, \bibinfo {author} {\bibfnamefont
  {N.}~\bibnamefont {Didier}}, \bibinfo {author} {\bibfnamefont {E.~S.}\
  \bibnamefont {Fried}}, \bibinfo {author} {\bibfnamefont {S.}~\bibnamefont
  {Hong}}, \bibinfo {author} {\bibfnamefont {P.}~\bibnamefont {Karalekas}},
  \bibinfo {author} {\bibfnamefont {C.~B.}\ \bibnamefont {Osborn}}, \bibinfo
  {author} {\bibfnamefont {A.}~\bibnamefont {Papageorge}}, \bibinfo {author}
  {\bibfnamefont {E.~C.}\ \bibnamefont {Peterson}}, \bibinfo {author}
  {\bibfnamefont {G.}~\bibnamefont {Prawiroatmodjo}}, \bibinfo {author}
  {\bibfnamefont {N.}~\bibnamefont {Rubin}}, \bibinfo {author} {\bibfnamefont
  {C.~A.}\ \bibnamefont {Ryan}}, \bibinfo {author} {\bibfnamefont
  {D.}~\bibnamefont {Scarabelli}}, \bibinfo {author} {\bibfnamefont
  {M.}~\bibnamefont {Scheer}}, \bibinfo {author} {\bibfnamefont {E.~A.}\
  \bibnamefont {Sete}}, \bibinfo {author} {\bibfnamefont {P.}~\bibnamefont
  {Sivarajah}}, \bibinfo {author} {\bibfnamefont {R.~S.}\ \bibnamefont
  {Smith}}, \bibinfo {author} {\bibfnamefont {A.}~\bibnamefont {Staley}},
  \bibinfo {author} {\bibfnamefont {N.}~\bibnamefont {Tezak}}, \bibinfo
  {author} {\bibfnamefont {W.~J.}\ \bibnamefont {Zeng}}, \bibinfo {author}
  {\bibfnamefont {A.}~\bibnamefont {Hudson}}, \bibinfo {author} {\bibfnamefont
  {B.~R.}\ \bibnamefont {Johnson}}, \bibinfo {author} {\bibfnamefont
  {M.}~\bibnamefont {Reagor}}, \bibinfo {author} {\bibfnamefont {M.~P.}\
  \bibnamefont {da~Silva}},\ and\ \bibinfo {author} {\bibfnamefont
  {C.}~\bibnamefont {Rigetti}},\ }\href {https://arxiv.org/abs/1712.05771}
  {\bibfield  {journal} {\bibinfo  {journal} {arXiv:1712.05771}\ } (\bibinfo
  {year} {2017})}\BibitemShut {NoStop}%
\bibitem [{\citenamefont {Pagano}\ \emph {et~al.}(2020)\citenamefont {Pagano},
  \citenamefont {Bapat}, \citenamefont {Becker}, \citenamefont {Collins},
  \citenamefont {De}, \citenamefont {Hess}, \citenamefont {Kaplan},
  \citenamefont {Kyprianidis}, \citenamefont {Tan}, \citenamefont {Baldwin},
  \citenamefont {Brady}, \citenamefont {Deshpande}, \citenamefont {Liu},
  \citenamefont {Jordan}, \citenamefont {Gorshkov},\ and\ \citenamefont
  {Monroe}}]{Pagano2020}%
  \BibitemOpen
  \bibfield  {author} {\bibinfo {author} {\bibfnamefont {G.}~\bibnamefont
  {Pagano}}, \bibinfo {author} {\bibfnamefont {A.}~\bibnamefont {Bapat}},
  \bibinfo {author} {\bibfnamefont {P.}~\bibnamefont {Becker}}, \bibinfo
  {author} {\bibfnamefont {K.~S.}\ \bibnamefont {Collins}}, \bibinfo {author}
  {\bibfnamefont {A.}~\bibnamefont {De}}, \bibinfo {author} {\bibfnamefont
  {P.~W.}\ \bibnamefont {Hess}}, \bibinfo {author} {\bibfnamefont {H.~B.}\
  \bibnamefont {Kaplan}}, \bibinfo {author} {\bibfnamefont {A.}~\bibnamefont
  {Kyprianidis}}, \bibinfo {author} {\bibfnamefont {W.~L.}\ \bibnamefont
  {Tan}}, \bibinfo {author} {\bibfnamefont {C.}~\bibnamefont {Baldwin}},
  \bibinfo {author} {\bibfnamefont {L.~T.}\ \bibnamefont {Brady}}, \bibinfo
  {author} {\bibfnamefont {A.}~\bibnamefont {Deshpande}}, \bibinfo {author}
  {\bibfnamefont {F.}~\bibnamefont {Liu}}, \bibinfo {author} {\bibfnamefont
  {S.}~\bibnamefont {Jordan}}, \bibinfo {author} {\bibfnamefont {A.~V.}\
  \bibnamefont {Gorshkov}},\ and\ \bibinfo {author} {\bibfnamefont
  {C.}~\bibnamefont {Monroe}},\ }\href
  {https://doi.org/10.1073/pnas.2006373117} {\bibfield  {journal} {\bibinfo
  {journal} {Proc. Natl. Acad. Sci.}\ }\textbf {\bibinfo {volume} {117}},\
  \bibinfo {pages} {25396} (\bibinfo {year} {2020})}\BibitemShut {NoStop}%
\bibitem [{\citenamefont {Harrigan}\ \emph {et~al.}(2021)\citenamefont
  {Harrigan}, \citenamefont {Sung}, \citenamefont {Neeley}, \citenamefont
  {Satzinger}, \citenamefont {Arute}, \citenamefont {Arya}, \citenamefont
  {Atalaya}, \citenamefont {Bardin}, \citenamefont {Barends}, \citenamefont
  {Boixo}, \citenamefont {Broughton}, \citenamefont {Buckley}, \citenamefont
  {Buell}, \citenamefont {Burkett}, \citenamefont {Bushnell}, \citenamefont
  {Chen}, \citenamefont {Chen}, \citenamefont {Chiaro}, \citenamefont
  {Collins}, \citenamefont {Courtney}, \citenamefont {Demura}, \citenamefont
  {Dunsworth}, \citenamefont {Eppens}, \citenamefont {Fowler}, \citenamefont
  {Foxen}, \citenamefont {Gidney}, \citenamefont {Giustina}, \citenamefont
  {Graff}, \citenamefont {Habegger}, \citenamefont {Ho}, \citenamefont {Hong},
  \citenamefont {Huang}, \citenamefont {Ioffe}, \citenamefont {Isakov},
  \citenamefont {Jeffrey}, \citenamefont {Jiang}, \citenamefont {Jones},
  \citenamefont {Kafri}, \citenamefont {Kechedzhi}, \citenamefont {Kelly},
  \citenamefont {Kim}, \citenamefont {Klimov}, \citenamefont {Korotkov},
  \citenamefont {Kostritsa}, \citenamefont {Landhuis}, \citenamefont {Laptev},
  \citenamefont {Lindmark}, \citenamefont {Leib}, \citenamefont {Martin},
  \citenamefont {Martinis}, \citenamefont {McClean}, \citenamefont {McEwen},
  \citenamefont {Megrant}, \citenamefont {Mi}, \citenamefont {Mohseni},
  \citenamefont {Mruczkiewicz}, \citenamefont {Mutus}, \citenamefont {Naaman},
  \citenamefont {Neill}, \citenamefont {Neukart}, \citenamefont {Niu},
  \citenamefont {O'Brien}, \citenamefont {O'Gorman}, \citenamefont {Ostby},
  \citenamefont {Petukhov}, \citenamefont {Putterman}, \citenamefont
  {Quintana}, \citenamefont {Roushan}, \citenamefont {Rubin}, \citenamefont
  {Sank}, \citenamefont {Skolik}, \citenamefont {Smelyanskiy}, \citenamefont
  {Strain}, \citenamefont {Streif}, \citenamefont {Szalay}, \citenamefont
  {Vainsencher}, \citenamefont {White}, \citenamefont {Yao}, \citenamefont
  {Yeh}, \citenamefont {Zalcman}, \citenamefont {Zhou}, \citenamefont {Neven},
  \citenamefont {Bacon}, \citenamefont {Lucero}, \citenamefont {Farhi},\ and\
  \citenamefont {Babbush}}]{Harrigan2021}%
  \BibitemOpen
  \bibfield  {author} {\bibinfo {author} {\bibfnamefont {M.~P.}\ \bibnamefont
  {Harrigan}}, \bibinfo {author} {\bibfnamefont {K.~J.}\ \bibnamefont {Sung}},
  \bibinfo {author} {\bibfnamefont {M.}~\bibnamefont {Neeley}}, \bibinfo
  {author} {\bibfnamefont {K.~J.}\ \bibnamefont {Satzinger}}, \bibinfo {author}
  {\bibfnamefont {F.}~\bibnamefont {Arute}}, \bibinfo {author} {\bibfnamefont
  {K.}~\bibnamefont {Arya}}, \bibinfo {author} {\bibfnamefont {J.}~\bibnamefont
  {Atalaya}}, \bibinfo {author} {\bibfnamefont {J.~C.}\ \bibnamefont {Bardin}},
  \bibinfo {author} {\bibfnamefont {R.}~\bibnamefont {Barends}}, \bibinfo
  {author} {\bibfnamefont {S.}~\bibnamefont {Boixo}}, \bibinfo {author}
  {\bibfnamefont {M.}~\bibnamefont {Broughton}}, \bibinfo {author}
  {\bibfnamefont {B.~B.}\ \bibnamefont {Buckley}}, \bibinfo {author}
  {\bibfnamefont {D.~A.}\ \bibnamefont {Buell}}, \bibinfo {author}
  {\bibfnamefont {B.}~\bibnamefont {Burkett}}, \bibinfo {author} {\bibfnamefont
  {N.}~\bibnamefont {Bushnell}}, \bibinfo {author} {\bibfnamefont
  {Y.}~\bibnamefont {Chen}}, \bibinfo {author} {\bibfnamefont {Z.}~\bibnamefont
  {Chen}}, \bibinfo {author} {\bibfnamefont {B.}~\bibnamefont {Chiaro}},
  \bibinfo {author} {\bibfnamefont {R.}~\bibnamefont {Collins}}, \bibinfo
  {author} {\bibfnamefont {W.}~\bibnamefont {Courtney}}, \bibinfo {author}
  {\bibfnamefont {S.}~\bibnamefont {Demura}}, \bibinfo {author} {\bibfnamefont
  {A.}~\bibnamefont {Dunsworth}}, \bibinfo {author} {\bibfnamefont
  {D.}~\bibnamefont {Eppens}}, \bibinfo {author} {\bibfnamefont
  {A.}~\bibnamefont {Fowler}}, \bibinfo {author} {\bibfnamefont
  {B.}~\bibnamefont {Foxen}}, \bibinfo {author} {\bibfnamefont
  {C.}~\bibnamefont {Gidney}}, \bibinfo {author} {\bibfnamefont
  {M.}~\bibnamefont {Giustina}}, \bibinfo {author} {\bibfnamefont
  {R.}~\bibnamefont {Graff}}, \bibinfo {author} {\bibfnamefont
  {S.}~\bibnamefont {Habegger}}, \bibinfo {author} {\bibfnamefont
  {A.}~\bibnamefont {Ho}}, \bibinfo {author} {\bibfnamefont {S.}~\bibnamefont
  {Hong}}, \bibinfo {author} {\bibfnamefont {T.}~\bibnamefont {Huang}},
  \bibinfo {author} {\bibfnamefont {L.~B.}\ \bibnamefont {Ioffe}}, \bibinfo
  {author} {\bibfnamefont {S.~V.}\ \bibnamefont {Isakov}}, \bibinfo {author}
  {\bibfnamefont {E.}~\bibnamefont {Jeffrey}}, \bibinfo {author} {\bibfnamefont
  {Z.}~\bibnamefont {Jiang}}, \bibinfo {author} {\bibfnamefont
  {C.}~\bibnamefont {Jones}}, \bibinfo {author} {\bibfnamefont
  {D.}~\bibnamefont {Kafri}}, \bibinfo {author} {\bibfnamefont
  {K.}~\bibnamefont {Kechedzhi}}, \bibinfo {author} {\bibfnamefont
  {J.}~\bibnamefont {Kelly}}, \bibinfo {author} {\bibfnamefont
  {S.}~\bibnamefont {Kim}}, \bibinfo {author} {\bibfnamefont {P.~V.}\
  \bibnamefont {Klimov}}, \bibinfo {author} {\bibfnamefont {A.~N.}\
  \bibnamefont {Korotkov}}, \bibinfo {author} {\bibfnamefont {F.}~\bibnamefont
  {Kostritsa}}, \bibinfo {author} {\bibfnamefont {D.}~\bibnamefont {Landhuis}},
  \bibinfo {author} {\bibfnamefont {P.}~\bibnamefont {Laptev}}, \bibinfo
  {author} {\bibfnamefont {M.}~\bibnamefont {Lindmark}}, \bibinfo {author}
  {\bibfnamefont {M.}~\bibnamefont {Leib}}, \bibinfo {author} {\bibfnamefont
  {O.}~\bibnamefont {Martin}}, \bibinfo {author} {\bibfnamefont {J.~M.}\
  \bibnamefont {Martinis}}, \bibinfo {author} {\bibfnamefont {J.~R.}\
  \bibnamefont {McClean}}, \bibinfo {author} {\bibfnamefont {M.}~\bibnamefont
  {McEwen}}, \bibinfo {author} {\bibfnamefont {A.}~\bibnamefont {Megrant}},
  \bibinfo {author} {\bibfnamefont {X.}~\bibnamefont {Mi}}, \bibinfo {author}
  {\bibfnamefont {M.}~\bibnamefont {Mohseni}}, \bibinfo {author} {\bibfnamefont
  {W.}~\bibnamefont {Mruczkiewicz}}, \bibinfo {author} {\bibfnamefont
  {J.}~\bibnamefont {Mutus}}, \bibinfo {author} {\bibfnamefont
  {O.}~\bibnamefont {Naaman}}, \bibinfo {author} {\bibfnamefont
  {C.}~\bibnamefont {Neill}}, \bibinfo {author} {\bibfnamefont
  {F.}~\bibnamefont {Neukart}}, \bibinfo {author} {\bibfnamefont {M.~Y.}\
  \bibnamefont {Niu}}, \bibinfo {author} {\bibfnamefont {T.~E.}\ \bibnamefont
  {O'Brien}}, \bibinfo {author} {\bibfnamefont {B.}~\bibnamefont {O'Gorman}},
  \bibinfo {author} {\bibfnamefont {E.}~\bibnamefont {Ostby}}, \bibinfo
  {author} {\bibfnamefont {A.}~\bibnamefont {Petukhov}}, \bibinfo {author}
  {\bibfnamefont {H.}~\bibnamefont {Putterman}}, \bibinfo {author}
  {\bibfnamefont {C.}~\bibnamefont {Quintana}}, \bibinfo {author}
  {\bibfnamefont {P.}~\bibnamefont {Roushan}}, \bibinfo {author} {\bibfnamefont
  {N.~C.}\ \bibnamefont {Rubin}}, \bibinfo {author} {\bibfnamefont
  {D.}~\bibnamefont {Sank}}, \bibinfo {author} {\bibfnamefont {A.}~\bibnamefont
  {Skolik}}, \bibinfo {author} {\bibfnamefont {V.}~\bibnamefont {Smelyanskiy}},
  \bibinfo {author} {\bibfnamefont {D.}~\bibnamefont {Strain}}, \bibinfo
  {author} {\bibfnamefont {M.}~\bibnamefont {Streif}}, \bibinfo {author}
  {\bibfnamefont {M.}~\bibnamefont {Szalay}}, \bibinfo {author} {\bibfnamefont
  {A.}~\bibnamefont {Vainsencher}}, \bibinfo {author} {\bibfnamefont
  {T.}~\bibnamefont {White}}, \bibinfo {author} {\bibfnamefont {Z.~J.}\
  \bibnamefont {Yao}}, \bibinfo {author} {\bibfnamefont {P.}~\bibnamefont
  {Yeh}}, \bibinfo {author} {\bibfnamefont {A.}~\bibnamefont {Zalcman}},
  \bibinfo {author} {\bibfnamefont {L.}~\bibnamefont {Zhou}}, \bibinfo {author}
  {\bibfnamefont {H.}~\bibnamefont {Neven}}, \bibinfo {author} {\bibfnamefont
  {D.}~\bibnamefont {Bacon}}, \bibinfo {author} {\bibfnamefont
  {E.}~\bibnamefont {Lucero}}, \bibinfo {author} {\bibfnamefont
  {E.}~\bibnamefont {Farhi}},\ and\ \bibinfo {author} {\bibfnamefont
  {R.}~\bibnamefont {Babbush}},\ }\href
  {https://doi.org/10.1038/s41567-020-01105-y} {\bibfield  {journal} {\bibinfo
  {journal} {Nat. Phys.}\ }\textbf {\bibinfo {volume} {17}},\ \bibinfo {pages}
  {332} (\bibinfo {year} {2021})}\BibitemShut {NoStop}%
\bibitem [{\citenamefont {Ebadi}\ \emph {et~al.}(2022)\citenamefont {Ebadi},
  \citenamefont {Keesling}, \citenamefont {Cain}, \citenamefont {Wang},
  \citenamefont {Levine}, \citenamefont {Bluvstein}, \citenamefont {Semeghini},
  \citenamefont {Omran}, \citenamefont {Liu}, \citenamefont {Samajdar},
  \citenamefont {Luo}, \citenamefont {Nash}, \citenamefont {Gao}, \citenamefont
  {Barak}, \citenamefont {Farhi}, \citenamefont {Sachdev}, \citenamefont
  {Gemelke}, \citenamefont {Zhou}, \citenamefont {Choi}, \citenamefont
  {Pichler}, \citenamefont {Wang}, \citenamefont {Greiner}, \citenamefont
  {Vuletić},\ and\ \citenamefont {Lukin}}]{ebadi2022quantum}%
  \BibitemOpen
  \bibfield  {author} {\bibinfo {author} {\bibfnamefont {S.}~\bibnamefont
  {Ebadi}}, \bibinfo {author} {\bibfnamefont {A.}~\bibnamefont {Keesling}},
  \bibinfo {author} {\bibfnamefont {M.}~\bibnamefont {Cain}}, \bibinfo {author}
  {\bibfnamefont {T.~T.}\ \bibnamefont {Wang}}, \bibinfo {author}
  {\bibfnamefont {H.}~\bibnamefont {Levine}}, \bibinfo {author} {\bibfnamefont
  {D.}~\bibnamefont {Bluvstein}}, \bibinfo {author} {\bibfnamefont
  {G.}~\bibnamefont {Semeghini}}, \bibinfo {author} {\bibfnamefont
  {A.}~\bibnamefont {Omran}}, \bibinfo {author} {\bibfnamefont {J.-G.}\
  \bibnamefont {Liu}}, \bibinfo {author} {\bibfnamefont {R.}~\bibnamefont
  {Samajdar}}, \bibinfo {author} {\bibfnamefont {X.-Z.}\ \bibnamefont {Luo}},
  \bibinfo {author} {\bibfnamefont {B.}~\bibnamefont {Nash}}, \bibinfo {author}
  {\bibfnamefont {X.}~\bibnamefont {Gao}}, \bibinfo {author} {\bibfnamefont
  {B.}~\bibnamefont {Barak}}, \bibinfo {author} {\bibfnamefont
  {E.}~\bibnamefont {Farhi}}, \bibinfo {author} {\bibfnamefont
  {S.}~\bibnamefont {Sachdev}}, \bibinfo {author} {\bibfnamefont
  {N.}~\bibnamefont {Gemelke}}, \bibinfo {author} {\bibfnamefont
  {L.}~\bibnamefont {Zhou}}, \bibinfo {author} {\bibfnamefont {S.}~\bibnamefont
  {Choi}}, \bibinfo {author} {\bibfnamefont {H.}~\bibnamefont {Pichler}},
  \bibinfo {author} {\bibfnamefont {S.-T.}\ \bibnamefont {Wang}}, \bibinfo
  {author} {\bibfnamefont {M.}~\bibnamefont {Greiner}}, \bibinfo {author}
  {\bibfnamefont {V.}~\bibnamefont {Vuletić}},\ and\ \bibinfo {author}
  {\bibfnamefont {M.~D.}\ \bibnamefont {Lukin}},\ }\href
  {https://doi.org/10.1126/science.abo6587} {\bibfield  {journal} {\bibinfo
  {journal} {Science}\ }\textbf {\bibinfo {volume} {376}},\ \bibinfo {pages}
  {1209} (\bibinfo {year} {2022})}\BibitemShut {NoStop}%
\bibitem [{\citenamefont {Graham}\ \emph {et~al.}(2022)\citenamefont {Graham},
  \citenamefont {Song}, \citenamefont {Scott}, \citenamefont {Poole},
  \citenamefont {Phuttitarn}, \citenamefont {Jooya}, \citenamefont {Eichler},
  \citenamefont {Jiang}, \citenamefont {Marra}, \citenamefont {Grinkemeyer},
  \citenamefont {Kwon}, \citenamefont {Ebert}, \citenamefont {Cherek},
  \citenamefont {Lichtman}, \citenamefont {Gillette}, \citenamefont {Gilbert},
  \citenamefont {Bowman}, \citenamefont {Ballance}, \citenamefont {Campbell},
  \citenamefont {Dahl}, \citenamefont {Crawford}, \citenamefont {Blunt},
  \citenamefont {Rogers}, \citenamefont {Noel},\ and\ \citenamefont
  {Saffman}}]{graham2021demonstration}%
  \BibitemOpen
  \bibfield  {author} {\bibinfo {author} {\bibfnamefont {T.~M.}\ \bibnamefont
  {Graham}}, \bibinfo {author} {\bibfnamefont {Y.}~\bibnamefont {Song}},
  \bibinfo {author} {\bibfnamefont {J.}~\bibnamefont {Scott}}, \bibinfo
  {author} {\bibfnamefont {C.}~\bibnamefont {Poole}}, \bibinfo {author}
  {\bibfnamefont {L.}~\bibnamefont {Phuttitarn}}, \bibinfo {author}
  {\bibfnamefont {K.}~\bibnamefont {Jooya}}, \bibinfo {author} {\bibfnamefont
  {P.}~\bibnamefont {Eichler}}, \bibinfo {author} {\bibfnamefont
  {X.}~\bibnamefont {Jiang}}, \bibinfo {author} {\bibfnamefont
  {A.}~\bibnamefont {Marra}}, \bibinfo {author} {\bibfnamefont
  {B.}~\bibnamefont {Grinkemeyer}}, \bibinfo {author} {\bibfnamefont
  {M.}~\bibnamefont {Kwon}}, \bibinfo {author} {\bibfnamefont {M.}~\bibnamefont
  {Ebert}}, \bibinfo {author} {\bibfnamefont {J.}~\bibnamefont {Cherek}},
  \bibinfo {author} {\bibfnamefont {M.~T.}\ \bibnamefont {Lichtman}}, \bibinfo
  {author} {\bibfnamefont {M.}~\bibnamefont {Gillette}}, \bibinfo {author}
  {\bibfnamefont {J.}~\bibnamefont {Gilbert}}, \bibinfo {author} {\bibfnamefont
  {D.}~\bibnamefont {Bowman}}, \bibinfo {author} {\bibfnamefont
  {T.}~\bibnamefont {Ballance}}, \bibinfo {author} {\bibfnamefont
  {C.}~\bibnamefont {Campbell}}, \bibinfo {author} {\bibfnamefont {E.~D.}\
  \bibnamefont {Dahl}}, \bibinfo {author} {\bibfnamefont {O.}~\bibnamefont
  {Crawford}}, \bibinfo {author} {\bibfnamefont {N.~S.}\ \bibnamefont {Blunt}},
  \bibinfo {author} {\bibfnamefont {B.}~\bibnamefont {Rogers}}, \bibinfo
  {author} {\bibfnamefont {T.}~\bibnamefont {Noel}},\ and\ \bibinfo {author}
  {\bibfnamefont {M.}~\bibnamefont {Saffman}},\ }\href
  {https://doi.org/10.1038/s41586-022-04603-6} {\bibfield  {journal} {\bibinfo
  {journal} {Nature}\ }\textbf {\bibinfo {volume} {604}},\ \bibinfo {pages}
  {457} (\bibinfo {year} {2022})}\BibitemShut {NoStop}%
\bibitem [{\citenamefont {Pelofske}\ \emph {et~al.}(2023)\citenamefont
  {Pelofske}, \citenamefont {B\"artschi},\ and\ \citenamefont
  {Eidenbenz}}]{Pelofske2023}%
  \BibitemOpen
  \bibfield  {author} {\bibinfo {author} {\bibfnamefont {E.}~\bibnamefont
  {Pelofske}}, \bibinfo {author} {\bibfnamefont {A.}~\bibnamefont
  {B\"artschi}},\ and\ \bibinfo {author} {\bibfnamefont {S.}~\bibnamefont
  {Eidenbenz}},\ }\href {https://arxiv.org/abs/2301.00520} {\bibfield
  {journal} {\bibinfo  {journal} {arXiv:2301.00520}\ } (\bibinfo {year}
  {2023})}\BibitemShut {NoStop}%
\bibitem [{\citenamefont {Moses}\ \emph {et~al.}(2023)\citenamefont {Moses},
  \citenamefont {Baldwin}, \citenamefont {Allman}, \citenamefont {Ancona},
  \citenamefont {Ascarrunz}, \citenamefont {Barnes}, \citenamefont
  {Bartolotta}, \citenamefont {Bjork}, \citenamefont {Blanchard}, \citenamefont
  {Bohn}, \citenamefont {Bohnet}, \citenamefont {Brown}, \citenamefont
  {Burdick}, \citenamefont {Burton}, \citenamefont {Campbell}, \citenamefont
  {III}, \citenamefont {Carron}, \citenamefont {Chambers}, \citenamefont
  {Chan}, \citenamefont {Chen}, \citenamefont {Chernoguzov}, \citenamefont
  {Chertkov}, \citenamefont {Colina}, \citenamefont {Curtis}, \citenamefont
  {Daniel}, \citenamefont {DeCross}, \citenamefont {Deen}, \citenamefont
  {Delaney}, \citenamefont {Dreiling}, \citenamefont {Ertsgaard}, \citenamefont
  {Esposito}, \citenamefont {Estey}, \citenamefont {Fabrikant}, \citenamefont
  {Figgatt}, \citenamefont {Foltz}, \citenamefont {Foss-Feig}, \citenamefont
  {Francois}, \citenamefont {Gaebler}, \citenamefont {Gatterman}, \citenamefont
  {Gilbreth}, \citenamefont {Giles}, \citenamefont {Glynn}, \citenamefont
  {Hall}, \citenamefont {Hankin}, \citenamefont {Hansen}, \citenamefont
  {Hayes}, \citenamefont {Higashi}, \citenamefont {Hoffman}, \citenamefont
  {Horning}, \citenamefont {Hout}, \citenamefont {Jacobs}, \citenamefont
  {Johansen}, \citenamefont {Jones}, \citenamefont {Karcz}, \citenamefont
  {Klein}, \citenamefont {Lauria}, \citenamefont {Lee}, \citenamefont {Liefer},
  \citenamefont {Lytle}, \citenamefont {Lu}, \citenamefont {Lucchetti},
  \citenamefont {Malm}, \citenamefont {Matheny}, \citenamefont {Mathewson},
  \citenamefont {Mayer}, \citenamefont {Miller}, \citenamefont {Mills},
  \citenamefont {Neyenhuis}, \citenamefont {Nugent}, \citenamefont {Olson},
  \citenamefont {Parks}, \citenamefont {Price}, \citenamefont {Price},
  \citenamefont {Pugh}, \citenamefont {Ransford}, \citenamefont {Reed},
  \citenamefont {Roman}, \citenamefont {Rowe}, \citenamefont {Ryan-Anderson},
  \citenamefont {Sanders}, \citenamefont {Sedlacek}, \citenamefont {Shevchuk},
  \citenamefont {Siegfried}, \citenamefont {Skripka}, \citenamefont {Spaun},
  \citenamefont {Sprenkle}, \citenamefont {Stutz}, \citenamefont {Swallows},
  \citenamefont {Tobey}, \citenamefont {Tran}, \citenamefont {Tran},
  \citenamefont {Vogt}, \citenamefont {Volin}, \citenamefont {Walker},
  \citenamefont {Zolot},\ and\ \citenamefont {Pino}}]{moses2023race}%
  \BibitemOpen
  \bibfield  {author} {\bibinfo {author} {\bibfnamefont {S.~A.}\ \bibnamefont
  {Moses}}, \bibinfo {author} {\bibfnamefont {C.~H.}\ \bibnamefont {Baldwin}},
  \bibinfo {author} {\bibfnamefont {M.~S.}\ \bibnamefont {Allman}}, \bibinfo
  {author} {\bibfnamefont {R.}~\bibnamefont {Ancona}}, \bibinfo {author}
  {\bibfnamefont {L.}~\bibnamefont {Ascarrunz}}, \bibinfo {author}
  {\bibfnamefont {C.}~\bibnamefont {Barnes}}, \bibinfo {author} {\bibfnamefont
  {J.}~\bibnamefont {Bartolotta}}, \bibinfo {author} {\bibfnamefont
  {B.}~\bibnamefont {Bjork}}, \bibinfo {author} {\bibfnamefont
  {P.}~\bibnamefont {Blanchard}}, \bibinfo {author} {\bibfnamefont
  {M.}~\bibnamefont {Bohn}}, \bibinfo {author} {\bibfnamefont {J.~G.}\
  \bibnamefont {Bohnet}}, \bibinfo {author} {\bibfnamefont {N.~C.}\
  \bibnamefont {Brown}}, \bibinfo {author} {\bibfnamefont {N.~Q.}\ \bibnamefont
  {Burdick}}, \bibinfo {author} {\bibfnamefont {W.~C.}\ \bibnamefont {Burton}},
  \bibinfo {author} {\bibfnamefont {S.~L.}\ \bibnamefont {Campbell}}, \bibinfo
  {author} {\bibfnamefont {J.~P.~C.}\ \bibnamefont {III}}, \bibinfo {author}
  {\bibfnamefont {C.}~\bibnamefont {Carron}}, \bibinfo {author} {\bibfnamefont
  {J.}~\bibnamefont {Chambers}}, \bibinfo {author} {\bibfnamefont {J.~W.}\
  \bibnamefont {Chan}}, \bibinfo {author} {\bibfnamefont {Y.~H.}\ \bibnamefont
  {Chen}}, \bibinfo {author} {\bibfnamefont {A.}~\bibnamefont {Chernoguzov}},
  \bibinfo {author} {\bibfnamefont {E.}~\bibnamefont {Chertkov}}, \bibinfo
  {author} {\bibfnamefont {J.}~\bibnamefont {Colina}}, \bibinfo {author}
  {\bibfnamefont {J.~P.}\ \bibnamefont {Curtis}}, \bibinfo {author}
  {\bibfnamefont {R.}~\bibnamefont {Daniel}}, \bibinfo {author} {\bibfnamefont
  {M.}~\bibnamefont {DeCross}}, \bibinfo {author} {\bibfnamefont
  {D.}~\bibnamefont {Deen}}, \bibinfo {author} {\bibfnamefont {C.}~\bibnamefont
  {Delaney}}, \bibinfo {author} {\bibfnamefont {J.~M.}\ \bibnamefont
  {Dreiling}}, \bibinfo {author} {\bibfnamefont {C.~T.}\ \bibnamefont
  {Ertsgaard}}, \bibinfo {author} {\bibfnamefont {J.}~\bibnamefont {Esposito}},
  \bibinfo {author} {\bibfnamefont {B.}~\bibnamefont {Estey}}, \bibinfo
  {author} {\bibfnamefont {M.}~\bibnamefont {Fabrikant}}, \bibinfo {author}
  {\bibfnamefont {C.}~\bibnamefont {Figgatt}}, \bibinfo {author} {\bibfnamefont
  {C.}~\bibnamefont {Foltz}}, \bibinfo {author} {\bibfnamefont
  {M.}~\bibnamefont {Foss-Feig}}, \bibinfo {author} {\bibfnamefont
  {D.}~\bibnamefont {Francois}}, \bibinfo {author} {\bibfnamefont {J.~P.}\
  \bibnamefont {Gaebler}}, \bibinfo {author} {\bibfnamefont {T.~M.}\
  \bibnamefont {Gatterman}}, \bibinfo {author} {\bibfnamefont {C.~N.}\
  \bibnamefont {Gilbreth}}, \bibinfo {author} {\bibfnamefont {J.}~\bibnamefont
  {Giles}}, \bibinfo {author} {\bibfnamefont {E.}~\bibnamefont {Glynn}},
  \bibinfo {author} {\bibfnamefont {A.}~\bibnamefont {Hall}}, \bibinfo {author}
  {\bibfnamefont {A.~M.}\ \bibnamefont {Hankin}}, \bibinfo {author}
  {\bibfnamefont {A.}~\bibnamefont {Hansen}}, \bibinfo {author} {\bibfnamefont
  {D.}~\bibnamefont {Hayes}}, \bibinfo {author} {\bibfnamefont
  {B.}~\bibnamefont {Higashi}}, \bibinfo {author} {\bibfnamefont {I.~M.}\
  \bibnamefont {Hoffman}}, \bibinfo {author} {\bibfnamefont {B.}~\bibnamefont
  {Horning}}, \bibinfo {author} {\bibfnamefont {J.~J.}\ \bibnamefont {Hout}},
  \bibinfo {author} {\bibfnamefont {R.}~\bibnamefont {Jacobs}}, \bibinfo
  {author} {\bibfnamefont {J.}~\bibnamefont {Johansen}}, \bibinfo {author}
  {\bibfnamefont {L.}~\bibnamefont {Jones}}, \bibinfo {author} {\bibfnamefont
  {J.}~\bibnamefont {Karcz}}, \bibinfo {author} {\bibfnamefont
  {T.}~\bibnamefont {Klein}}, \bibinfo {author} {\bibfnamefont
  {P.}~\bibnamefont {Lauria}}, \bibinfo {author} {\bibfnamefont
  {P.}~\bibnamefont {Lee}}, \bibinfo {author} {\bibfnamefont {D.}~\bibnamefont
  {Liefer}}, \bibinfo {author} {\bibfnamefont {C.}~\bibnamefont {Lytle}},
  \bibinfo {author} {\bibfnamefont {S.~T.}\ \bibnamefont {Lu}}, \bibinfo
  {author} {\bibfnamefont {D.}~\bibnamefont {Lucchetti}}, \bibinfo {author}
  {\bibfnamefont {A.}~\bibnamefont {Malm}}, \bibinfo {author} {\bibfnamefont
  {M.}~\bibnamefont {Matheny}}, \bibinfo {author} {\bibfnamefont
  {B.}~\bibnamefont {Mathewson}}, \bibinfo {author} {\bibfnamefont
  {K.}~\bibnamefont {Mayer}}, \bibinfo {author} {\bibfnamefont {D.~B.}\
  \bibnamefont {Miller}}, \bibinfo {author} {\bibfnamefont {M.}~\bibnamefont
  {Mills}}, \bibinfo {author} {\bibfnamefont {B.}~\bibnamefont {Neyenhuis}},
  \bibinfo {author} {\bibfnamefont {L.}~\bibnamefont {Nugent}}, \bibinfo
  {author} {\bibfnamefont {S.}~\bibnamefont {Olson}}, \bibinfo {author}
  {\bibfnamefont {J.}~\bibnamefont {Parks}}, \bibinfo {author} {\bibfnamefont
  {G.~N.}\ \bibnamefont {Price}}, \bibinfo {author} {\bibfnamefont
  {Z.}~\bibnamefont {Price}}, \bibinfo {author} {\bibfnamefont
  {M.}~\bibnamefont {Pugh}}, \bibinfo {author} {\bibfnamefont {A.}~\bibnamefont
  {Ransford}}, \bibinfo {author} {\bibfnamefont {A.~P.}\ \bibnamefont {Reed}},
  \bibinfo {author} {\bibfnamefont {C.}~\bibnamefont {Roman}}, \bibinfo
  {author} {\bibfnamefont {M.}~\bibnamefont {Rowe}}, \bibinfo {author}
  {\bibfnamefont {C.}~\bibnamefont {Ryan-Anderson}}, \bibinfo {author}
  {\bibfnamefont {S.}~\bibnamefont {Sanders}}, \bibinfo {author} {\bibfnamefont
  {J.}~\bibnamefont {Sedlacek}}, \bibinfo {author} {\bibfnamefont
  {P.}~\bibnamefont {Shevchuk}}, \bibinfo {author} {\bibfnamefont
  {P.}~\bibnamefont {Siegfried}}, \bibinfo {author} {\bibfnamefont
  {T.}~\bibnamefont {Skripka}}, \bibinfo {author} {\bibfnamefont
  {B.}~\bibnamefont {Spaun}}, \bibinfo {author} {\bibfnamefont {R.~T.}\
  \bibnamefont {Sprenkle}}, \bibinfo {author} {\bibfnamefont {R.~P.}\
  \bibnamefont {Stutz}}, \bibinfo {author} {\bibfnamefont {M.}~\bibnamefont
  {Swallows}}, \bibinfo {author} {\bibfnamefont {R.~I.}\ \bibnamefont {Tobey}},
  \bibinfo {author} {\bibfnamefont {A.}~\bibnamefont {Tran}}, \bibinfo {author}
  {\bibfnamefont {T.}~\bibnamefont {Tran}}, \bibinfo {author} {\bibfnamefont
  {E.}~\bibnamefont {Vogt}}, \bibinfo {author} {\bibfnamefont {C.}~\bibnamefont
  {Volin}}, \bibinfo {author} {\bibfnamefont {J.}~\bibnamefont {Walker}},
  \bibinfo {author} {\bibfnamefont {A.~M.}\ \bibnamefont {Zolot}},\ and\
  \bibinfo {author} {\bibfnamefont {J.~M.}\ \bibnamefont {Pino}},\ }\href
  {https://arxiv.org/abs/2305.03828} {\bibfield  {journal} {\bibinfo  {journal}
  {arXiv:2305.03828}\ } (\bibinfo {year} {2023})}\BibitemShut {NoStop}%
\bibitem [{\citenamefont {Shaydulin}\ \emph {et~al.}(2023)\citenamefont
  {Shaydulin}, \citenamefont {Li}, \citenamefont {Chakrabarti}, \citenamefont
  {DeCross}, \citenamefont {Herman}, \citenamefont {Kumar}, \citenamefont
  {Larson}, \citenamefont {Lykov}, \citenamefont {Minssen}, \citenamefont
  {Sun}, \citenamefont {Alexeev}, \citenamefont {Dreiling}, \citenamefont
  {Gaebler}, \citenamefont {Gatterman}, \citenamefont {Gerber}, \citenamefont
  {Gilmore}, \citenamefont {Gresh}, \citenamefont {Hewitt}, \citenamefont
  {Horst}, \citenamefont {Hu}, \citenamefont {Johansen}, \citenamefont
  {Matheny}, \citenamefont {Mengle}, \citenamefont {Mills}, \citenamefont
  {Moses}, \citenamefont {Neyenhuis}, \citenamefont {Siegfried}, \citenamefont
  {Yalovetzky},\ and\ \citenamefont {Pistoia}}]{Shaydulin2023}%
  \BibitemOpen
  \bibfield  {author} {\bibinfo {author} {\bibfnamefont {R.}~\bibnamefont
  {Shaydulin}}, \bibinfo {author} {\bibfnamefont {C.}~\bibnamefont {Li}},
  \bibinfo {author} {\bibfnamefont {S.}~\bibnamefont {Chakrabarti}}, \bibinfo
  {author} {\bibfnamefont {M.}~\bibnamefont {DeCross}}, \bibinfo {author}
  {\bibfnamefont {D.}~\bibnamefont {Herman}}, \bibinfo {author} {\bibfnamefont
  {N.}~\bibnamefont {Kumar}}, \bibinfo {author} {\bibfnamefont
  {J.}~\bibnamefont {Larson}}, \bibinfo {author} {\bibfnamefont
  {D.}~\bibnamefont {Lykov}}, \bibinfo {author} {\bibfnamefont
  {P.}~\bibnamefont {Minssen}}, \bibinfo {author} {\bibfnamefont
  {Y.}~\bibnamefont {Sun}}, \bibinfo {author} {\bibfnamefont {Y.}~\bibnamefont
  {Alexeev}}, \bibinfo {author} {\bibfnamefont {J.~M.}\ \bibnamefont
  {Dreiling}}, \bibinfo {author} {\bibfnamefont {J.~P.}\ \bibnamefont
  {Gaebler}}, \bibinfo {author} {\bibfnamefont {T.~M.}\ \bibnamefont
  {Gatterman}}, \bibinfo {author} {\bibfnamefont {J.~A.}\ \bibnamefont
  {Gerber}}, \bibinfo {author} {\bibfnamefont {K.}~\bibnamefont {Gilmore}},
  \bibinfo {author} {\bibfnamefont {D.}~\bibnamefont {Gresh}}, \bibinfo
  {author} {\bibfnamefont {N.}~\bibnamefont {Hewitt}}, \bibinfo {author}
  {\bibfnamefont {C.~V.}\ \bibnamefont {Horst}}, \bibinfo {author}
  {\bibfnamefont {S.}~\bibnamefont {Hu}}, \bibinfo {author} {\bibfnamefont
  {J.}~\bibnamefont {Johansen}}, \bibinfo {author} {\bibfnamefont
  {M.}~\bibnamefont {Matheny}}, \bibinfo {author} {\bibfnamefont
  {T.}~\bibnamefont {Mengle}}, \bibinfo {author} {\bibfnamefont
  {M.}~\bibnamefont {Mills}}, \bibinfo {author} {\bibfnamefont {S.~A.}\
  \bibnamefont {Moses}}, \bibinfo {author} {\bibfnamefont {B.}~\bibnamefont
  {Neyenhuis}}, \bibinfo {author} {\bibfnamefont {P.}~\bibnamefont
  {Siegfried}}, \bibinfo {author} {\bibfnamefont {R.}~\bibnamefont
  {Yalovetzky}},\ and\ \bibinfo {author} {\bibfnamefont {M.}~\bibnamefont
  {Pistoia}},\ }\href {https://arxiv.org/abs/2308.02342} {\bibfield  {journal}
  {\bibinfo  {journal} {arXiv:2308.02342}\ } (\bibinfo {year}
  {2023})}\BibitemShut {NoStop}%
\bibitem [{\citenamefont {Sack}\ and\ \citenamefont {Egger}(2023)}]{Sack2023}%
  \BibitemOpen
  \bibfield  {author} {\bibinfo {author} {\bibfnamefont {S.~H.}\ \bibnamefont
  {Sack}}\ and\ \bibinfo {author} {\bibfnamefont {D.~J.}\ \bibnamefont
  {Egger}},\ }\href {https://arxiv.org/abs/2307.14427} {\bibfield  {journal}
  {\bibinfo  {journal} {arXiv:2307.14427}\ } (\bibinfo {year}
  {2023})}\BibitemShut {NoStop}%
\bibitem [{\citenamefont {Maciejewski}\ \emph {et~al.}(2023)\citenamefont
  {Maciejewski}, \citenamefont {Hadfield}, \citenamefont {Hall}, \citenamefont
  {Hodson}, \citenamefont {Dupont}, \citenamefont {Evert}, \citenamefont {Sud},
  \citenamefont {Alam}, \citenamefont {Wang}, \citenamefont {Jeffrey},
  \citenamefont {Sundar}, \citenamefont {Lott}, \citenamefont {Grabbe},
  \citenamefont {Rieffel}, \citenamefont {Reagor},\ and\ \citenamefont
  {Venturelli}}]{Maciejewski2023}%
  \BibitemOpen
  \bibfield  {author} {\bibinfo {author} {\bibfnamefont {F.~B.}\ \bibnamefont
  {Maciejewski}}, \bibinfo {author} {\bibfnamefont {S.}~\bibnamefont
  {Hadfield}}, \bibinfo {author} {\bibfnamefont {B.}~\bibnamefont {Hall}},
  \bibinfo {author} {\bibfnamefont {M.}~\bibnamefont {Hodson}}, \bibinfo
  {author} {\bibfnamefont {M.}~\bibnamefont {Dupont}}, \bibinfo {author}
  {\bibfnamefont {B.}~\bibnamefont {Evert}}, \bibinfo {author} {\bibfnamefont
  {J.}~\bibnamefont {Sud}}, \bibinfo {author} {\bibfnamefont {M.~S.}\
  \bibnamefont {Alam}}, \bibinfo {author} {\bibfnamefont {Z.}~\bibnamefont
  {Wang}}, \bibinfo {author} {\bibfnamefont {S.}~\bibnamefont {Jeffrey}},
  \bibinfo {author} {\bibfnamefont {B.}~\bibnamefont {Sundar}}, \bibinfo
  {author} {\bibfnamefont {P.~A.}\ \bibnamefont {Lott}}, \bibinfo {author}
  {\bibfnamefont {S.}~\bibnamefont {Grabbe}}, \bibinfo {author} {\bibfnamefont
  {E.~G.}\ \bibnamefont {Rieffel}}, \bibinfo {author} {\bibfnamefont {M.~J.}\
  \bibnamefont {Reagor}},\ and\ \bibinfo {author} {\bibfnamefont
  {D.}~\bibnamefont {Venturelli}},\ }\href {https://arxiv.org/abs/2308.12423}
  {\bibfield  {journal} {\bibinfo  {journal} {arXiv:2308.12423}\ } (\bibinfo
  {year} {2023})}\BibitemShut {NoStop}%
\bibitem [{\citenamefont {Zhu}\ \emph {et~al.}(2022)\citenamefont {Zhu},
  \citenamefont {Zhang}, \citenamefont {Sundar}, \citenamefont {Green},
  \citenamefont {Alderete}, \citenamefont {Nguyen}, \citenamefont {Hazzard},\
  and\ \citenamefont {Linke}}]{zhu2022multi}%
  \BibitemOpen
  \bibfield  {author} {\bibinfo {author} {\bibfnamefont {Y.}~\bibnamefont
  {Zhu}}, \bibinfo {author} {\bibfnamefont {Z.}~\bibnamefont {Zhang}}, \bibinfo
  {author} {\bibfnamefont {B.}~\bibnamefont {Sundar}}, \bibinfo {author}
  {\bibfnamefont {A.~M.}\ \bibnamefont {Green}}, \bibinfo {author}
  {\bibfnamefont {C.~H.}\ \bibnamefont {Alderete}}, \bibinfo {author}
  {\bibfnamefont {N.~H.}\ \bibnamefont {Nguyen}}, \bibinfo {author}
  {\bibfnamefont {K.~R.~A.}\ \bibnamefont {Hazzard}},\ and\ \bibinfo {author}
  {\bibfnamefont {N.~M.}\ \bibnamefont {Linke}},\ }\href
  {https://doi.org/10.1088/2058-9565/ac91ef} {\bibfield  {journal} {\bibinfo
  {journal} {Quantum Sci. Technol.}\ }\textbf {\bibinfo {volume} {8}},\
  \bibinfo {pages} {015007} (\bibinfo {year} {2022})}\BibitemShut {NoStop}%
\bibitem [{\citenamefont {Farhi}\ \emph {et~al.}(2017)\citenamefont {Farhi},
  \citenamefont {Goldstone}, \citenamefont {Gutmann},\ and\ \citenamefont
  {Neven}}]{Farhi2017}%
  \BibitemOpen
  \bibfield  {author} {\bibinfo {author} {\bibfnamefont {E.}~\bibnamefont
  {Farhi}}, \bibinfo {author} {\bibfnamefont {J.}~\bibnamefont {Goldstone}},
  \bibinfo {author} {\bibfnamefont {S.}~\bibnamefont {Gutmann}},\ and\ \bibinfo
  {author} {\bibfnamefont {H.}~\bibnamefont {Neven}},\ }\href
  {https://arxiv.org/abs/1703.06199} {\bibfield  {journal} {\bibinfo  {journal}
  {arXiv:1703.06199}\ } (\bibinfo {year} {2017})}\BibitemShut {NoStop}%
\bibitem [{\citenamefont {Wang}\ \emph {et~al.}(2018)\citenamefont {Wang},
  \citenamefont {Hadfield}, \citenamefont {Jiang},\ and\ \citenamefont
  {Rieffel}}]{PhysRevA.97.022304}%
  \BibitemOpen
  \bibfield  {author} {\bibinfo {author} {\bibfnamefont {Z.}~\bibnamefont
  {Wang}}, \bibinfo {author} {\bibfnamefont {S.}~\bibnamefont {Hadfield}},
  \bibinfo {author} {\bibfnamefont {Z.}~\bibnamefont {Jiang}},\ and\ \bibinfo
  {author} {\bibfnamefont {E.~G.}\ \bibnamefont {Rieffel}},\ }\href
  {https://doi.org/10.1103/PhysRevA.97.022304} {\bibfield  {journal} {\bibinfo
  {journal} {Phys. Rev. A}\ }\textbf {\bibinfo {volume} {97}},\ \bibinfo
  {pages} {022304} (\bibinfo {year} {2018})}\BibitemShut {NoStop}%
\bibitem [{\citenamefont {Marwaha}(2021)}]{Marwaha2021localclassicalmax}%
  \BibitemOpen
  \bibfield  {author} {\bibinfo {author} {\bibfnamefont {K.}~\bibnamefont
  {Marwaha}},\ }\href {https://doi.org/10.22331/q-2021-04-20-437} {\bibfield
  {journal} {\bibinfo  {journal} {{Quantum}}\ }\textbf {\bibinfo {volume}
  {5}},\ \bibinfo {pages} {437} (\bibinfo {year} {2021})}\BibitemShut {NoStop}%
\bibitem [{\citenamefont {Wurtz}\ and\ \citenamefont
  {Love}(2021)}]{PhysRevA.103.042612}%
  \BibitemOpen
  \bibfield  {author} {\bibinfo {author} {\bibfnamefont {J.}~\bibnamefont
  {Wurtz}}\ and\ \bibinfo {author} {\bibfnamefont {P.}~\bibnamefont {Love}},\
  }\href {https://doi.org/10.1103/PhysRevA.103.042612} {\bibfield  {journal}
  {\bibinfo  {journal} {Phys. Rev. A}\ }\textbf {\bibinfo {volume} {103}},\
  \bibinfo {pages} {042612} (\bibinfo {year} {2021})}\BibitemShut {NoStop}%
\bibitem [{\citenamefont {Basso}\ \emph {et~al.}(2022)\citenamefont {Basso},
  \citenamefont {Farhi}, \citenamefont {Marwaha}, \citenamefont {Villalonga},\
  and\ \citenamefont {Zhou}}]{basso_et_al}%
  \BibitemOpen
  \bibfield  {author} {\bibinfo {author} {\bibfnamefont {J.}~\bibnamefont
  {Basso}}, \bibinfo {author} {\bibfnamefont {E.}~\bibnamefont {Farhi}},
  \bibinfo {author} {\bibfnamefont {K.}~\bibnamefont {Marwaha}}, \bibinfo
  {author} {\bibfnamefont {B.}~\bibnamefont {Villalonga}},\ and\ \bibinfo
  {author} {\bibfnamefont {L.}~\bibnamefont {Zhou}},\ }in\ \href
  {https://doi.org/10.4230/LIPIcs.TQC.2022.7} {\emph {\bibinfo {booktitle}
  {17th Conference on the Theory of Quantum Computation, Communication and
  Cryptography (TQC 2022)}}},\ \bibinfo {series} {Leibniz International
  Proceedings in Informatics (LIPIcs)}, Vol.\ \bibinfo {volume} {232},\
  \bibinfo {editor} {edited by\ \bibinfo {editor} {\bibfnamefont
  {F.}~\bibnamefont {Le~Gall}}\ and\ \bibinfo {editor} {\bibfnamefont
  {T.}~\bibnamefont {Morimae}}}\ (\bibinfo  {publisher} {Schloss Dagstuhl --
  Leibniz-Zentrum f{\"u}r Informatik},\ \bibinfo {address} {Dagstuhl,
  Germany},\ \bibinfo {year} {2022})\ pp.\ \bibinfo {pages}
  {7:1--7:21}\BibitemShut {NoStop}%
\bibitem [{\citenamefont {Farhi}\ \emph {et~al.}(2022)\citenamefont {Farhi},
  \citenamefont {Goldstone}, \citenamefont {Gutmann},\ and\ \citenamefont
  {Zhou}}]{Farhi2022quantumapproximate}%
  \BibitemOpen
  \bibfield  {author} {\bibinfo {author} {\bibfnamefont {E.}~\bibnamefont
  {Farhi}}, \bibinfo {author} {\bibfnamefont {J.}~\bibnamefont {Goldstone}},
  \bibinfo {author} {\bibfnamefont {S.}~\bibnamefont {Gutmann}},\ and\ \bibinfo
  {author} {\bibfnamefont {L.}~\bibnamefont {Zhou}},\ }\href
  {https://doi.org/10.22331/q-2022-07-07-759} {\bibfield  {journal} {\bibinfo
  {journal} {{Quantum}}\ }\textbf {\bibinfo {volume} {6}},\ \bibinfo {pages}
  {759} (\bibinfo {year} {2022})}\BibitemShut {NoStop}%
\bibitem [{\citenamefont {Marwaha}\ and\ \citenamefont
  {Hadfield}(2022)}]{marwaha2022bounds}%
  \BibitemOpen
  \bibfield  {author} {\bibinfo {author} {\bibfnamefont {K.}~\bibnamefont
  {Marwaha}}\ and\ \bibinfo {author} {\bibfnamefont {S.}~\bibnamefont
  {Hadfield}},\ }\href@noop {} {\bibfield  {journal} {\bibinfo  {journal}
  {Quantum}\ }\textbf {\bibinfo {volume} {6}},\ \bibinfo {pages} {757}
  (\bibinfo {year} {2022})}\BibitemShut {NoStop}%
\bibitem [{\citenamefont {Ayanzadeh}\ \emph
  {et~al.}(2022{\natexlab{a}})\citenamefont {Ayanzadeh}, \citenamefont
  {Alavisamani}, \citenamefont {Das},\ and\ \citenamefont
  {Qureshi}}]{Ayanzadeh2022}%
  \BibitemOpen
  \bibfield  {author} {\bibinfo {author} {\bibfnamefont {R.}~\bibnamefont
  {Ayanzadeh}}, \bibinfo {author} {\bibfnamefont {N.}~\bibnamefont
  {Alavisamani}}, \bibinfo {author} {\bibfnamefont {P.}~\bibnamefont {Das}},\
  and\ \bibinfo {author} {\bibfnamefont {M.}~\bibnamefont {Qureshi}},\ }\href
  {https://arxiv.org/abs/2210.17037} {\bibfield  {journal} {\bibinfo  {journal}
  {arXiv:2210.17037}\ } (\bibinfo {year} {2022}{\natexlab{a}})}\BibitemShut
  {NoStop}%
\bibitem [{\citenamefont {Bravyi}\ \emph {et~al.}(2020)\citenamefont {Bravyi},
  \citenamefont {Kliesch}, \citenamefont {Koenig},\ and\ \citenamefont
  {Tang}}]{PhysRevLett.125.260505}%
  \BibitemOpen
  \bibfield  {author} {\bibinfo {author} {\bibfnamefont {S.}~\bibnamefont
  {Bravyi}}, \bibinfo {author} {\bibfnamefont {A.}~\bibnamefont {Kliesch}},
  \bibinfo {author} {\bibfnamefont {R.}~\bibnamefont {Koenig}},\ and\ \bibinfo
  {author} {\bibfnamefont {E.}~\bibnamefont {Tang}},\ }\href
  {https://doi.org/10.1103/PhysRevLett.125.260505} {\bibfield  {journal}
  {\bibinfo  {journal} {Phys. Rev. Lett.}\ }\textbf {\bibinfo {volume} {125}},\
  \bibinfo {pages} {260505} (\bibinfo {year} {2020})}\BibitemShut {NoStop}%
\bibitem [{\citenamefont {Wagner}\ \emph {et~al.}(2023)\citenamefont {Wagner},
  \citenamefont {N\"u{\ss}lein},\ and\ \citenamefont {Liers}}]{Wagner2023}%
  \BibitemOpen
  \bibfield  {author} {\bibinfo {author} {\bibfnamefont {F.}~\bibnamefont
  {Wagner}}, \bibinfo {author} {\bibfnamefont {J.}~\bibnamefont
  {N\"u{\ss}lein}},\ and\ \bibinfo {author} {\bibfnamefont {F.}~\bibnamefont
  {Liers}},\ }\href {https://arxiv.org/abs/2302.05493} {\bibfield  {journal}
  {\bibinfo  {journal} {arXiv:2302.05493}\ } (\bibinfo {year}
  {2023})}\BibitemShut {NoStop}%
\bibitem [{\citenamefont {Ayanzadeh}\ \emph
  {et~al.}(2022{\natexlab{b}})\citenamefont {Ayanzadeh}, \citenamefont
  {Dorband}, \citenamefont {Halem},\ and\ \citenamefont {Finin}}]{Ramin2022}%
  \BibitemOpen
  \bibfield  {author} {\bibinfo {author} {\bibfnamefont {R.}~\bibnamefont
  {Ayanzadeh}}, \bibinfo {author} {\bibfnamefont {J.}~\bibnamefont {Dorband}},
  \bibinfo {author} {\bibfnamefont {M.}~\bibnamefont {Halem}},\ and\ \bibinfo
  {author} {\bibfnamefont {T.}~\bibnamefont {Finin}},\ }in\ \href
  {https://doi.org/10.1109/IGARSS46834.2022.9884795} {\emph {\bibinfo
  {booktitle} {IGARSS 2022 - 2022 IEEE International Geoscience and Remote
  Sensing Symposium}}}\ (\bibinfo {year} {2022})\ pp.\ \bibinfo {pages}
  {4911--4914}\BibitemShut {NoStop}%
\bibitem [{\citenamefont {Sherrington}\ and\ \citenamefont
  {Kirkpatrick}(1975)}]{PhysRevLett.35.1792}%
  \BibitemOpen
  \bibfield  {author} {\bibinfo {author} {\bibfnamefont {D.}~\bibnamefont
  {Sherrington}}\ and\ \bibinfo {author} {\bibfnamefont {S.}~\bibnamefont
  {Kirkpatrick}},\ }\href {https://doi.org/10.1103/PhysRevLett.35.1792}
  {\bibfield  {journal} {\bibinfo  {journal} {Phys. Rev. Lett.}\ }\textbf
  {\bibinfo {volume} {35}},\ \bibinfo {pages} {1792} (\bibinfo {year}
  {1975})}\BibitemShut {NoStop}%
\bibitem [{\citenamefont {Montanari}(2018)}]{Montanari2018}%
  \BibitemOpen
  \bibfield  {author} {\bibinfo {author} {\bibfnamefont {A.}~\bibnamefont
  {Montanari}},\ }\href {https://arxiv.org/abs/1812.10897} {\bibfield
  {journal} {\bibinfo  {journal} {arXiv:1812.10897}\ } (\bibinfo {year}
  {2018})}\BibitemShut {NoStop}%
\bibitem [{\citenamefont {Weidenfeller}\ \emph {et~al.}(2022)\citenamefont
  {Weidenfeller}, \citenamefont {Valor}, \citenamefont {Gacon}, \citenamefont
  {Tornow}, \citenamefont {Bello}, \citenamefont {Woerner},\ and\ \citenamefont
  {Egger}}]{Weidenfeller2022}%
  \BibitemOpen
  \bibfield  {author} {\bibinfo {author} {\bibfnamefont {J.}~\bibnamefont
  {Weidenfeller}}, \bibinfo {author} {\bibfnamefont {L.~C.}\ \bibnamefont
  {Valor}}, \bibinfo {author} {\bibfnamefont {J.}~\bibnamefont {Gacon}},
  \bibinfo {author} {\bibfnamefont {C.}~\bibnamefont {Tornow}}, \bibinfo
  {author} {\bibfnamefont {L.}~\bibnamefont {Bello}}, \bibinfo {author}
  {\bibfnamefont {S.}~\bibnamefont {Woerner}},\ and\ \bibinfo {author}
  {\bibfnamefont {D.~J.}\ \bibnamefont {Egger}},\ }\href
  {https://doi.org/10.22331/q-2022-12-07-870} {\bibfield  {journal} {\bibinfo
  {journal} {{Quantum}}\ }\textbf {\bibinfo {volume} {6}},\ \bibinfo {pages}
  {870} (\bibinfo {year} {2022})}\BibitemShut {NoStop}%
\bibitem [{\citenamefont {Boettcher}(2005)}]{Boettcher2005}%
  \BibitemOpen
  \bibfield  {author} {\bibinfo {author} {\bibfnamefont {S.}~\bibnamefont
  {Boettcher}},\ }\href {https://doi.org/10.1140/epjb/e2005-00280-6} {\bibfield
   {journal} {\bibinfo  {journal} {Eur. Phys. J. B}\ }\textbf {\bibinfo
  {volume} {46}},\ \bibinfo {pages} {501} (\bibinfo {year} {2005})}\BibitemShut
  {NoStop}%
\bibitem [{\citenamefont {Parisi}(1979)}]{PhysRevLett.43.1754}%
  \BibitemOpen
  \bibfield  {author} {\bibinfo {author} {\bibfnamefont {G.}~\bibnamefont
  {Parisi}},\ }\href {https://doi.org/10.1103/PhysRevLett.43.1754} {\bibfield
  {journal} {\bibinfo  {journal} {Phys. Rev. Lett.}\ }\textbf {\bibinfo
  {volume} {43}},\ \bibinfo {pages} {1754} (\bibinfo {year}
  {1979})}\BibitemShut {NoStop}%
\bibitem [{\citenamefont {Schmidt}(2008)}]{Schmidt2008}%
  \BibitemOpen
  \bibfield  {author} {\bibinfo {author} {\bibfnamefont {M.~J.}\ \bibnamefont
  {Schmidt}},\ }\emph {\bibinfo {title} {Replica {S}ymmetry {B}reaking at {L}ow
  {T}emperatures}},\ \href@noop {} {Ph.D. thesis},\ \bibinfo  {school} {Julius
  Maximilians-Universit\"at W\"urzburg.} (\bibinfo {year} {2008})\BibitemShut
  {NoStop}%
\bibitem [{\citenamefont {Aizenman}\ \emph {et~al.}(1987)\citenamefont
  {Aizenman}, \citenamefont {Lebowitz},\ and\ \citenamefont
  {Ruelle}}]{Aizenman1987}%
  \BibitemOpen
  \bibfield  {author} {\bibinfo {author} {\bibfnamefont {M.}~\bibnamefont
  {Aizenman}}, \bibinfo {author} {\bibfnamefont {J.~L.}\ \bibnamefont
  {Lebowitz}},\ and\ \bibinfo {author} {\bibfnamefont {D.}~\bibnamefont
  {Ruelle}},\ }\href {https://doi.org/10.1007/BF01217677} {\bibfield  {journal}
  {\bibinfo  {journal} {Commun. Math. Phys.}\ }\textbf {\bibinfo {volume}
  {112}},\ \bibinfo {pages} {3} (\bibinfo {year} {1987})}\BibitemShut {NoStop}%
\bibitem [{\citenamefont {Montanari}\ and\ \citenamefont
  {Sen}(2015)}]{Montanari2015}%
  \BibitemOpen
  \bibfield  {author} {\bibinfo {author} {\bibfnamefont {A.}~\bibnamefont
  {Montanari}}\ and\ \bibinfo {author} {\bibfnamefont {S.}~\bibnamefont
  {Sen}},\ }\href {https://arxiv.org/abs/1504.05910} {\bibfield  {journal}
  {\bibinfo  {journal} {arXiv:1504.05910}\ } (\bibinfo {year}
  {2015})}\BibitemShut {NoStop}%
\bibitem [{\citenamefont {Afonso S.~Bandeira}(2019)}]{Bandeira2019}%
  \BibitemOpen
  \bibfield  {author} {\bibinfo {author} {\bibfnamefont {A.~S.~W.}\
  \bibnamefont {Afonso S.~Bandeira}, \bibfnamefont {Dmitriy~Kunisky}},\ }\href
  {https://arxiv.org/abs/1902.07324} {\bibfield  {journal} {\bibinfo  {journal}
  {arXiv:1902.07324}\ } (\bibinfo {year} {2019})}\BibitemShut {NoStop}%
\bibitem [{\citenamefont {Cai}\ \emph {et~al.}(2022)\citenamefont {Cai},
  \citenamefont {Babbush}, \citenamefont {Benjamin}, \citenamefont {Endo},
  \citenamefont {Huggins}, \citenamefont {Li}, \citenamefont {McClean},\ and\
  \citenamefont {O'Brien}}]{Cai2022}%
  \BibitemOpen
  \bibfield  {author} {\bibinfo {author} {\bibfnamefont {Z.}~\bibnamefont
  {Cai}}, \bibinfo {author} {\bibfnamefont {R.}~\bibnamefont {Babbush}},
  \bibinfo {author} {\bibfnamefont {S.~C.}\ \bibnamefont {Benjamin}}, \bibinfo
  {author} {\bibfnamefont {S.}~\bibnamefont {Endo}}, \bibinfo {author}
  {\bibfnamefont {W.~J.}\ \bibnamefont {Huggins}}, \bibinfo {author}
  {\bibfnamefont {Y.}~\bibnamefont {Li}}, \bibinfo {author} {\bibfnamefont
  {J.~R.}\ \bibnamefont {McClean}},\ and\ \bibinfo {author} {\bibfnamefont
  {T.~E.}\ \bibnamefont {O'Brien}},\ }\href {https://arxiv.org/abs/2210.00921}
  {\bibfield  {journal} {\bibinfo  {journal} {arXiv:2210.00921}\ } (\bibinfo
  {year} {2022})}\BibitemShut {NoStop}%
\bibitem [{\citenamefont {Dupont}\ and\ \citenamefont
  {Sundar}(2023)}]{Dupont2023}%
  \BibitemOpen
  \bibfield  {author} {\bibinfo {author} {\bibfnamefont {M.}~\bibnamefont
  {Dupont}}\ and\ \bibinfo {author} {\bibfnamefont {B.}~\bibnamefont
  {Sundar}},\ }\href {https://arxiv.org/abs/2307.05821} {\bibfield  {journal}
  {\bibinfo  {journal} {arXiv:2307.05821}\ } (\bibinfo {year}
  {2023})}\BibitemShut {NoStop}%
\bibitem [{\citenamefont {Mohseni}\ \emph {et~al.}(2022)\citenamefont
  {Mohseni}, \citenamefont {McMahon},\ and\ \citenamefont
  {Byrnes}}]{Mohseni2022}%
  \BibitemOpen
  \bibfield  {author} {\bibinfo {author} {\bibfnamefont {N.}~\bibnamefont
  {Mohseni}}, \bibinfo {author} {\bibfnamefont {P.~L.}\ \bibnamefont
  {McMahon}},\ and\ \bibinfo {author} {\bibfnamefont {T.}~\bibnamefont
  {Byrnes}},\ }\href {https://doi.org/10.1038/s42254-022-00440-8} {\bibfield
  {journal} {\bibinfo  {journal} {Nat. Rev. Phys.}\ }\textbf {\bibinfo {volume}
  {4}},\ \bibinfo {pages} {363} (\bibinfo {year} {2022})}\BibitemShut {NoStop}%
\bibitem [{\citenamefont {Marsh}\ and\ \citenamefont
  {Wang}(2019)}]{marsh2019quantum}%
  \BibitemOpen
  \bibfield  {author} {\bibinfo {author} {\bibfnamefont {S.}~\bibnamefont
  {Marsh}}\ and\ \bibinfo {author} {\bibfnamefont {J.}~\bibnamefont {Wang}},\
  }\href@noop {} {\bibfield  {journal} {\bibinfo  {journal} {Quantum Inf.
  Process.}\ }\textbf {\bibinfo {volume} {18}},\ \bibinfo {pages} {1} (\bibinfo
  {year} {2019})}\BibitemShut {NoStop}%
\bibitem [{\citenamefont {B\"artschi}\ and\ \citenamefont
  {Eidenbenz}(2020)}]{bartschi2020grover}%
  \BibitemOpen
  \bibfield  {author} {\bibinfo {author} {\bibfnamefont {A.}~\bibnamefont
  {B\"artschi}}\ and\ \bibinfo {author} {\bibfnamefont {S.}~\bibnamefont
  {Eidenbenz}},\ }in\ \href {https://doi.org/10.1109/QCE49297.2020.00020}
  {\emph {\bibinfo {booktitle} {2020 IEEE International Conference on Quantum
  Computing and Engineering (QCE)}}}\ (\bibinfo {year} {2020})\ pp.\ \bibinfo
  {pages} {72--82}\BibitemShut {NoStop}%
\bibitem [{\citenamefont {Herman}\ \emph {et~al.}(2022)\citenamefont {Herman},
  \citenamefont {Shaydulin}, \citenamefont {Sun}, \citenamefont {Chakrabarti},
  \citenamefont {Hu}, \citenamefont {Minssen}, \citenamefont {Rattew},
  \citenamefont {Yalovetzky},\ and\ \citenamefont {Pistoia}}]{Herman2022}%
  \BibitemOpen
  \bibfield  {author} {\bibinfo {author} {\bibfnamefont {D.}~\bibnamefont
  {Herman}}, \bibinfo {author} {\bibfnamefont {R.}~\bibnamefont {Shaydulin}},
  \bibinfo {author} {\bibfnamefont {Y.}~\bibnamefont {Sun}}, \bibinfo {author}
  {\bibfnamefont {S.}~\bibnamefont {Chakrabarti}}, \bibinfo {author}
  {\bibfnamefont {S.}~\bibnamefont {Hu}}, \bibinfo {author} {\bibfnamefont
  {P.}~\bibnamefont {Minssen}}, \bibinfo {author} {\bibfnamefont
  {A.}~\bibnamefont {Rattew}}, \bibinfo {author} {\bibfnamefont
  {R.}~\bibnamefont {Yalovetzky}},\ and\ \bibinfo {author} {\bibfnamefont
  {M.}~\bibnamefont {Pistoia}},\ }\href {https://arxiv.org/abs/2209.15024}
  {\bibfield  {journal} {\bibinfo  {journal} {arXiv:2209.15024}\ } (\bibinfo
  {year} {2022})}\BibitemShut {NoStop}%
\bibitem [{\citenamefont {Tilly}\ \emph {et~al.}(2022)\citenamefont {Tilly},
  \citenamefont {Chen}, \citenamefont {Cao}, \citenamefont {Picozzi},
  \citenamefont {Setia}, \citenamefont {Li}, \citenamefont {Grant},
  \citenamefont {Wossnig}, \citenamefont {Rungger}, \citenamefont {Booth},\
  and\ \citenamefont {Tennyson}}]{TILLY20221}%
  \BibitemOpen
  \bibfield  {author} {\bibinfo {author} {\bibfnamefont {J.}~\bibnamefont
  {Tilly}}, \bibinfo {author} {\bibfnamefont {H.}~\bibnamefont {Chen}},
  \bibinfo {author} {\bibfnamefont {S.}~\bibnamefont {Cao}}, \bibinfo {author}
  {\bibfnamefont {D.}~\bibnamefont {Picozzi}}, \bibinfo {author} {\bibfnamefont
  {K.}~\bibnamefont {Setia}}, \bibinfo {author} {\bibfnamefont
  {Y.}~\bibnamefont {Li}}, \bibinfo {author} {\bibfnamefont {E.}~\bibnamefont
  {Grant}}, \bibinfo {author} {\bibfnamefont {L.}~\bibnamefont {Wossnig}},
  \bibinfo {author} {\bibfnamefont {I.}~\bibnamefont {Rungger}}, \bibinfo
  {author} {\bibfnamefont {G.~H.}\ \bibnamefont {Booth}},\ and\ \bibinfo
  {author} {\bibfnamefont {J.}~\bibnamefont {Tennyson}},\ }\href
  {https://doi.org/https://doi.org/10.1016/j.physrep.2022.08.003} {\bibfield
  {journal} {\bibinfo  {journal} {Phys. Rep.}\ }\textbf {\bibinfo {volume}
  {986}},\ \bibinfo {pages} {1} (\bibinfo {year} {2022})},\ \bibinfo {note}
  {the Variational Quantum Eigensolver: a review of methods and best
  practices}\BibitemShut {NoStop}%
\bibitem [{\citenamefont {Morningstar}\ and\ \citenamefont
  {Weinstein}(1996)}]{PhysRevD.54.4131}%
  \BibitemOpen
  \bibfield  {author} {\bibinfo {author} {\bibfnamefont {C.~J.}\ \bibnamefont
  {Morningstar}}\ and\ \bibinfo {author} {\bibfnamefont {M.}~\bibnamefont
  {Weinstein}},\ }\href {https://doi.org/10.1103/PhysRevD.54.4131} {\bibfield
  {journal} {\bibinfo  {journal} {Phys. Rev. D}\ }\textbf {\bibinfo {volume}
  {54}},\ \bibinfo {pages} {4131} (\bibinfo {year} {1996})}\BibitemShut
  {NoStop}%
\bibitem [{\citenamefont {Abrams}\ \emph {et~al.}(2020)\citenamefont {Abrams},
  \citenamefont {Didier}, \citenamefont {Johnson}, \citenamefont {Silva},\ and\
  \citenamefont {Ryan}}]{Abrams2020}%
  \BibitemOpen
  \bibfield  {author} {\bibinfo {author} {\bibfnamefont {D.~M.}\ \bibnamefont
  {Abrams}}, \bibinfo {author} {\bibfnamefont {N.}~\bibnamefont {Didier}},
  \bibinfo {author} {\bibfnamefont {B.~R.}\ \bibnamefont {Johnson}}, \bibinfo
  {author} {\bibfnamefont {M.~P.~d.}\ \bibnamefont {Silva}},\ and\ \bibinfo
  {author} {\bibfnamefont {C.~A.}\ \bibnamefont {Ryan}},\ }\href
  {https://doi.org/10.1038/s41928-020-00498-1} {\bibfield  {journal} {\bibinfo
  {journal} {Nat. Electron.}\ }\textbf {\bibinfo {volume} {3}},\ \bibinfo
  {pages} {744} (\bibinfo {year} {2020})}\BibitemShut {NoStop}%
\bibitem [{\citenamefont {Knill}\ \emph {et~al.}(2008)\citenamefont {Knill},
  \citenamefont {Leibfried}, \citenamefont {Reichle}, \citenamefont {Britton},
  \citenamefont {Blakestad}, \citenamefont {Jost}, \citenamefont {Langer},
  \citenamefont {Ozeri}, \citenamefont {Seidelin},\ and\ \citenamefont
  {Wineland}}]{PhysRevA.77.012307}%
  \BibitemOpen
  \bibfield  {author} {\bibinfo {author} {\bibfnamefont {E.}~\bibnamefont
  {Knill}}, \bibinfo {author} {\bibfnamefont {D.}~\bibnamefont {Leibfried}},
  \bibinfo {author} {\bibfnamefont {R.}~\bibnamefont {Reichle}}, \bibinfo
  {author} {\bibfnamefont {J.}~\bibnamefont {Britton}}, \bibinfo {author}
  {\bibfnamefont {R.~B.}\ \bibnamefont {Blakestad}}, \bibinfo {author}
  {\bibfnamefont {J.~D.}\ \bibnamefont {Jost}}, \bibinfo {author}
  {\bibfnamefont {C.}~\bibnamefont {Langer}}, \bibinfo {author} {\bibfnamefont
  {R.}~\bibnamefont {Ozeri}}, \bibinfo {author} {\bibfnamefont
  {S.}~\bibnamefont {Seidelin}},\ and\ \bibinfo {author} {\bibfnamefont
  {D.~J.}\ \bibnamefont {Wineland}},\ }\href
  {https://doi.org/10.1103/PhysRevA.77.012307} {\bibfield  {journal} {\bibinfo
  {journal} {Phys. Rev. A}\ }\textbf {\bibinfo {volume} {77}},\ \bibinfo
  {pages} {012307} (\bibinfo {year} {2008})}\BibitemShut {NoStop}%
\bibitem [{\citenamefont {Erhard}\ \emph {et~al.}(2019)\citenamefont {Erhard},
  \citenamefont {Wallman}, \citenamefont {Postler}, \citenamefont {Meth},
  \citenamefont {Stricker}, \citenamefont {Martinez}, \citenamefont
  {Schindler}, \citenamefont {Monz}, \citenamefont {Emerson},\ and\
  \citenamefont {Blatt}}]{erhard_characterizing_2019}%
  \BibitemOpen
  \bibfield  {author} {\bibinfo {author} {\bibfnamefont {A.}~\bibnamefont
  {Erhard}}, \bibinfo {author} {\bibfnamefont {J.~J.}\ \bibnamefont {Wallman}},
  \bibinfo {author} {\bibfnamefont {L.}~\bibnamefont {Postler}}, \bibinfo
  {author} {\bibfnamefont {M.}~\bibnamefont {Meth}}, \bibinfo {author}
  {\bibfnamefont {R.}~\bibnamefont {Stricker}}, \bibinfo {author}
  {\bibfnamefont {E.~A.}\ \bibnamefont {Martinez}}, \bibinfo {author}
  {\bibfnamefont {P.}~\bibnamefont {Schindler}}, \bibinfo {author}
  {\bibfnamefont {T.}~\bibnamefont {Monz}}, \bibinfo {author} {\bibfnamefont
  {J.}~\bibnamefont {Emerson}},\ and\ \bibinfo {author} {\bibfnamefont
  {R.}~\bibnamefont {Blatt}},\ }\href
  {https://doi.org/10.1038/s41467-019-13068-7} {\bibfield  {journal} {\bibinfo
  {journal} {Nat. Commun.}\ }\textbf {\bibinfo {volume} {10}},\ \bibinfo
  {pages} {5347} (\bibinfo {year} {2019})}\BibitemShut {NoStop}%
\bibitem [{\citenamefont {Barenco}\ \emph {et~al.}(1995)\citenamefont
  {Barenco}, \citenamefont {Bennett}, \citenamefont {Cleve}, \citenamefont
  {DiVincenzo}, \citenamefont {Margolus}, \citenamefont {Shor}, \citenamefont
  {Sleator}, \citenamefont {Smolin},\ and\ \citenamefont
  {Weinfurter}}]{barenco1995elementary}%
  \BibitemOpen
  \bibfield  {author} {\bibinfo {author} {\bibfnamefont {A.}~\bibnamefont
  {Barenco}}, \bibinfo {author} {\bibfnamefont {C.~H.}\ \bibnamefont
  {Bennett}}, \bibinfo {author} {\bibfnamefont {R.}~\bibnamefont {Cleve}},
  \bibinfo {author} {\bibfnamefont {D.~P.}\ \bibnamefont {DiVincenzo}},
  \bibinfo {author} {\bibfnamefont {N.}~\bibnamefont {Margolus}}, \bibinfo
  {author} {\bibfnamefont {P.}~\bibnamefont {Shor}}, \bibinfo {author}
  {\bibfnamefont {T.}~\bibnamefont {Sleator}}, \bibinfo {author} {\bibfnamefont
  {J.~A.}\ \bibnamefont {Smolin}},\ and\ \bibinfo {author} {\bibfnamefont
  {H.}~\bibnamefont {Weinfurter}},\ }\href
  {https://doi.org/10.1103/PhysRevA.52.3457} {\bibfield  {journal} {\bibinfo
  {journal} {Phys. Rev. A}\ }\textbf {\bibinfo {volume} {52}},\ \bibinfo
  {pages} {3457} (\bibinfo {year} {1995})}\BibitemShut {NoStop}%
\bibitem [{\citenamefont {Huang}\ \emph {et~al.}(2021)\citenamefont {Huang},
  \citenamefont {Wang}, \citenamefont {Wu}, \citenamefont {Ding}, \citenamefont
  {Ye}, \citenamefont {Kong}, \citenamefont {Zhang}, \citenamefont {Ni},
  \citenamefont {Song}, \citenamefont {Shi} \emph {et~al.}}]{huang2021quantum}%
  \BibitemOpen
  \bibfield  {author} {\bibinfo {author} {\bibfnamefont {C.}~\bibnamefont
  {Huang}}, \bibinfo {author} {\bibfnamefont {T.}~\bibnamefont {Wang}},
  \bibinfo {author} {\bibfnamefont {F.}~\bibnamefont {Wu}}, \bibinfo {author}
  {\bibfnamefont {D.}~\bibnamefont {Ding}}, \bibinfo {author} {\bibfnamefont
  {Q.}~\bibnamefont {Ye}}, \bibinfo {author} {\bibfnamefont {L.}~\bibnamefont
  {Kong}}, \bibinfo {author} {\bibfnamefont {F.}~\bibnamefont {Zhang}},
  \bibinfo {author} {\bibfnamefont {X.}~\bibnamefont {Ni}}, \bibinfo {author}
  {\bibfnamefont {Z.}~\bibnamefont {Song}}, \bibinfo {author} {\bibfnamefont
  {Y.}~\bibnamefont {Shi}}, \emph {et~al.},\ }\href
  {https://arxiv.org/abs/2105.06074} {\bibfield  {journal} {\bibinfo  {journal}
  {arXiv:2105.06074}\ } (\bibinfo {year} {2021})}\BibitemShut {NoStop}%
\bibitem [{\citenamefont {Vidal}(2004)}]{PhysRevLett.93.040502}%
  \BibitemOpen
  \bibfield  {author} {\bibinfo {author} {\bibfnamefont {G.}~\bibnamefont
  {Vidal}},\ }\href {https://doi.org/10.1103/PhysRevLett.93.040502} {\bibfield
  {journal} {\bibinfo  {journal} {Phys. Rev. Lett.}\ }\textbf {\bibinfo
  {volume} {93}},\ \bibinfo {pages} {040502} (\bibinfo {year}
  {2004})}\BibitemShut {NoStop}%
\bibitem [{\citenamefont {Dupont}\ \emph
  {et~al.}(2022{\natexlab{a}})\citenamefont {Dupont}, \citenamefont {Didier},
  \citenamefont {Hodson}, \citenamefont {Moore},\ and\ \citenamefont
  {Reagor}}]{PRXQuantum.3.040339}%
  \BibitemOpen
  \bibfield  {author} {\bibinfo {author} {\bibfnamefont {M.}~\bibnamefont
  {Dupont}}, \bibinfo {author} {\bibfnamefont {N.}~\bibnamefont {Didier}},
  \bibinfo {author} {\bibfnamefont {M.~J.}\ \bibnamefont {Hodson}}, \bibinfo
  {author} {\bibfnamefont {J.~E.}\ \bibnamefont {Moore}},\ and\ \bibinfo
  {author} {\bibfnamefont {M.~J.}\ \bibnamefont {Reagor}},\ }\href
  {https://doi.org/10.1103/PRXQuantum.3.040339} {\bibfield  {journal} {\bibinfo
   {journal} {PRX Quantum}\ }\textbf {\bibinfo {volume} {3}},\ \bibinfo {pages}
  {040339} (\bibinfo {year} {2022}{\natexlab{a}})}\BibitemShut {NoStop}%
\bibitem [{\citenamefont {Dupont}\ \emph
  {et~al.}(2022{\natexlab{b}})\citenamefont {Dupont}, \citenamefont {Didier},
  \citenamefont {Hodson}, \citenamefont {Moore},\ and\ \citenamefont
  {Reagor}}]{PhysRevA.106.022423}%
  \BibitemOpen
  \bibfield  {author} {\bibinfo {author} {\bibfnamefont {M.}~\bibnamefont
  {Dupont}}, \bibinfo {author} {\bibfnamefont {N.}~\bibnamefont {Didier}},
  \bibinfo {author} {\bibfnamefont {M.~J.}\ \bibnamefont {Hodson}}, \bibinfo
  {author} {\bibfnamefont {J.~E.}\ \bibnamefont {Moore}},\ and\ \bibinfo
  {author} {\bibfnamefont {M.~J.}\ \bibnamefont {Reagor}},\ }\href
  {https://doi.org/10.1103/PhysRevA.106.022423} {\bibfield  {journal} {\bibinfo
   {journal} {Phys. Rev. A}\ }\textbf {\bibinfo {volume} {106}},\ \bibinfo
  {pages} {022423} (\bibinfo {year} {2022}{\natexlab{b}})}\BibitemShut
  {NoStop}%
\bibitem [{\citenamefont {Glover}\ and\ \citenamefont
  {Laguna}(1998)}]{Glover1998}%
  \BibitemOpen
  \bibfield  {author} {\bibinfo {author} {\bibfnamefont {F.}~\bibnamefont
  {Glover}}\ and\ \bibinfo {author} {\bibfnamefont {M.}~\bibnamefont
  {Laguna}},\ }\bibinfo {title} {Tabu search},\ in\ \href
  {https://doi.org/10.1007/978-1-4613-0303-9_33} {\emph {\bibinfo {booktitle}
  {Handbook of {C}ombinatorial {O}ptimization: {V}olume 1--3}}},\ \bibinfo
  {editor} {edited by\ \bibinfo {editor} {\bibfnamefont {D.-Z.}\ \bibnamefont
  {Du}}\ and\ \bibinfo {editor} {\bibfnamefont {P.~M.}\ \bibnamefont
  {Pardalos}}}\ (\bibinfo  {publisher} {Springer US},\ \bibinfo {address}
  {Boston, MA},\ \bibinfo {year} {1998})\ pp.\ \bibinfo {pages}
  {2093--2229}\BibitemShut {NoStop}%
\bibitem [{\citenamefont {Kirkpatrick}\ \emph {et~al.}(1983)\citenamefont
  {Kirkpatrick}, \citenamefont {Gelatt},\ and\ \citenamefont
  {Vecchi}}]{Kirkpatrick1983}%
  \BibitemOpen
  \bibfield  {author} {\bibinfo {author} {\bibfnamefont {S.}~\bibnamefont
  {Kirkpatrick}}, \bibinfo {author} {\bibfnamefont {C.~D.}\ \bibnamefont
  {Gelatt}},\ and\ \bibinfo {author} {\bibfnamefont {M.~P.}\ \bibnamefont
  {Vecchi}},\ }\href {https://doi.org/10.1126/science.220.4598.671} {\bibfield
  {journal} {\bibinfo  {journal} {Science}\ }\textbf {\bibinfo {volume}
  {220}},\ \bibinfo {pages} {671} (\bibinfo {year} {1983})}\BibitemShut
  {NoStop}%
\bibitem [{\citenamefont {Kirkpatrick}(1984)}]{Kirkpatrick1984}%
  \BibitemOpen
  \bibfield  {author} {\bibinfo {author} {\bibfnamefont {S.}~\bibnamefont
  {Kirkpatrick}},\ }\href {https://doi.org/10.1007/BF01009452} {\bibfield
  {journal} {\bibinfo  {journal} {J. Stat. Phys.}\ }\textbf {\bibinfo {volume}
  {34}},\ \bibinfo {pages} {975} (\bibinfo {year} {1984})}\BibitemShut
  {NoStop}%
\bibitem [{\citenamefont {LaRose}\ \emph {et~al.}(2022)\citenamefont {LaRose},
  \citenamefont {Rieffel},\ and\ \citenamefont {Venturelli}}]{larose2022mixer}%
  \BibitemOpen
  \bibfield  {author} {\bibinfo {author} {\bibfnamefont {R.}~\bibnamefont
  {LaRose}}, \bibinfo {author} {\bibfnamefont {E.}~\bibnamefont {Rieffel}},\
  and\ \bibinfo {author} {\bibfnamefont {D.}~\bibnamefont {Venturelli}},\
  }\href {https://doi.org/10.1007/s42484-022-00069-x} {\bibfield  {journal}
  {\bibinfo  {journal} {Quantum Mach. Intell.}\ }\textbf {\bibinfo {volume}
  {4}},\ \bibinfo {pages} {17} (\bibinfo {year} {2022})}\BibitemShut {NoStop}%
\end{thebibliography}%
\let\addcontentsline\oldaddcontentsline

\clearpage

\onecolumngrid
\setlength{\belowcaptionskip}{0pt}
\setcounter{secnumdepth}{3}
\setcounter{page}{1}
\setcounter{figure}{0}
\setcounter{table}{0}
\setcounter{equation}{0}
\setcounter{section}{0}
\renewcommand{\thepage}{S\arabic{page}}
\renewcommand{\thesection}{S\arabic{section}}
\renewcommand{\thetable}{S\arabic{table}}
\renewcommand{\thefigure}{S\arabic{figure}}
\renewcommand{\theequation}{S\arabic{equation}}
\counterwithout*{equation}{section}
\newpage\clearpage

\begin{center}
    \Large\textbf{Supplementary Information for\\Quantum-Enhanced Greedy Combinatorial Optimization Solver}
\end{center}

{\hypersetup{linkcolor=black}\tableofcontents}

\section{Decomposition of Two-qubit Gates with \texorpdfstring{$\sqrt{i\textrm{SWAP}}$}{√iSWAP}}

In Fig.~\ref{fig:gate_decomposition} of the main text, we showed decompositions of the parametric two-qubit gates $\textrm{Rzz}(\varphi)$ and $\textrm{Rzz}(\varphi)\times\textrm{SWAP}$ in terms of the two-qubit gate $\sqrt{i\textrm{SWAP}}$ and one-qubit gates $U_{1...14}$. Here, we give explicit forms of the $U_i$ in terms of hardware-native one-qubit gates $\textrm{Rz}(\theta\in\mathbb{R})=\exp(-iZ\theta/2)$ and $\textrm{Rx}(\phi\in\mathbb{Z})=\exp(-iX\phi\pi/4)$ where $X$ and $Z$ are Pauli operators. The one-qubit gates in the decomposition of $\textrm{Rzz}(\varphi)$ from Fig.~\ref{fig:gate_decomposition}a are,
\begin{align}\label{eq:Rzz_decomp}
    & U_1 = \textrm{Rz}(\pi)\textrm{Rx}(1)\textrm{Rz}(\alpha)\textrm{Rx}(1)\textrm{Rz}(\pi/2),\nonumber\\
    & U_2 = \textrm{Rz}(\pi/2)\textrm{Rx}(1)\textrm{Rz}(\pi/2),\nonumber\\
    & U_3 = \textrm{Rx}(1)\textrm{Rz}(\beta)\textrm{Rx}(1),\nonumber\\
    & U_4 = \textrm{I},\nonumber\\
    & U_5 = \textrm{Rz}(\gamma)\textrm{Rx}(1)\textrm{Rz}(\alpha)\textrm{Rx}(1)\textrm{Rz}(\pi),\nonumber\\
    & U_6 = \textrm{Rz}(\gamma)\textrm{Rx}(1)\textrm{Rz}(\pi/2),
\end{align}
where $\alpha = \sin^{-1}\left(\tan\frac{\tilde{\varphi}}{2}\right)$, $\beta = 2\arccos\left(\sqrt{2}\sin\frac{\tilde{\varphi}}{2}\right)$, $\gamma = \frac{\pi}{2}{\rm sgn}\left(\frac{\pi}{2}-|\varphi|\right)$, $\tilde{\varphi} = {\rm mod}\left(\varphi + \frac{\pi}{2},\pi\right) - \frac{\pi}{2}$, and $\textrm{I}$ is the identity. Note that gates in each line in Eq.~\eqref{eq:Rzz_decomp} are applied right to left. The one-qubit gates in the decomposition of $\textrm{Rzz}(\varphi)\times\textrm{SWAP}$ from Fig.~\ref{fig:gate_decomposition}B are,
\begin{align}
    & U_7 = \textrm{Rz}(\kappa_1)\textrm{Rx}(1)\textrm{Rz}(\kappa_1),\nonumber\\
    & U_8 = \textrm{Rz}(\pi/2)\textrm{Rx}(1)\textrm{Rz}(\pi/2),\nonumber\\
    & U_9 = \textrm{Rz}(\kappa_1)\textrm{Rx}(1)\textrm{Rz}(\zeta_1),\nonumber\\
    & U_{10} = \textrm{Rz}(\pi/2)\textrm{Rx}(1)\textrm{Rz}(\zeta_2),\nonumber\\
    & U_{11} = \textrm{Rz}(\kappa_1)\textrm{Rx}(1)\textrm{Rz}(\alpha)\textrm{Rx}(1)\textrm{Rz}(\kappa_1),\nonumber\\
    & U_{12} = \textrm{Rz}(\pi/2)\textrm{Rx}(1)\textrm{Rz}(\beta)\textrm{Rx}(1)\textrm{Rz}(\pi/2),\nonumber\\
    & U_{13} = \textrm{Rz}(\kappa_2)\textrm{Rx}(1)\textrm{Rz}(\zeta_3)\textrm{Rx}(1)\textrm{Rz}(\kappa_1),\nonumber\\
    & U_{14} = \textrm{Rz}(\kappa_3)\textrm{Rx}(1)\textrm{Rz}(\zeta_4)\textrm{Rx}(1)\textrm{Rz}(\pi/2).
\end{align}
For simplicity, below we give the values for the gate angles only for $0 \leq \varphi \leq \frac{\pi}{2}$:
\begin{align}
& \alpha = \arccos\left[\frac{\cos\varphi + \sin\varphi - 1 + \sqrt{\sin2\varphi}}{\sqrt{2}} {\rm sign}\left(\frac{\pi}{4} - \varphi\right) \right] - \pi, \nonumber\\
& \beta = \arccos\left(\frac{\cos\varphi + \sin\varphi - 1 - \sqrt{\sin2\varphi}}{\sqrt{2}} \right) - \pi, \nonumber\\
& \lambda = -\arccos\sqrt{ \frac{ 1 + \tan\frac{\varphi}{2}}{2}}, \nonumber\\
& \psi = \arccos\sqrt{ \frac{1}{ 1 + \tan\frac{\varphi}{2}}}, \nonumber\\
& \zeta_1 = \left(\lambda + \psi\right) + \frac{\pi}{2}, \nonumber\\
& \zeta_2 = \left(\lambda - \psi\right) + \frac{\pi}{2}, \nonumber\\
& \zeta_3 = \left(\lambda + \psi\right) - \pi, \nonumber\\
& \zeta_4 = \left(\lambda - \psi\right) - \pi, \nonumber\\
& \kappa_1 = \kappa_2 = \kappa_3 = \frac{\pi}{2}.
\end{align}

\section{Number of Native Gates Executed on Hardware}

\begin{figure}[!ht]
    \centering
    \includegraphics[width=0.55\columnwidth]{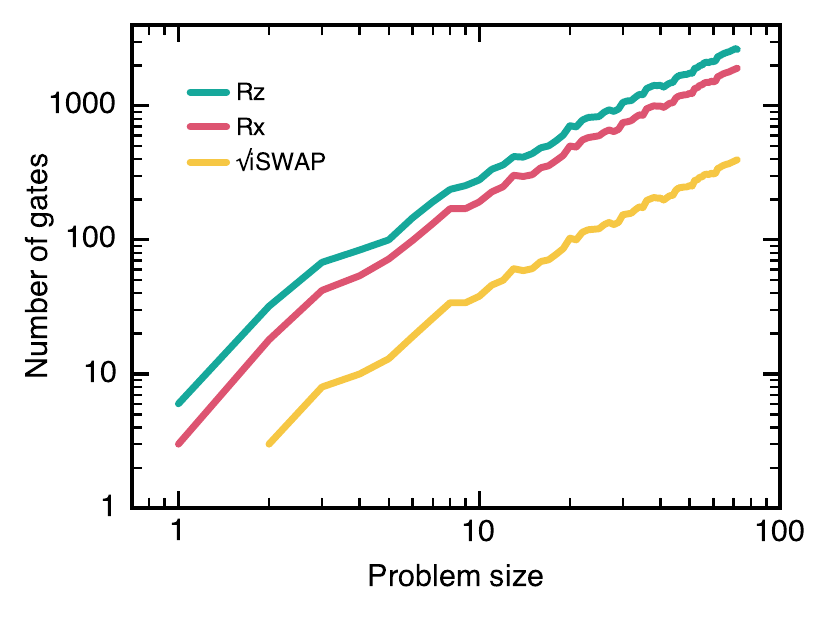} 
    \caption{\textbf{Number of native gates executed on hardware.} Number of native one-qubit $\textrm{Rz}$ and $\textrm{Rx}$ gates and two-qubit $\sqrt{i\textrm{SWAP}}$ gates executed on the hardware as a function of the problem size (related to the iteration step).}
    \label{fig:number_of_gates_vs_step}
\end{figure}

In Fig.~\ref{fig:number_of_gates_vs_step}, we show the number of native one-qubit $\textrm{Rz}$ and $\textrm{Rx}$ gates and two-qubit $\sqrt{i\textrm{SWAP}}$ gates executed on the hardware as a function of the problem size up to $N=72$. The number of gates scales linearly with the problem size.

\section{Random Sampling and Extreme Value Distribution}

In the main text we compared the approximation ratio of the quantum-enhanced greedy algorithm with the approximation ratio of bit strings drawn randomly from a uniform distribution. The cost for a bit string $\mathbf{B}=(B_1,\cdots, B_N)$ is $C=\sum^N_{i=1,j<i} w_{ij} Z_i Z_j$ where $Z_i = (-1)^{B_i}$. The average value of $C$ is $0$ for random $B_i$, therefore the average approximation ratio of random samples is $r=1/2$ (see Methods). Here, we elaborate on the distribution of $C$ for random $B_i$, and the expected \textit{best} random guess from \textit{many} random samples.

For random bits $B_i$ and weights $w_{ij}$, each term in the cost is a Bernoulli random variable with mean $0$ and variance $1$. The cost $C$, which is the sum of these random variables, is a normally distributed random variable with mean $0$ and variance $N(N-1)/2$ as $N\to+\infty$ [see Fig.~\ref{fig:distribution_cost_sk}].

Suppose we draw $M$ random bit strings and keep the one which has the minimum cost $C$. The probability density for the best random guess converges for sufficient samples to the type-I generalized extreme value distribution, also called the Gumbel distribution, $C\sim\textrm{Gumbel}(\mu_M,\sigma_M)$, where $\mu_M=\sqrt{N(N-1)/2}\cdot\Phi^{-1}(1-1/M)$ and $\sigma_M=\sqrt{N(N-1)/2}\cdot\Phi^{-1}(1-1/eM)-\mu_M$. $\Phi^{-1}$ is the inverse of the cumulative distribution function of the normal distribution. The mean value of the Gumbel distribution is,
\begin{equation}
    C_\textrm{best-rnd}\left(N,M\right)\simeq-\left(1+\frac{\gamma}{\ln M}\right)\sqrt{\frac{N(N-1)}{2}\ln\left(\frac{ M^2}{2\pi \ln\frac{M^2}{2\pi}}\right)},
    \label{eq:best_random_guess}
\end{equation}
where $\gamma\simeq 0.577215...$ is the Euler-Mascheroni constant. Eq.~\eqref{eq:best_random_guess} is expected to hold for $1\ll M\lesssim 2^N$. The upper limit for $M$ is set by the fact that when $M$ saturates this limit, the best random guesses sample from the tail of the distribution of $C$, where the assumption that $C$ is normally distributed is no longer valid.

\begin{figure}[!ht]
    \centering
    \includegraphics[width=0.55\columnwidth]{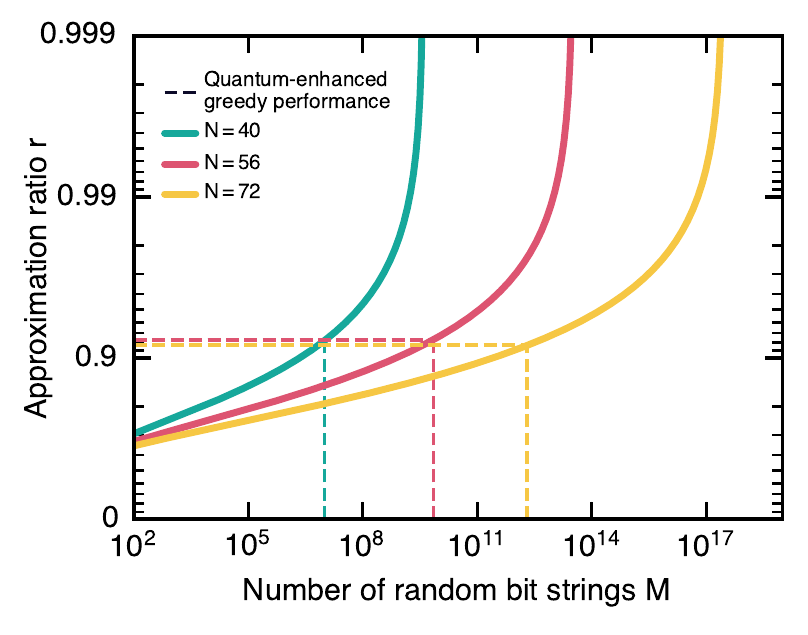} 
    \caption{\textbf{Performance of random sampling through the lens of extreme value statistics.} Approximation ratio $r$ based on Eq.~\eqref{eq:best_random_guess} as a function of the number $M$ of uniformly drawn random bit strings for different problem sizes $N=40$, $56$, and $72$. Dashed lines report the performance for the quantum-enhanced greedy algorithm and the corresponding number of bitstrings needed to be drawn at random on average to match that performance.}
    \label{fig:random_sampling}
\end{figure}

The quantum-enhanced algorithm has a total of $N$ iterative steps. For each of these steps, the embedded QAOA scanned a grid of angles of size $16\times 16$ and sampled $256$ bit strings for each pair of angles. This leads to a total of $N\times 16\times 16\times 256=2^{16}N$ bit strings generated in practice for a given problem instance. This is about $M\simeq 10^6$ bit strings generated for problem sizes $N=40$, $56$, and $72$. Generating at random this many bit strings and keeping the best one would lead to an average approximation ratio $r\simeq 0.8$-$0.9$ according to Eq.~\eqref{eq:best_random_guess} and as shown in Fig.~\ref{fig:random_sampling}. For comparison, achieving the average approximation of $r\simeq 0.92$ of the quantum-enhanced algorithm with random sampling would require generating an average $M\simeq 10^7$, $10^{10}$, and $M=10^{12}$ random bit strings for $N=40$, $56$, and $72$, respectively [Fig.~\ref{fig:random_sampling}].

\section{Finite-Size Classical Greedy Baseline for SK Problem Instances}

When provided with random bit strings, the quantum-enhanced greedy algorithm maps to a classical greedy algorithm for solving the optimization problem of interest. We proved in Methods that for an infinite-size Sherrington-Kirkpatrick (SK) instance, the classical greedy algorithm has an approximation ratio of $r\simeq 0.848497...$. Here, we run the algorithm on finite-size SK problems to get a more realistic number with respect to the finite sizes studied in this work.

\begin{figure}[!ht]
    \centering
    \includegraphics[width=0.9\columnwidth]{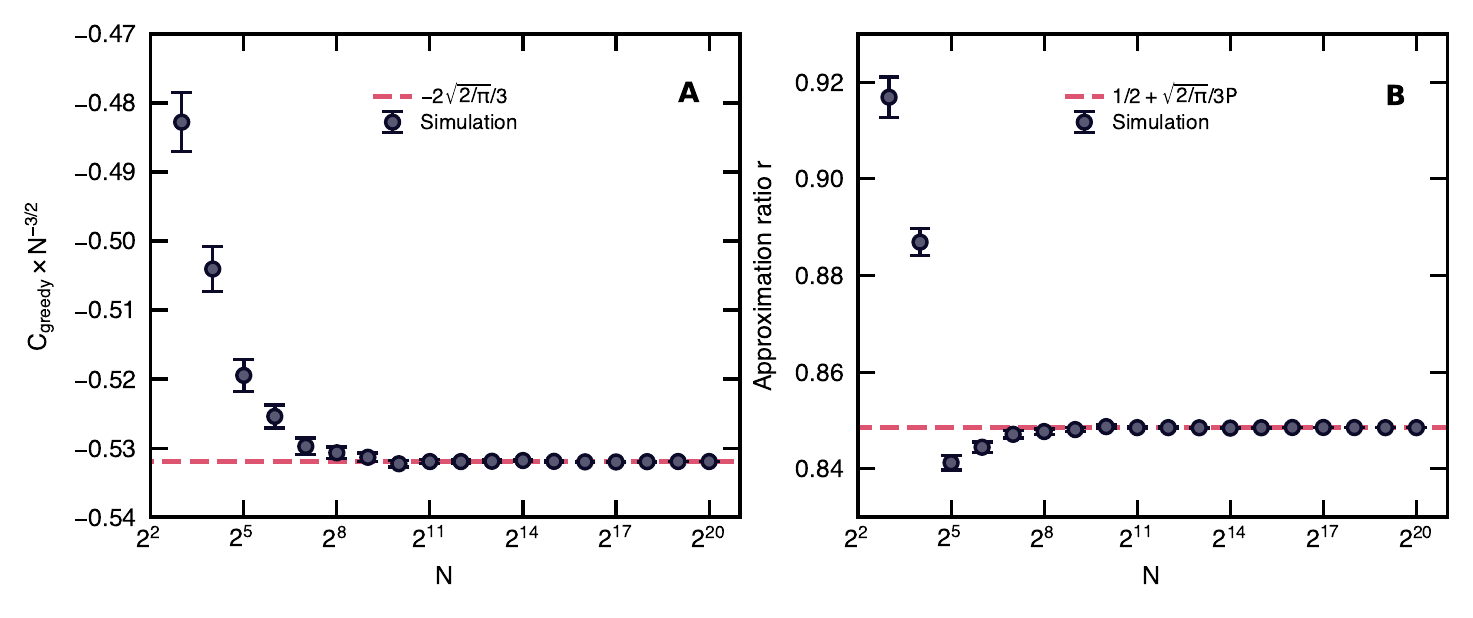} 
    \caption{\textbf{Performance of the classical greedy baseline.} \textsf{\textbf{A.}} Average ground state energy density obtained by the classical greedy algorithm on $1,000$ SK problem instances. As $N\to+\infty$, it converges to $-2\sqrt{2/\pi}/3$. \textsf{\textbf{B.}} Average approximation ratio $r$ versus the problem size $N$ for the classical greedy algorithm. Each data point is averaged over $1,000$ random SK problem instances. For $N\to+\infty$, we expect that  $r=1/2+\sqrt{2/\pi}/3\mathsf{P}\simeq 0.848497...$ where $\mathsf{P}$ is the Parisi constant (see Methods). For $N\leq 16$, the exact ground state is computed by brute force. For larger sizes, the ground state is approximated as described in Methods. Error bars indicate one standard deviation.}
    \label{fig:classical_greedy_vs_N}
\end{figure}

\section{Average Spectrum of Sherrington-Kirkpatrick Problem Instances}

We consider SK problem instances with random weights $\pm 1$. For each problem instance of size $N$, we iterate over all possible $2^N$ bit strings and store the corresponding cost. The nature of the weights $\pm 1$ strongly restricts the possible values of the cost, making it possible to store the costs $C\in\mathbb{Z}$ as a map $'\textrm{map}[C]=\mathbb{N}'$ with $\mathbb{N}$ counting the number of bit strings with the given cost, and which is updated during the iteration over the bit strings. The set of all costs is called the spectrum.

\begin{figure}[!ht]
    \centering
    \includegraphics[width=0.6\columnwidth]{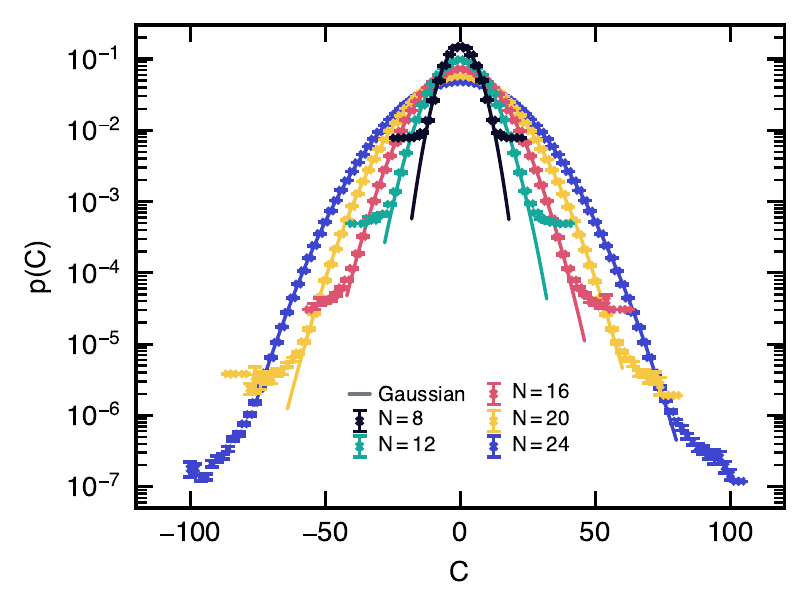} 
    \caption{\textbf{Spectrum of Sherrington-Kirkpatrick models.} For each problem size $N=8$, $12$, $16$, $20$, and $24$, we compute the spectra for $1,000$ randomly generated problems. The spectra are then averaged and plotted as a probability distribution. Error bars indicate one standard deviation.}
    \label{fig:distribution_cost_sk}
\end{figure}

For various problem sizes, we generate $1,000$ random problem instances. We compute their individual spectra, average them, and plot the result as a probability distribution $P(C)$ in Fig.~\ref{fig:distribution_cost_sk}. We find that it is symmetric around $C=0$ and that the bulk are fitted well by a Gaussian. This shows that, as expected, $C_\textrm{min}\simeq -C_\textrm{max}$ and that a randomly generated bit string would be symmetrically distributed between the two extrema of the spectrum, leading to an average approximation ratio $r\simeq 1/2$ for solving SK models by random sampling.

\begin{figure}[!ht]
    \centering
    \includegraphics[width=0.6\columnwidth]{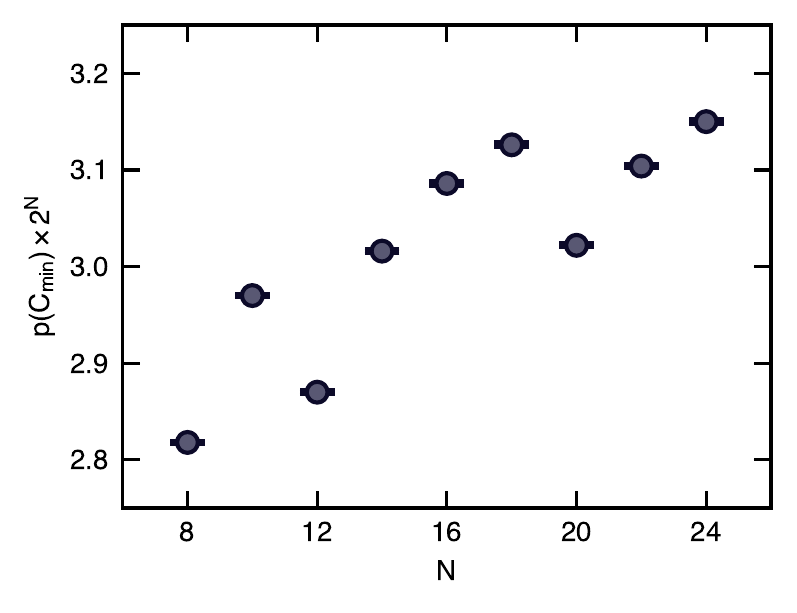} 
    \caption{\textbf{Degeneracy of the optimal solutions.} Probability of finding an optimal solution multiplied by the total number of valid solutions as a function of the problem size $N$. In other words, this is the average degeneracy of the optimal solution versus $N$. Each data point is averaged over $1,000$ randomly generated problem instances. Error bars indicate one standard deviation.}
    \label{fig:probability_opt_solution_vs_N}
\end{figure}

For various problem sizes, we generate $1,000$ random problem instances. For each of them, we count the number of bit strings having the optimal cost $C_\textrm{min}$. As shown in Fig.~\ref{fig:probability_opt_solution_vs_N}, the average (which is $\approx 3$) is roughly independent of $N$. Therefore, the ground states of SK $\pm 1$ problems are not massively degenerate and appear to occupy a fraction of the total space that is exponentially suppressed in $N$. Note that it is not possible to infer anything about the exact scaling with $N$ for large $N$ due to the small sizes accessible.

\section{Additional Details Regarding Matrix Product State Simulations}

\begin{figure}[!ht]
    \centering
    \includegraphics[width=0.45\columnwidth]{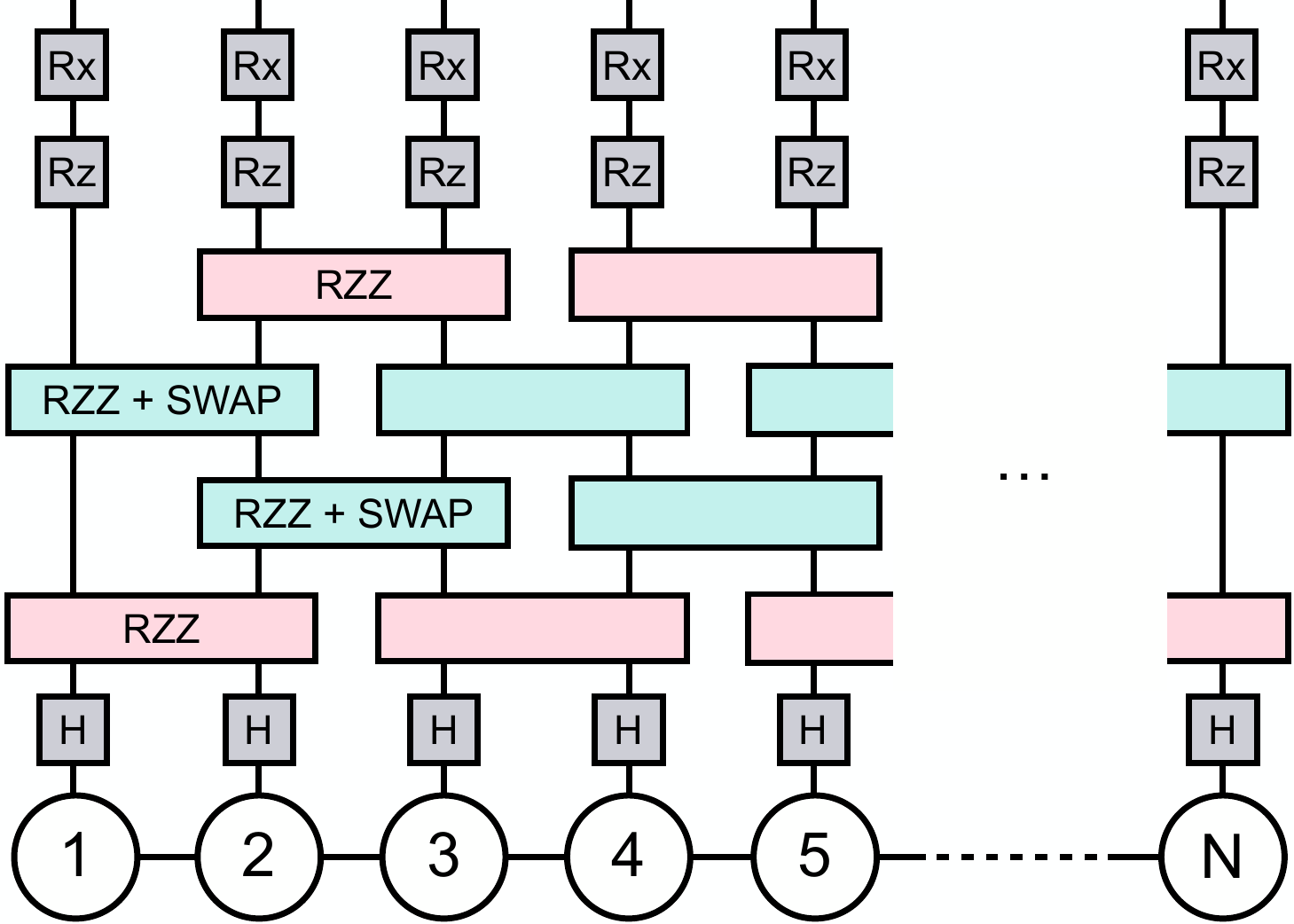} 
    \caption{\textbf{Quantum circuit simulated with matrix product states.} A graph problem of $N$ nodes is embedded at random onto the linear topology of the matrix product state simulating $N$ qubits~\cite{PhysRevLett.93.040502}. A total of four layers of two-qubit gates are used in a brick wall pattern on even/odd edges.}
    \label{fig:mps_circuit}
\end{figure}

The quantum-inspired classical algorithm of the main text simulates shallow circuits up to $N=72$ qubits using a tensor network approach based on matrix product states~\cite{PhysRevLett.93.040502}. Such circuits are  truncated one-layer QAOA circuit with two swap cycles embedded into a one-dimensional lattice with open boundary conditions, as shown in Fig.~\ref{fig:mps_circuit}. The circuits are shallow enough to be executed exactly with a relatively low bond dimension (the bond dimension is a control parameter of a matrix product state simulator), independent of the number of qubits involved. We embed the graph problem randomly onto the one-dimensional topology of the matrix product state. The truncated one-layer QAOA circuit of Fig.~\ref{fig:mps_circuit} loads at random a total of $2(N-1)$ edges of the graph problem (to be compared with the total number of edges $N(N-1)/2$). It has been shown that simulating, even approximately, generic QAOA circuits with matrix product states is hard~\cite{PRXQuantum.3.040339,PhysRevA.106.022423}.

\section{Data: Approximation Ratio Versus Problem Size}

\begin{table}[!ht]
    \centering
    \begin{tabular}{lcccccc}
        \thead{\textbf{Problem size}} & \thead{\textbf{Quantum}} & \thead{\textbf{Q-inspired}} & \thead{\textbf{Random}} & \thead{\textbf{QAOA$_1$}~\cite{Farhi2022quantumapproximate}} & \thead{\textbf{Greedy}} & \thead{\textbf{SDP}~\cite{Aizenman1987,Montanari2015,Bandeira2019}}\\
        \hline\\[-0.8em]
        \makecell{$N=8$} & \makecell{$0.97(2)$} & \makecell{$0.989(9)$} & \makecell{} & \makecell{} & \makecell{} & \makecell{}\\[0.3em]
        \hline\\[-0.8em]
        \makecell{$N=24$} & \makecell{$0.97(2)$} & \makecell{$0.959(5)$} & \makecell{} & \makecell{} & \makecell{} & \makecell{}\\[0.3em]
        \hline\\[-0.8em]
        \makecell{$N=40$} & \makecell{$0.93(2)$} & \makecell{$0.964(4)$} & \makecell{} & \makecell{} & \makecell{} & \makecell{}\\[0.3em]
        \hline\\[-0.8em]
        \makecell{$N=56$} & \makecell{$0.93(2)$} & \makecell{$0.963(4)$} & \makecell{} & \makecell{} & \makecell{} & \makecell{}\\[0.3em]
        \hline\\[-0.8em]
        \makecell{$N=72$} & \makecell{$0.92(1)$} & \makecell{$0.954(3)$} & \makecell{} & \makecell{} & \makecell{} & \makecell{}\\[0.3em]
        \hline\\[-0.8em]
        \makecell{$N=\infty$} & \makecell{} & \makecell{} & \makecell{$1/2$} & \makecell{$0.698688...$} & \makecell{$0.848497...$} & \makecell{$0.917090...$}\\[0.3em]
        \hline\\[-0.8em]
    \end{tabular}
    \caption{\textbf{Approximation ratio.} Approximation ratio $r$ for different methods and different problem sizes $N$. The data is the same as in the figure of the main text showing the approximation ratio versus the problem size.}
    \label{tab:data_approximation_ratio}
\end{table}

We provide in Table~\ref{tab:data_approximation_ratio} the data from the figure of the main text showing the approximation ratio versus the problem size.

\section{Exact Small-Scale Simulations with Nontruncated QAOA Circuits}

We worked with truncated QAOA circuits in the main text due to hardware limitations. Here, we provide additional data for the quantum-enhanced greedy algorithm based on exact small-scale classical simulations of the QAOA with a complete (nontruncated) circuit.

\begin{figure}[!ht]
    \centering
    \includegraphics[width=0.7\columnwidth]{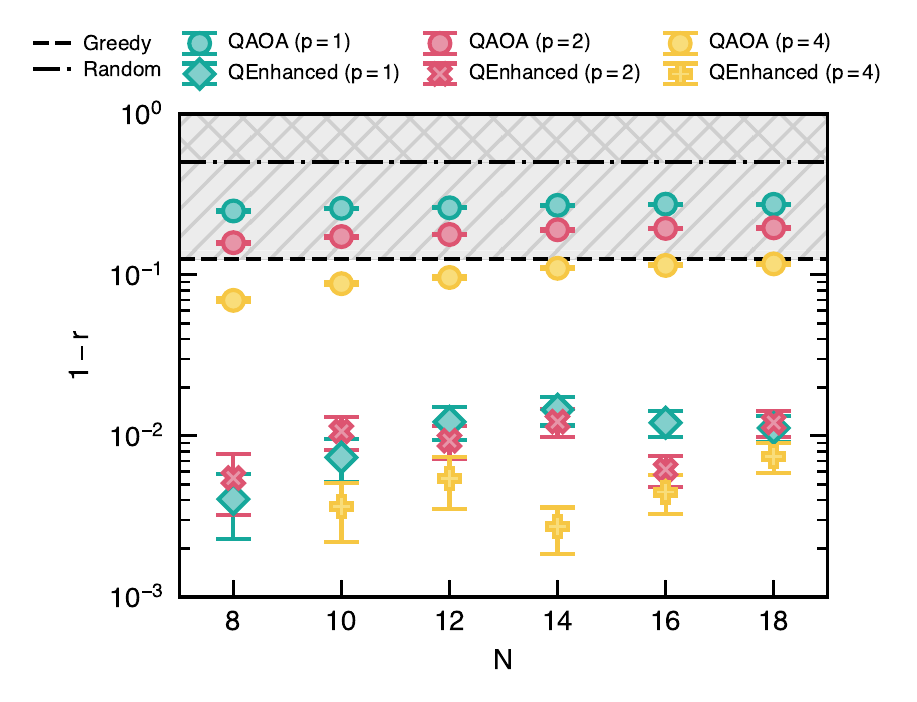} 
    \caption{\textbf{Performance of the quantum-enhanced greedy algorithm based on exact small-scale QAOA simulations.} $1-r$, with $r$ the average approximation ratio over $100$ random SK instances, as a function of the problem size $N$. For each of these instances, the complete (nontruncated) QAOA is executed with various number of layers $p=1$, $2$, and $4$. Results are shown for the QAOA and the quantum-enhanced greedy algorithm. Two baselines are represented by the hatched regions: The classical greedy baseline with $1-r\simeq 0.12$ for the largest sizes considered (see Fig.~\ref{fig:classical_greedy_vs_N}) and the random sampling baseline with $1-r=1/2$. Error bars indicate one standard deviation.}
    \label{fig:exact_small_simulation_vs_N}
\end{figure}

Data are plotted in Fig.~\ref{fig:exact_small_simulation_vs_N}. The performance of the QAOA increases with the number of layers $p$. It is systematically above the random sampling baseline $1-r=1/2$, but barely passes the classical greedy baseline $1-r\simeq 0.12$ at $p=4$, and achieves systematically worse performance for $p<4$. On the other hand, the quantum-enhanced greedy algorithm performs better than the classical greedy baseline. In addition, we observe that improving the performance of the underlying QAOA by increasing the number of layers $p$ leads to an improvement in the performance of the quantum-enhanced greedy algorithm. The improvement over the QAOA or the classical greedy baseline is between $\times 10$ and $\times 50$.

\section{Quantumness in the Second Versus the Third Step of the Iterative Algorithm}

In the main text, we studied the case of the quantum-enhanced algorithm with quantum-generated bit strings, and the randomized classical greedy algorithm with randomly generated bit strings. Here, we consider a mix-and-match scenario for step $2$ (selection) and step $3$ (decision) of the algorithm, where in each of these steps, the bit strings may be generated randomly or from a quantum computer. Specifically, we used quantumly generated bit strings in both steps 2 and 3 in the quantum algorithm in the main text, and using randomly generated bit strings in both steps 2 and 3 maps to the randomized classical greedy algorithm in the main text. Here, we additionally consider two mores cases: one case where we use quantumly generated bit strings for the selection step and randomly generated bit strings for the freezing step, and another case with vice versa.

\begin{figure}[!ht]
    \centering
    \includegraphics[width=0.6\columnwidth]{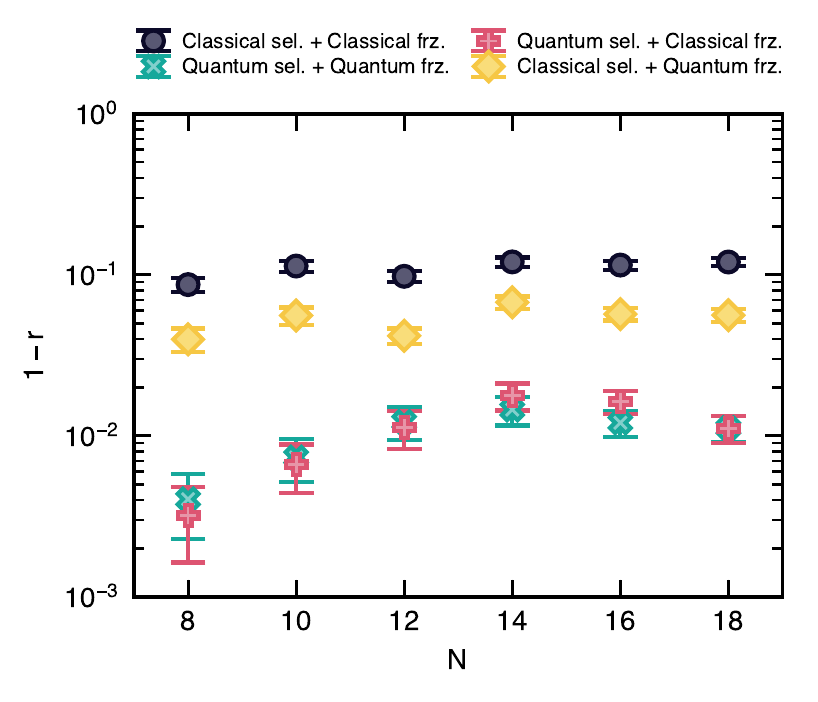} 
    \caption{\textbf{Contribution of bit string selection and freezing steps to the performance of the quantum-enhanced greedy algorithm.} The algorithm includes steps which select and freeze bit strings which can randomly generated or be the output of a quantum computer. The greedy baseline corresponds to the bit strings being random in both steps while the algorithm works with the output of a quantum computer in both steps. Here, the intermediate scenarios are considered. The data are $1-r$, with $r$ the average approximation ratio over a $100$ random SK instances, as a function of the problem size $N$. For each of these instances, the complete (nontruncated) QAOA is emulated classically a single layer ($p=1$). Error bars indicate one standard deviation.}
    \label{fig:step_contribution_vs_N}
\end{figure}

We consider exact small-scale simulations using QAOA with one layer ($p=1$). Data are plotted in Fig.~\ref{fig:step_contribution_vs_N} and show that the use of quantum-generated bit strings is comparatively very beneficial to inform the selection step. Using random or quantum-generated bit strings in the freezing step give nearly equal answers when the selection step is done with quantum-generated bit strings. We explain this as follows. If a variable $i$ is selected, and we use random bit strings in the freezing step, expectation values of all observables that do not involve $Z_i$ are 0. Therefore, the only relevant nonzero term in the cost is $v_i Z_i$, thus $Z_i$ is frozen to $Z_i = -{\rm sign}(v_i)$. If instead, quantum-generated bit strings are used in the freezing step, the relevant terms in the cost are both one-body term $v_i Z_i$ and two-body terms $|w_{ij} Z_i \langle Z_j\rangle|$. The two-body terms may be nonzero; However, they tend to be small relative to $|v_i|$ in the intermediate iterative steps. Therefore, predominantly, the freezing again leads to $Z_i = -{\rm sign}(v_i)$. This need not be true at arbitrary $p$, but appears to be true at $p=1$.

\section{Performance Versus Runtime}

We investigate the performance versus runtime of different algorithms for a single typical $N=72$ variables Sherrington-Kirkpatrick instance with random $\pm 1$ weights.

We implement classical heuristics such as tabu search~\cite{Glover1998} and simulated annealing~\cite{Kirkpatrick1983,Kirkpatrick1984}, a classical semidefinite programming solver~\cite{Aizenman1987,Montanari2015,Bandeira2019}, the classical randomized greedy algorithm developed in this work as well as its quantum-enhanced version executed on quantum hardware and executed through classical matrix product state simulations (see the main text and the Methods). We also report data for the truncated one-layer QAOA executed at the first iterative step of the quantum-enhanced greedy algorithm.

\begin{figure}[!ht]
    \centering
    \includegraphics[width=0.6\columnwidth]{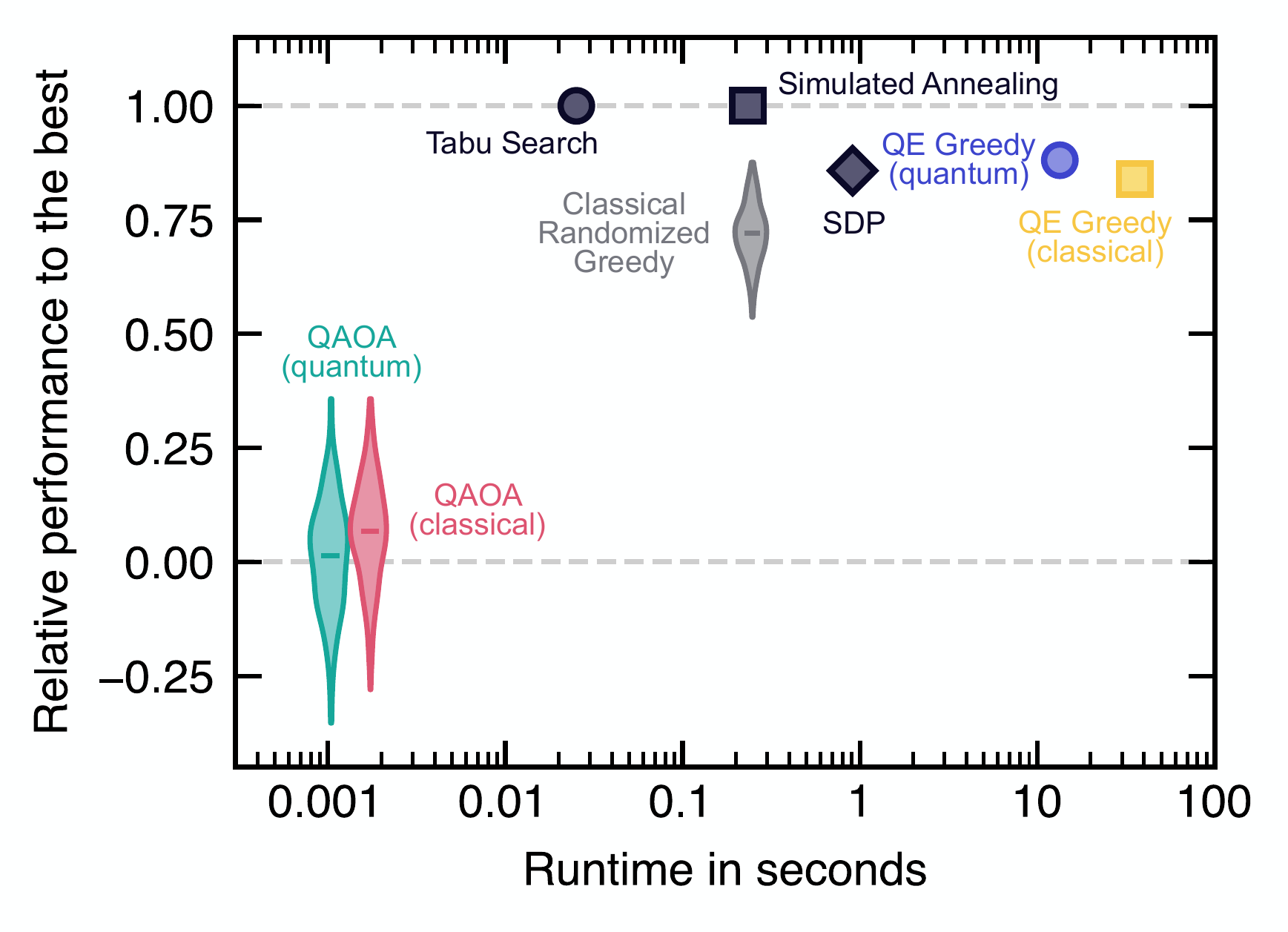} 
    \caption{\textbf{Relative performance versus runtime for different approaches.} A typical $N=72$ variables Sherrington-Kirkpatrick instance with random $\pm 1$ weights is considered. Classical execution is performed on a single core of an Intel Skylake E$5$-$2686$ v$5$ $3.1$GHz processor.}
    \label{fig:performance_vs_runtime}
\end{figure}

Data are plotted in Fig.~\ref{fig:performance_vs_runtime}. Tabu search and simulated annealing find the best answer in the least amount of time, with standard parameterization. The classical randomized greedy algorithm is a randomized algorithm that will return a different solution at every run. The distribution (over $256$ independent runs) shows the distribution of solutions and the runtime is that of obtaining a single solution. For the truncated one-layer QAOA, we also display the distribution of solutions (over $256$ shots), assuming optimal angles. The runtime is that of obtaining a single bit string by running a single QAOA circuit at optimal angles. Quantum refers to running on Rigetti Aspen-M-3 superconducting processor and classical to quantum-inspired classical simulations based on matrix product states (see the Methods). Finally, the quantum-enhanced greedy algorithm is displayed: The iterative process with the freezing of a single variable at a time requires running the QAOA $N=72$ times to get to the final solution. At each step, we assume we know the optimal angles. The runtime is that of collecting $256$ shots for each of the $72$ iterative steps at optimal angles. We note that for this specific problem instance, the quantum run returned a better solution than the classical (quantum-inspired) simulation.

A gap of a few orders of magnitude need to be closed for the quantum-enhanced greedy algorithm to be competitive with state-of-the-art classical methods. As noted in the main text, the complexity of the quantum-enhanced greedy algorithm is $O[(N/K)N_\textrm{edges}]$ with $N_\textrm{edges}$ the number of two-body terms in the graph problem and $K$ the number of variables frozen at each iteration step. Here, we emphasize that minimizing runtime was not a goal of this work. Reducing the runtime of the underlying QAOA is one direction: One could imagine an adaptive scheme for setting the number of shots at each iterative step. Another direction would seek to reduce the length of the iterative process by, e.g., implementing freezing strategies considering multiple variables at a time ($K$ of them at once) and brute-forcing the problem once it becomes small enough. We believe this would bring for $N=72$ the quantum-enhanced greedy algorithm in the $0.1$ to $1$ seconds range of runtime. Moreover, SK problem instances have the largest possible number of edges $N_\textrm{edges}\sim O(N^2)$ because of the all-to-all connectivity. Sparser graphs with, e.g., $N_\textrm{edges}\sim O(N)$, might lead to a better absolute and relative runtime for the quantum-enhanced greedy algorithm.

\section{Dealing with Hard Constraints by Example: Portfolio Optimization}

\subsection{Problem Definition}

We take as an example the portfolio optimization problems of Ref.~\cite{Herman2022}. The goal is to build a portfolio maximizing the potential return while minimizing the volatility from a given basket of $N$ assets. Whether an asset is selected for the portfolio is encoded as a binary variable. Without entering into the details (refer to Ref.~\cite{Herman2022}), the problem takes the form of minimizing an objective function with binary variables and nonzero scalar parameters $v_i$ and $w_{ij}$, as per the definition of the cost function (Eq.~\eqref{eq:objective_f}) in the main text. In addition to the minimization, there are two hard constraints on what makes a valid portfolio. They are of the form,
\begin{equation}
    \sum\nolimits_{i=1}^NZ_i \leq A,~~~~\textrm{and}~~~~\sum\nolimits_{i=1}^N\mu_iZ_i \geq B,~~~~\textrm{with}~A,B,\mu_i\in\mathbb{R}.
    \label{eq:portfolio_constraints}
\end{equation}
The first inequality constrains the maximum size of the portfolio and the second one the minimum expected return from the portfolio. The parameters $A$, $B$, and $\mu_i$ are provided as part of the problem.

\subsection{Hard Constraints in the Quantum-Enhanced Greedy Algorithm}

The hard constraints of Eq.~\eqref{eq:portfolio_constraints} can be enforced classically as part of the quantum-enhanced greedy algorithm, thus ensuring that the final bit string is necessarily a valid solution. Moreover, it is not required to introduce auxiliary slack variables to map the inequalities into equalities, making the algorithm even more friendly for near term quantum devices.

The algorithm presented in the main text is modified as follows to account for the constraints. In the freezing step, a variable otherwise identified as being a candidate for freezing, will not be frozen if it would take the potential final solution out of the space of valid bit strings. However, this may only be possible for certain types of constraints, including those of Eq.~\eqref{eq:portfolio_constraints}. Different strategies can be envisioned to adapt the freezing decision process. For instance, if freezing to a classical value is not possible, then the other value can be selected despite its higher cost. One could also go back to the election step and select another variable for freezing.

Whereas the hard-constraints will be fulfilled independently of the underlying quantum algorithm (e.g., QAOA or adiabatic quantum evolution), it might be advantageous for these algorithms to favor in-constraint bit strings for their output. A possible approach is to design quantum circuits working within the in-constraint space, as discussed in general in Ref.~\cite{Hadfield2019} and more specifically for Hamming constrained problems in Ref.~\cite{larose2022mixer}, but it is practically infeasible on current quantum hardware due to noise. Another typical approach uses penalty terms that will disfavor the appearance of out-of-constraint bit strings, but implementing them on near-term devices and tuning their strength can be challenging. Another strategy might be to classically filter the out-of-constraint bit strings returned by the quantum algorithm, if any. However, this requires that the density of valid bit strings is sufficiently high.

\subsection{Results}

\begin{figure}[!ht]
    \centering
    \includegraphics[width=0.6\columnwidth]{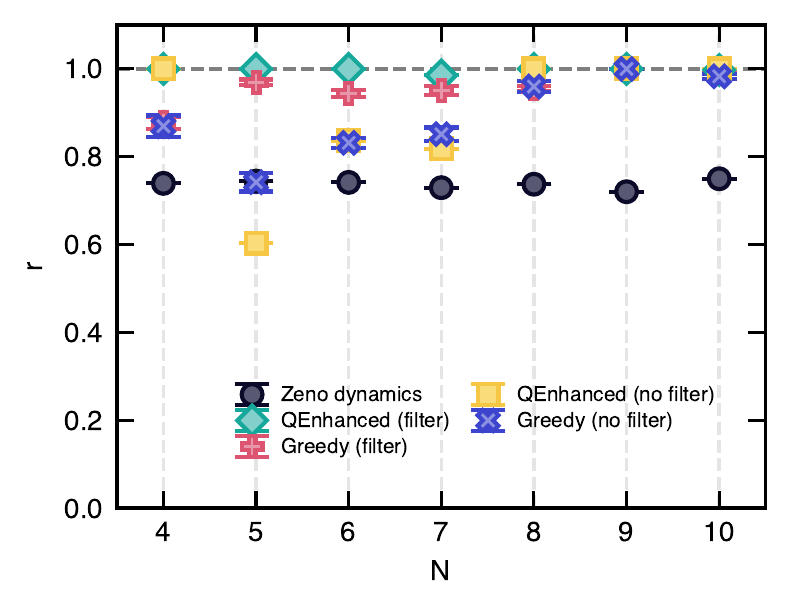} 
    \caption{\textbf{Performance of different approaches for constrained portfolio optimization problems from Ref.~\cite{Herman2022}.} Average approximation ratio for different problem sizes $N$ computed from classical emulations. Each size $N$ corresponds to a given problem provided by Ref.~\cite{Herman2022}. The Zeno dynamics data points were extracted from Ref.~\cite{Herman2022} and serve as a reference. The performance of the quantum-enhanced greedy algorithm (based on a one-layer QAOA circuit) and its classical greedy counterpart are also displayed, with and without filtering. The classical greedy baseline data points are averaged over $100$ trials. Error bars indicate one standard deviation.}
    \label{fig:portfolio_optimization}
\end{figure}

Ref.~\cite{Herman2022} introduced Zeno dynamics embedded into QAOA to deal with hard-constraints in the form of Eq.~\eqref{eq:portfolio_constraints}. The authors generated a problem of each size $N=4,\ldots,10$ and solved it using a one-layer QAOA circuit with Zeno dynamics. The corresponding data points are reported in Fig.~\ref{fig:portfolio_optimization}, with an approximation ratio of about $r\simeq 0.74$, independent of $N$. Here, the approximation ratio is defined with respect to the best and worst solutions of the in-constraint space.

On the exact same set of problems, we run the quantum-enhanced greedy algorithm. We use a one-layer QAOA circuit. We use two-body expectation values to inform the selection process. The freezing decision is such that if setting a variable to a classical value makes the solution invalid, the other value is selected despite its higher cost. The constraints on all problem instances are relatively loose since bit strings drawn at random have more than a $50\%$ chance of being valid. This makes it possible to classically filter bit strings outputted from the QAOA. Here, filtering means discarding bit strings not fulfilling the hard constraints of Eq.~\eqref{eq:portfolio_constraints}. For example, an $N-$bit string $\{1,1,1,1,\ldots1\}$ will not fulfill the first constraint of Eq.~\eqref{eq:portfolio_constraints} $\sum_{i=1}^NZ_i\leq A$ for $A=N/2$ as $\sum_{i=1}^NZ_i=N$. Thu, such a bit string would be discarded.

We try the filtered and unfiltered versions as well as the classical greedy baseline where bit strings are generated at random. Results are plotted in Fig.~\ref{fig:portfolio_optimization}. We find that filtering bit strings leads to better performance. We also find that the classical greedy baseline systematically outperforms the one-layer QAOA circuit with Zeno dynamics data and that it has a very high approximation ratio $r\gtrsim 0.95$ for the sizes considered.

\subsection{Outlook}

It would be precipitate to draw conclusions about the absolute performance of the quantum-enhanced algorithm in the presence of constraints since: (i) the classical greedy baseline performs very well, making it difficult to discern any significant improvements; and (ii) the problem set is very limited, with just one instance per problem size and all problems being relatively small ($N\leq 10$). Further work is required to extend the analysis to a statistically significant problem set. Moreover, it would be interesting to explore the different parameters entering the quantum-enhanced algorithm. Tighter constraints would not enable filtering out bit strings and it would be worthwhile to explore problems (not necessarily related to portfolio optimization) for which sampling the valid space of solutions is difficult due to the problem size. Finally, we note that a QAOA circuit with Zeno dynamics could be embedded in the quantum-enhanced algorithm presented here.

\section{Classical Greedy Baseline for Other Graphs}

The randomized greedy algorithm is iterative from step $\ell=1$ to $\ell=N$ for a graph with $N$ nodes. At step $\ell$, we select an active variable at random. This means that there are $N-1$ other variables in the problem, including $\ell-1$ frozen ones and $N-\ell$ active ones. The goal is to find the classical value to which freeze this variable to minimize the objective function [Eq.~\eqref{eq:objective_f}]. In the following, we show how one can derive the classical greedy baseline analytically for other problems than the SK instances considered in the main text. The list is nonexhaustive.

\subsection{Ring Graph with Random \texorpdfstring{$\pm 1$}{±1} weights}

We consider a ring graph with random weights $w_{i,i+1}=\pm 1$ on the edges [Eq.~\eqref{eq:objective_f}]. We now compute the average approximation of the classical greedy baseline for this problem following the same ideas developed for SK problem instances in Methods. For the freezing decision of a randomly selected node, we need to know the probabilities that (a) none of its nearest-neighbors are frozen, that (b) one of its nearest-neighbor is frozen, and that (c) both of its nearest-neighbors are frozen. At step $\ell$, these probabilities read,
\begin{align}
    p_a\bigl(N,\ell\bigr)&=\left(\frac{N-\ell}{N-1}\right)\left(\frac{N-\ell-1}{N-2}\right),~~(\ell\geq 1)\\
    p_b\bigl(N,\ell\bigr)&=2\left(\frac{\ell-1}{N-1}\right)\left(\frac{N-\ell}{N-2}\right),~~(\ell\geq 2)\\
    p_c\bigl(N,\ell\bigr)&=\left(\frac{\ell-1}{N-1}\right)\left(\frac{\ell-2}{N-2}\right)~~(\ell\geq 3),
\end{align}
where $p_a+p_b+p_c=1$. The final value of the objective function takes the general form,
\begin{equation}
    C_\textrm{greedy}=\sum\nolimits_{\ell=1}^NC_\ell=C_1+C_2+\sum\nolimits_{\ell=3}^NC_\ell,
\end{equation}
where, on average, $C_1=0$, $C_2=-p_b(N,\ell=2)$, and $C_{\ell\geq 3}= - p_b(N,\ell) - p_c\bigl(N,\ell\bigr)$. Hence, on average, the expectation value obtained for the objective function with the greedy algorithm on the ring graph is,
\begin{equation}
    C_\textrm{greedy}=-p_b(N,\ell=2) - \sum\nolimits_{\ell=3}^N\Biggl[p_b(N,\ell) + p_c\bigl(N,\ell\bigr)\Biggr]=-\frac{2N}{3}.
\end{equation}
Since the problem is trivially solved with $C_\textrm{min}=-C_\textrm{max}=-N$, this leads to an average approximation ratio of,
\begin{equation}
    r=\frac{1}{2}\left(1+\frac{C_\textrm{greedy}}{C_\textrm{min}}\right)=\frac{5}{6}\simeq 0.833333...
\end{equation}

\subsection{Random \texorpdfstring{$3$}{3}-Regular Graphs with Random \texorpdfstring{$\pm 1$}{±1} weights}

We employ the same strategy than for the ring graph to solve random $3$-regular graphs with uniform random weights $w_{ij}=\pm 1$ on the edges $\{(i,j)\}$ [Eq.~\eqref{eq:objective_f}]. In a $3$-regular graph, each node has three neighbors, and we assume $N\geq 3$. Therefore, for the freezing decision of a randomly selected node, we need to know the probabilities that (a) none of its nearest-neighbors are frozen, that (b) one of its nearest-neighbor is frozen, that (c) two of its nearest-neighbors are frozen, and that (d) all of its nearest-neighbors are frozen. At step $\ell$, these probabilities read,
\begin{align}
    p_a\bigl(N,\ell\bigr)&=\left(\frac{N-\ell}{N-1}\right)\left(\frac{N-\ell-1}{N-2}\right)\left(\frac{N-\ell-2}{N-3}\right),~~(\ell\geq 1)\\
    p_b\bigl(N,\ell\bigr)&=3\left(\frac{\ell-1}{N-1}\right)\left(\frac{N-\ell}{N-2}\right)\left(\frac{N-\ell-1}{N-3}\right),~~(\ell\geq 2)\\
    p_c\bigl(N,\ell\bigr)&=3\left(\frac{\ell-1}{N-1}\right)\left(\frac{\ell-2}{N-2}\right)\left(\frac{N-\ell}{N-3}\right),~~(\ell\geq 3)\\
    p_d\bigl(N,\ell\bigr)&=\left(\frac{\ell-1}{N-1}\right)\left(\frac{\ell-2}{N-2}\right)\left(\frac{\ell-3}{N-3}\right)~~(\ell\geq 4),
\end{align}
where $p_a+p_b+p_c+p_d=1$. The final value of the objective function takes the general form,
\begin{equation}
    C_\textrm{greedy}=\sum\nolimits_{\ell=1}^NC_\ell=C_1+C_2+C_3+\sum\nolimits_{\ell=4}^NC_\ell,
\end{equation}
where, on average, $C_1=0$, $C_2=-p_b(N,\ell=2)$, $C_3= - p_b(N,\ell=3) - p_c\bigl(N,\ell=3\bigr)$, and $C_{\ell\geq 4}= - p_b(N,\ell) - p_c\bigl(N,\ell\bigr) - 3p_d\bigl(N,\ell\bigr)/2$. Hence, on average, the expectation value obtained for the objective function with the greedy algorithm on random $3$-regular graphs is,
\begin{align}
    C_\textrm{greedy}=&- p_b(N,\ell=2) - p_b(N,\ell=3) - p_c(N,\ell=3)\nonumber\\
    &-\sum\nolimits_{\ell=4}^N\Biggl[p_b(N,\ell) + p_c\bigl(N,\ell\bigr) + 3p_d\bigl(N,\ell\bigr)/2\Biggr]=-\frac{7N}{8}
\end{align}

\subsection{Numerical simulations}

\begin{figure}[!ht]
    \centering
    \includegraphics[width=1\columnwidth]{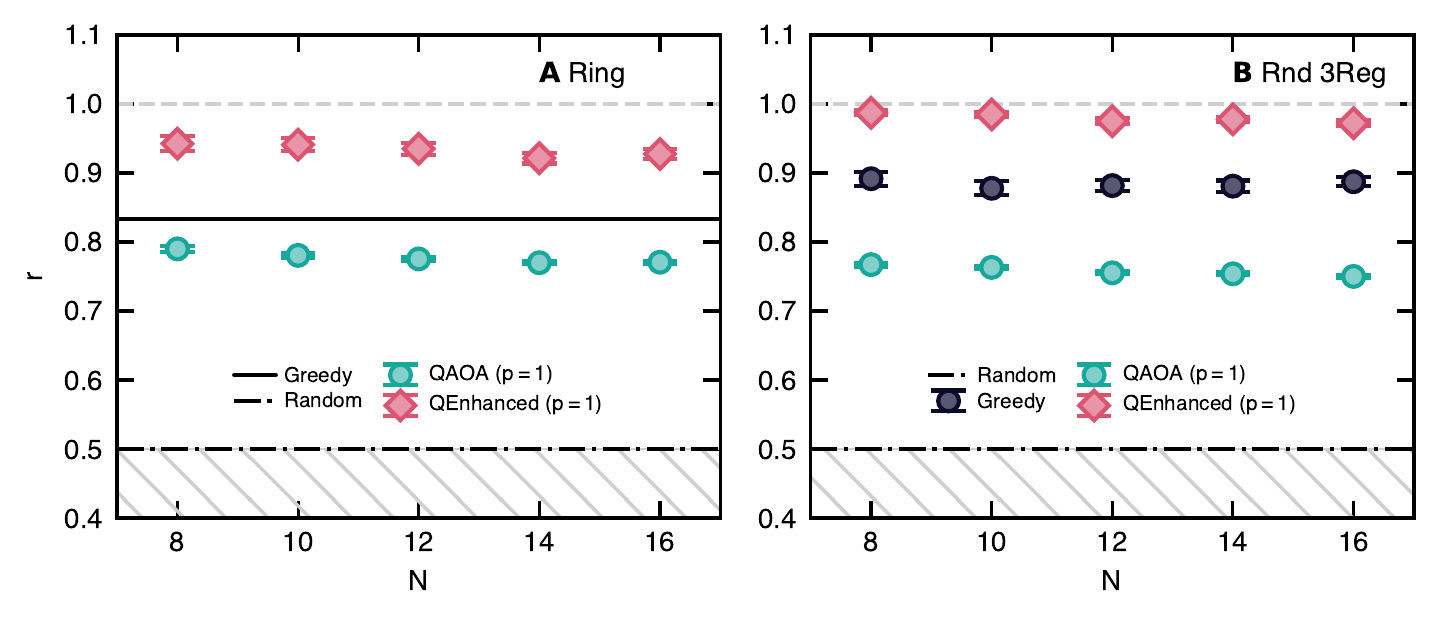} 
    \caption{\textbf{Exact small-scale simulations for ring and random $3$-regular graphs.} Approximation ratio $r$, averaged over $100$ random problem instances, as a function of the problem size $N$. For each of these instances, a nontruncated one-layer $(p=1)$ QAOA is executed. Error bars indicate one standard deviation. \textsf{\textbf{A.}} Ring graphs with random $\pm 1$ weights. \textsf{\textbf{B}} Random $3$-regular graphs with random $\pm 1$ weights.}
    \label{fig:other_graphs_vs_N}
\end{figure}

We perform small-scale noiseless simulations of the quantum-enhanced greedy algorithm for the ring and random $3$-regular graphs with random $\pm 1$ weights. The simulations are based on a nontruncated one-layer ($p=1$) QAOA circuit and averaged over $100$ randomly generated problem instances. The classical randomized greedy baseline is also displayed. Results are reported in Fig.~\ref{fig:other_graphs_vs_N} with the quantum-enhanced greedy algorithm systematically outperforming the classical greedy algorithm.

\end{document}